\documentclass[twocolumn]{autart}

\usepackage{hyperref}                     
\usepackage{mathtools,array} 
\usepackage{flushend}
\usepackage{amssymb,amsmath}
\usepackage{enumitem}
\usepackage{etoolbox}
\usepackage{algorithm}
\usepackage{graphicx} 
\usepackage[square,numbers]{natbib}
\usepackage{mathtools}
\usepackage{subcaption}

\usepackage{todonotes}

\usepackage{stackengine}
\def\delequal{\mathrel{\ensurestackMath{\stackon[1pt]{=}{\scriptstyle\Delta}}}}

\errorcontextlines\maxdimen
\makeatletter

\newtheorem{definition}{Definition}
\newtheorem{theorem}{Theorem}
\newtheorem{remark}{Remark}

\newtheorem{lemma}{Lemma}
\newtheorem{proposition}{Proposition}
\newtheorem{corollary}{Corollary}
\newtheorem{example}{Example}

\newcommand*{\rom}[1]{\expandafter\@slowromancap\romannumeral #1@}
\newcommand\norm[1]{\left\lVert#1\right\rVert}
\newcommand*\xoverline[2][0.9]{
    \sbox{\myboxA}{$\m@th#2$}
    \setbox\myboxB\null
    \ht\myboxB=\ht\myboxA
    \dp\myboxB=\dp\myboxA
    \wd\myboxB=#1\wd\myboxA
    \sbox\myboxB{$\m@th\overline{\copy\myboxB}$} 
    
    \setlength\mylenA{\the\wd\myboxA}
    
    \addtolength\mylenA{-\the\wd\myboxB}
    \ifdim\wd\myboxB<\wd\myboxA
      \rlap{\hskip 0.5\mylenA\usebox\myboxB}{\usebox\myboxA}
    \else
        \hskip -0.5\mylenA\rlap{\usebox\myboxA}{\hskip 0.5\mylenA\usebox\myboxB}
    \fi}
\makeatother

\begin{document}

\begin{frontmatter}

\title{Attack-Resilient State Estimation with Intermittent Data Authentication\thanksref{footnoteinfo}} 

\thanks[footnoteinfo]{This work is sponsored in part by the ONR agreements N00014-17-1-2504 and N00014-20-1-2745, AFOSR award FA9550-19-1-0169, and the NSF CNS-1652544 award.
This paper was not presented at any IFAC meeting; preliminary version of some of these results were presented in~\cite{khazraei_acc20}. 
}
\author[a]{Amir Khazraei}\ead{amir.khazraei@duke.edu},   
\author[a]{Miroslav Pajic}\ead{miroslav.pajic@duke.edu}

\address[a]{Department of Electrical and Computer Engineering, Duke University, Durham, NC 27708, USA}

\begin{keyword}                          
Security of control systems; Cyber-physical systems; Attack-resilient state estimation; Perfect attackability;         
\end{keyword}      

\begin{abstract}                         
Network-based attacks on control systems may alter sensor data delivered to the controller, effectively causing degradation in control performance. As a result, having access to accurate state estimates, even in the presence of attacks on sensor measurements, is of critical importance. 
In this paper, we analyze performance of resilient state estimators (RSEs) when \emph{any} subset of sensors may be compromised by a \emph{stealthy attacker}. Specifically, we consider systems with the well-known $l_0$-based RSE and two commonly used sound intrusion detectors (IDs).   
For linear time-invariant plants with bounded noise, we define the notion of perfect attackability (PA) when attacks may result in unbounded estimation errors while remaining undetected by the employed ID (i.e., stealthy).
%
%
We derive necessary and sufficient PA conditions, 
showing that a system can be perfectly attackable even if the plant is stable. 
While PA can be prevented with the use the standard cryptographic mechanisms (e.g., message authentication) that ensure data integrity under network-based attacks, their continuous use imposes significant communication and computational overhead.
Consequently, we also study the impact that even intermittent use of data authentication has on RSE performance guarantees in the presence of stealthy attacks.
We show that if messages from some of the sensors are even intermittently authenticated, stealthy attacks could not result in unbounded state estimation~errors.
\end{abstract}

\end{frontmatter}

\section{Introduction}  \label{sec:intro} 
The challenge of securing control systems has recently attracted significant attention due to high profile attacks, such as the attack on Ukrainian power
grid~\cite{zetter2016inside} and the StuxNet attack~\cite{langner2011stuxnet}. In such incidents, the attacker can affect a physical plant by altering actuation commands or sensory measurements, or affecting execution of the controller.  
%
One approach to address this problem has been to exploit a dynamical model of the plant for attack detection and attack-resilient control (e.g., \cite{mo2009secure,mo2010false, smith2015covert,fawzi2014secure,shoukry2017secure,pajic_tcns17,pajic_csm17,teixeira2015secure,khazraei2017new}). 

For instance, consider the problem of attack-resilient control when measurements from a subset of the plant sensors may be compromised. 
One line of work employs a widely used (non-resilient) Kalman filter, with a standard residual-based probabilistic detector (e.g., $\mathcal{X}^2$ detector) triggering alarm in the presence of attack~\cite{mo2010false,kwon2014analysis,jovanov_tac19}. These Kalman filter-based controllers of linear time-invariant (LTI)  plants may be vulnerable to stealthy (i.e., undetected) attacks resulting in unbounded state-estimation errors; thus, such systems are referred to as \emph{perfectly attackable} (PA)~\cite{mo2010false,kwon2014analysis,jovanov_tac19}. 
Specifically, for LTI systems with Gaussian noise and Kalman filter-based controllers, the notion of perfect attackability (PA)\footnote{For conciseness, we use PA for \emph{perfect attackability} or \emph{perfectly attackable}, when the meaning is clear from the context.} is introduced in~\cite{mo2010false}. 
In particular,~\cite{mo2010false}, and~\cite{jovanov_tac19} for larger classes of intrusion detectors (IDs), show that the system is 
PA if and only if the plant is unstable and the set of compromised sensors satisfies that no unstable eigenvector lie in the kernel of their observation~matrix.


Resilient (i.e., secure) state estimation is another approach to achieve attack-resilient control; here, the objective is to estimate the system state when a subset of the sensors is corrupted~\cite{fawzi2014secure,pasqualetti2013attack}. This allows for the use of standard feedback controllers to provide strong control guarantees in the presence of attacks. 
%
A common approach is to use a batch-processing resilient state estimator (RSE) to estimate the system state and attack vectors (e.g., \cite{fawzi2014secure,pajic_csm17,pajic_tcns17,shoukry2017secure}). For LTI systems without noise, 
the state and attack vectors can be obtained by solving an $l_0$, or under more restrictive condition $l_1$, optimization problem~\cite{fawzi2014secure}. These results are extended to systems with bounded noise~\cite{pajic_tcns17}, showing that the worst case state estimation error is a linear with the noise size; thus, the attacker cannot exploit the noise to introduce unbounded state-estimation errors, unless a sufficiently large number of sensors is corrupted. 
%
SMT- and graph-based estimators from~\cite{shoukry2017secure} and~\cite{luo2019scalable} improve computational efficiency of the estimators. 
However, all these methods employ a common restrictive assumption that the maximal number of corrupted sensors is bounded; at best, less than half of sensors can be compromised.~Moreover, to the best of our knowledge, the impact of stealthy attacks on the RSEs has not been considered, either in the general case or under such restrictive assumptions. 
 
However, the assumption that measurements from only a subset of sensors are compromised cannot be justified in the common scenarios where the attacker has access to the network used to transmit data from sensors to the controller. Thus, it is important to analyze impact of such Man-in-the-Middle attacks on performance of the RSEs. A common defense against network-based attacks is the use of cryptographic tools, such as adding Message Authentication Codes (MACs) to measurement messages to guarantee their integrity. 
%
%
Yet, continuous use of security primitives such as MACs, can cause computation and communication overhead, which limits its applicability in resource-constrained control systems~\cite{lesi_tecs17,lesi_tcps20}. 
To overcome this, intermittent data authentication can be used for control systems~\cite{jovanov_tac19};  specifically, LTI systems with Gaussian noise and a Kalman filter-based controller, 
cannot be PA if message authentication is at least intermittently employed. On the other hand, no such guarantees have been shown in systems with bounded-size noise and RSE-based controllers. 


Consequently, in this work, we focus on performance of LTI systems with bounded-size noise, employing
an RSE-based controller, under stealthy attacks on an \emph{arbitrary number} of sensors.
Specifically, we consider a system with an $l_0$-based RSE, due to the strongest resiliency guarantees, and one of two previously reported intrusion detectors (IDs) for systems with set-based noise. 
Due to the batch-processing nature of RSEs, we introduce two notions of PA for such systems -- at a single time point and over time, where a stealthy attacker may introduce arbitrarily large estimation errors. 
Then, we provide necessary and sufficient conditions for both notions of PA. 
We show that unlike PA in the Kalman filter-based estimators, 
a system may be PA over time even if the physical plant is not unstable.
%
Furthermore, we show that even intermittent data authentication guarantee can help against such perfect attacks for some types of IDs. 
Unlike~\cite{jovanov_tac19}, we show that using authentication only once in every bounded time interval ensures bounded estimation errors under any stealthy~attack. 
 
This paper is organized as follows. Section~\ref{sec:motive} formalizes the problem including the system and attack models. 
In Section \ref{sec:perfect}, we define the concept of perfectly attackable systems and find the necessary and sufficient conditions 
for PA. 
In Section~\ref{sec:intermittent}, we study effects of intermittent message authentication on performance guarantees under attack. 
Finally, our results are illustrated in case studies in Section~\ref{sec:numerical}, 
before concluding remarks in Section~\ref{sec:concl}.

\noindent\textbf{Notation.}
$\mathbb{B}$ and $\mathbb{R}$ denote the set of Boolean and real numbers, respectively, and $\mathbb{I}(.)$ is the indicator function. For a matrix $A$, $\mathcal{N}(A)$ denotes its null space, $A^T$ its transpose, 
$A^\dagger$ its Moore-Penrose pseudoinverse, and $||A||$ the $l_2$ norm of the matrix. 
For a vector $x\in{\mathbb{R}^n}$, we denote by $||x||_p$ the $p$-norm of $x$; when $p$ is not specified, the 2-norm is implied. 
In addition, we use $x_i$ to denote the $i^{\text{th}}$ element of $x$, while $\text{supp}(x)$ denotes the indices of nonzero elements of $x$ -- i.e., $\text{supp}(x)=\{i~|~i\in\{1,...,n\}, x_i\neq0\}$.

Projection vector $e_{i}$ is the unit vector 
where a 1 in its $i^{th}$ position is the only nonzero element of the vector. 
For set $\mathcal{S}$, $|\mathcal{S}|$ denotes the cardinality of the set and $\mathcal{S}^\complement$  its complement. $\mathcal{P}_{\mathcal{K}}x$ is the projection from the set $\mathcal{S}$ to set $\mathcal{K}$ ($\mathcal{K}\subseteq \mathcal{S}$) by keeping only elements of $x$ with indices from $\mathcal{K}$; formally, $\mathcal{P}_{\mathcal{K}}=\left[\begin{smallmatrix}
e_{j_1} | \ldots | e_{j_{|\mathcal{K}|}}\end{smallmatrix}\right]^T$, where $\mathcal{K}=\{s_{j_1},...,s_{j_{|\mathcal{K}|}}\}\subseteq{S}$ and $j_1<j_2<...<j_{|\mathcal{K}|}$.
If e.g., $\mathcal{S}=\{1,2,3,4\}$ and $\mathcal{K}=\{2,4\}$, then $\mathcal{P}_\mathcal{K}x=\left[\begin{smallmatrix} x_2&x_4
\end{smallmatrix}\right]^T$. 

\vspace{-2pt}
\section{Problem Description}\label{sec:motive} 
\vspace{-8pt}

\begin{figure} [!t]
\begin{center}
\includegraphics[width=0.422\textwidth]{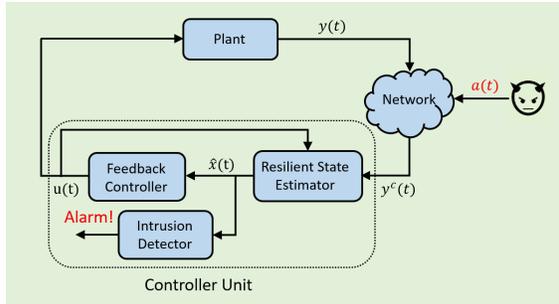}     
\caption{Control architecture under network-based attacks.} 
\label{fig:architecture}
\end{center}                
\end{figure}

We start by introducing the system (Fig.~\ref{fig:architecture}) and attack model, before formalizing the considered problem. 

\vspace{-6pt}
\subsection{System and Attack Model} \label{sec:A} 
\vspace{-6pt}


We now describe each system component from Fig.~\ref{fig:architecture}.

\noindent\textbf{Plant Model.}
We assume that the plant is an observable linear time-invariant (LTI) dynamical system that can be modeled 
in the standard state-space form as
\begin{equation}
\label{eq:plant}
\begin{split}
x(t+1) &= Ax(t)+Bu(t)+v_P(t),\\
y(t) &= Cx(t)+v_M(t).
\end{split}
\end{equation}
Here, $x \in {\mathbb{R}^n}$, $u \in {\mathbb{R}^m}$, $y \in {\mathbb{R}^p}$ denote the state, input and output vectors, respectively. The plant output vector captures measurements from the set of plant sensors 
$\mathcal{S}=\{s_1,s_2,...,s_p\}$.
\footnote{To simplify our notation, unless otherwise stated, we will use $i$ instead of $s_i$ to denote the $i$-th sensor.} 
In addition, $v_P \in {\mathbb{R}^n}$ and $v_M \in {\mathbb{R}^p}$ are bounded process and measurement noise vectors 
-- i.e., there exist $\delta_{v_P},\delta_{v_M}\in\mathbb{R}$ such that for all $t\geq{0}$, 
\begin{equation}
\label{eq:noise_bound}
\norm{v_P(t)}_2\leq{\delta_{v_P}}, \quad\norm{v_M (t)}_2\leq{\delta_{v_M}}. 
\end{equation}
Note that we make no assumptions about the distributions of the sensor and measurement noise models.

\noindent\textbf{Attack Model.}
We assume that the attacker was able to compromise information flow from a subset of sensors  $\mathcal{K}\subseteq{\mathcal{S}}$;\footnote{To simplify our presentation, we refer to these sensors as compromised since the effects of network-based attack are mathematically equivalent to compromising the sensors~\cite{ncs_attack_model}.} 
however, we make no assumption about the set~$\mathcal{K}$ (e.g., its size or elements). 
Hence, the sensor measurements delivered to the controller can be modeled~as
\begin{equation}\label{eq:network}
y^c(t) = y(t)+a(t).
\end{equation}
Here, $a(t)\in {\mathbb{R}^p}$ denotes the sparse attack signal injected by the attacker at time $t$ via the compromised information flows (i.e., sensors) from $\mathcal{K}$; hence, $\mathcal{K}=\text{supp}(a(t))$. 

We use a commonly adopted threat model (e.g.,~\cite{jovanov_tac19})~where:
\begin{enumerate} [label=(\roman*)]
    \item the attacker has the full knowledge of the system, its dynamics and design (e.g., controller and ID), as well as the employed security mechanisms -- e.g., the times when authentication is used, 
    \item the attacker has the required computation power to calculate suitable attack signals to inject via the set $\mathcal{K}$, while planning ahead as needed,
    \item the attacker's \underline{goal} is to design attack signal $a(t)$ such that it always remain \emph{stealthy} (i.e., undetected by the ID), while \emph{maximizing control degradation}. 
\end{enumerate}
The notions of \emph{stealthiness} and \emph{control performance degradation}  depend on the controller, and thus will be formally defined after the controller design is~introduced.


\noindent\textbf{Controller Design.}
The controller employs 
an RSE whose output is used for standard feedback control, and an 
ID (Fig.~\ref{fig:architecture}). 
To simplify our notation  while describing the RSE, 
the model~\eqref{eq:plant} can be considered in the form 
\begin{equation}
\label{eq:plant_withoutB}
\begin{split}
x(t+1) &= Ax(t),\\
y(t) = y^c(t) &= Cx(t) + w(t)+a(t);
\end{split}
\end{equation}
specifically, we can ignore the contribution of $u(t)$ as it is a known signal (no attacks on actuator are considered in this work) and thus has no effect on resilient state estimation. 
As shown in~\cite{pajic_iccps14,pajic_tcns17}, the bounds on the size of measurement noise $w$ in~\eqref{eq:plant_withoutB} can be related to the bounds on the size of process and measurement noise vectors $v_P$ and $v_M$; 
i.e., there exists $\delta_w>0$ such that
\begin{equation}
\label{eq:noisew}
||w(t)||\leq \delta_w, \qquad \text{for all } ~t\geq 0.
\end{equation}
%
%
%
%
\textbf{Resilient State Estimator.}
%
The goal of an RSE is to reconstruct the system state $x(t)$ from $N$ 
sensor measurements $\{y(t),..., y(t+N-1)\}$. 
We assume that $N=n$; however, the results can be extended to the case $N<n$, or $N>n$. 
To formally capture RSE requirements, we rewrite the system model from~\eqref{eq:plant_withoutB} as
\begin{equation}
\label{eq:outputs}
{\mathbf{y}}(t)=\mathbf{O}x(t)+\mathbf{a}(t)+{\mathbf{w}}(t),
\end{equation}
where $\mathbf{O}=[\mathbf{O}_{1}^T~|~...~|~\mathbf{O}_{p}^T]^T$. For each sensor $i$ and a subset of sensors $\mathcal{K}$, we define the matrices $\mathbf{O}_{i}$ and $\mathbf{O}_{\mathcal{K}}$~as 
\vspace{-6pt}
\begin{equation}
\mathbf{O}_{\mathcal{K}}=
\begin{bmatrix}
(\mathcal{P}_\mathcal{K}C)^T&(\mathcal{P}_\mathcal{K}CA)^T&\dots&(\mathcal{P}_\mathcal{K}CA^{N-1})^T
\end{bmatrix}^T,
\end{equation}
with $\mathbf{O}_{i}=\mathbf{O}_{\{s_i\}}$. 
Also, each of the block vectors $\mathbf{a}$, ${\mathbf{y}}$, ${\mathbf{w}}\in{\mathbb{R}^{pN}}$, satisfies $\mathbf{a}(t)=[\mathbf{a}_1^T(t)~|~...~|~\mathbf{a}_p^T(t)]^T$, ${\mathbf{y}}(t)=[{\mathbf{y}}_1^T(t)~|~...~|~{\mathbf{y}}_p^T(t)]^T$ and ${\mathbf{w}}(t)=[{\mathbf{w}}_1^T(t)~|~...~|~{\mathbf{w}}_p^T(t)]^T$. 
Now, for each sensor $i\in{\mathcal{S}}$, it holds that
\vspace{-2pt}
\begin{equation}
\label{eq:sensor_i}
{\mathbf{y}}_i(t)=\mathbf{O}_{i}x(t)+\mathbf{a}_i(t)+{\mathbf{w}_i}(t)
\end{equation}
with ${\mathbf{a}}_i(t)=[a_i(t)~|~a_i(t+1)~|...|~a_i(t+N-1)]^T\in{\mathbb{R}^N}$ denoting the values injected via $i^{\text{th}}$ sensor 
at time steps $t,...,t+N-1$, with ${\mathbf{a}}_i(t)=0$ if $i\notin{\mathcal{K}}$. Finally, ${{\mathbf{y}}}_i(t)=[y_i(t)~|~y_i(t+1)~|...|~y_i(t+N-1)]^T\in{\mathbb{R}^N}$ and ${{\mathbf{w}}}_i(t)=[w_i(t)~|~w_i(t+1)~|...|~w_i(t+N-1)]^T\in{\mathbb{R}^N}$ are the values of sensor~$i$ measurements and its noise.

In general, the RSE functionality 
can be captured as~\cite{fawzi2014secure}
\begin{equation}
\label{eq:RSEgeneral}
\mathcal{E}: \mathbb{R}^{Np} \mapsto \mathbb{R}^{n}\times\mathbb{R}^{Np}~~s.t.\quad
\mathcal{E}({\mathbf{y}}(t)) = \big(\hat{x}(t),{\hat{\mathbf{a}}}(t)\big).
\end{equation}
Here, $\hat{x}(t)$ and ${\hat{\mathbf{a}}}(t)$ are the state and attack vectors estimated 
from the delivered sensor measurements. 
The estimation error of an RSE is defined as
\vspace{-2pt}
\begin{equation}
\label{eq:DeltaX}
\Delta x(t)=\hat{x}(t)-x(t).
\end{equation}
A conventional RSE is the $l_0$-based  decoder~\cite{pajic_tcns17}, or its equivalent forms (e.g.,~\cite{fawzi2014secure,shoukry2017secure}), defined as optimization 
\begin{equation}
\label{eq:RSE}
\begin{split}
&\min\limits_{\hat{x}(t),{\hat{\mathbf{a}}}(t)} \sum\nolimits_{i=1}^{p}\mathbb{I}\big(\norm{{\hat{\mathbf{a}}}_i(t)}>0\big)\\
& s.\,\, t. \,\,\,\,\, {\mathbf{y}}(t)=\mathbf{O}\hat{x}(t)+\hat{{\mathbf{w}}}(t)+\hat{\mathbf{a}}(t)\\
&\,\,\,\,\,\,\,\,\,\,\,\,\,\,\,{\hat{{\mathbf{w}}}(t)} \in {\Omega}.
\end{split}
\end{equation}
Here, $\Omega$ denotes the feasible set of noise vectors, determined by the noise bounds from~\eqref{eq:noisew}. 
The vectors ${\hat{{\mathbf{w}}}(t)}$ and ${\hat{{\mathbf{a}}}(t)}$ are estimated at time $t$ independently from the 
estimated vectors at time step $t-1$. Hence, we denote ${\hat{{\mathbf{w}}}(t)}=[{\hat{\mathbf{w}}}_1^T(t)~|~...~|~\hat{{\mathbf{w}}}_p^T(t)]^T$, ${\hat{{\mathbf{a}}}(t)}=[{\hat{\mathbf{a}}}_1^T(t)~|~...~|~{\hat{\mathbf{a}}}_p^T(t)]^T$, with ${\hat{\mathbf{w}}}_i(t)=[\hat{w}^{(t)}_i(t)|...|\hat{w}^{(t)}_i(t+N-1)]^T$ and ${\hat{\mathbf{a}}}_i(t)=[\hat{a}^{(t)}_i(t)|...|\hat{a}^{(t)}_i(t+N-1)]^T$, in which $\hat{a}^{(t)}_i(k)$ and $\hat{w}^{(t)}_i(k)$ are the estimated noise and attack vectors at time $k$, as computed at time $t$, for $k=t,...,t+N-1$.

When not more than $s$ sensors are compromised in a $2s$-sparse observable system~\cite{shoukry2016event}, 
the estimation error 
of  the RSE~\eqref{eq:RSE}
is bounded~\cite{pajic_tcns17}; $2s$-sparse observable depends on the properties of the observability~matrix~of~$(A,C)$.

\noindent\textbf{Intrusion Detector.} 
We consider two 
ID used to detect the presence of any system anomaly (including attacks):
\begin{enumerate}
    \item \text{ID}$_{\textit{\rom{1}}}$: We capture the \text{ID$_\rom{1}$} functionality in the general form as mapping $\mathcal{D}_{\textit{\rom{1}}}: \mathbb{R}^{Np} \mapsto \mathbb{B}$ defined as
%
\vspace{-4pt}
\begin{equation}
\label{eq:ID1}
\mathcal{D}_{\textit{\rom{1}}}(\hat{\mathbf{a}}(t))=\mathbb{I}(\norm{\hat{\mathbf{a}}(t)}>0);
\end{equation}
i.e., if the estimated attack vector is non-zero, \textit{ID}$_{\textit{\rom{1}}}$ raises alarm. Note that our goal is \emph{not} to identify the exact set of attacked sensors, which would result in a nonzero threshold in~\eqref{eq:ID1}, as shown in~\cite{pajic_tcns17}. 

\item \text{ID$_{\textit{\rom{2}}}$}: We define the \text{ID$_{\textit{\rom{2}}}$} as 
$\mathcal{D}_{\textit{\rom{2}}}: \mathbb{R}^{Np+2n} \mapsto \mathbb{B}$ 
with
\vspace{-4pt}
\begin{equation}
\label{eq:ID2}
\begin{split}
&\mathcal{D}_{\textit{\rom{2}}}(\hat{\mathbf{a}}(t),\hat{x}(t),\hat{x}(t-1))=\\
& \mathbb{I}\big(\norm{\hat{\mathbf{a}}(t)} > 0\big)   \lor   \mathbb{I}\big(||\hat{x}(t)-A\hat{x}(t-1)|| > {d}\big);
\end{split}
\end{equation}
here, $\lor$ is Boolean OR and 
$d$ is defined by~Prop.~\ref{Pro:bound}.  
\end{enumerate}

We use $\mathcal{D}_{\textit{\rom{1}}}(\hat{\mathbf{a}})$ and $\mathcal{D}_{\textit{\rom{2}}}(\hat{\mathbf{a}},\hat{x})$ instead of $\mathcal{D}_{\textit{\rom{1}}}(\hat{\mathbf{a}}(t))$ and $\mathcal{D}_{\textit{\rom{2}}}(\hat{\mathbf{a}}(t), \hat{x}(t), \hat{x}(t-1))$, respectively. We also denote the system~\eqref{eq:plant_withoutB} with \textit{ID$_{\textit{{i}}}$} ($\textit{{i}}\in\{$ {\rom{1}}, {\rom{2}}$\}$) 
as $\Sigma_{\textit{i}}(A,C,\delta_w,\mathcal{K})$. Yet, if results hold for both IDs, we remove the subscript~\textit{i}.

\begin{proposition}
\label{Pro:bound}
For the system without attack, it holds that 
%
$||\hat{x}(t)-A\hat{x}(t-1)||\leq d=2\sqrt{N}\delta_w||\mathbf{O}^\dagger||_2(1+||A||_2)$.
\end{proposition}
\vspace{-12pt}
\begin{pf}
Constraints in~\eqref{eq:RSE} at time $t$ and $t-1$~imply 
\vspace{-4pt}
\begin{equation}
\begin{split}
\mathbf{O}{x}(t)+{\mathbf{w}}(t)=&\mathbf{O}\hat{x}(t)+\hat{{\mathbf{w}}}(t)\\
\mathbf{O}{x}(t-1)+{\mathbf{w}}(t-1)=&\mathbf{O}\hat{x}(t-1)+\hat{{\mathbf{w}}}(t-1)
\end{split}
\end{equation}
For $\Delta{\mathbf{w}}(t)={\mathbf{w}}(t)-\hat{{\mathbf{w}}}(t)$, since $(A,C)$ is observable, 
\begin{equation}\label{eq:16}
\begin{split}
\hat{x}(t)=&x(t)-\mathbf{O}^\dagger\Delta{{\mathbf{w}}}(t)\\
\hat{x}(t-1)=&x(t-1)-\mathbf{O}^\dagger\Delta{{\mathbf{w}}}(t-1).
\end{split}
\end{equation}
Hence, from~\eqref{eq:plant_withoutB}, 
$
||\hat{x}(t)-A\hat{x}(t-1)||=
||A\mathbf{O}^\dagger\Delta{{\mathbf{w}}}(t-1)-\mathbf{O}^\dagger\Delta{{\mathbf{w}}}(t)||
\leq ||\mathbf{O}^\dagger||_2||\Delta{{\mathbf{w}}}(t)||+||A||_2||\mathbf{O}^\dagger||_2||\Delta{{\mathbf{w}}}(t-1)||.
$
On the other hand, $||\Delta{{\mathbf{w}}}(t)||_2\leq 2\sqrt{N}\delta_w$, which also holds for $\Delta{{\mathbf{w}}}(t-1)$, and thus concludes the proof.
\end{pf}

\subsection{Problem Formulation}
In this work, we focus on the following  two problems.

\textit{Problem 1}: Under which conditions, a \emph{stealthy} attacker could introduce arbitrarily large estimation errors~\eqref{eq:DeltaX}? 
From~\eqref{eq:ID1},~\eqref{eq:ID2}, the stealthiness conditions for \text{ID}$_{\textit{\rom{1}}}$, \text{ID$_{\textit{\rom{2}}}$}
are
\begin{equation}
\label{eq:stealthiness}
\mathcal{D}_{\textit{\rom{1}}}(\hat{\mathbf{a}})=0, \qquad \mathcal{D}_{\textit{\rom{2}}}(\hat{\mathbf{a}},\hat{x})=0.
\end{equation}
Note that 
if an attack is stealthy from \text{ID$_{\textit{\rom{2}}}$} it cannot be detected by \text{ID$_{\textit{\rom{1}}}$} either. Due to the batch-processing nature of the RSE and bounded-size noise, the approach and conditions from~\cite{mo2010false,kwon2014analysis} cannot be used. Hence, we introduce PA for LTI systems with bounded-size noise. 

\textit{Problem 2}: As we show in next section, for a large class of systems $\Sigma_{\textit{\rom{1}}}(A,C,\delta_w,\mathcal{K})$, an unbounded state estimation error can be inserted by compromising a subset of sensors. Although the use of~\text{ID$_{\textit{\rom{2}}}$} (i.e., for systems $\Sigma_{\textit{\rom{2}}}(A,C,\delta_w,\mathcal{K})$) restricts these conditions, unstable plants are vulnerable to perfect attacks (i.e., stealthy attacks that cause unbounded estimation errors). 
On the other hand, the use of security mechanisms, such as message authentication, could ensure
integrity of the received sensor measurements. Thus, a stealthy attack vector has to satisfy $a_i(t)=0$ when the measurement of sensor $s_i$ is  authenticated at time $t$, and ${a}(t)=0$ if integrity of all sensors is enforced at time $t$. 
Since authentication comes with additional computational and communication cost, we study the effects of intermittent data authentication on attack impact. Our goal is to find conditions that the authentication policy (i.e., times when authentication is used) should satisfy so that the systems $\Sigma_{\textit{\rom{1}}}(A,C,\delta_w,\mathcal{K})$,  $\Sigma_{\textit{\rom{2}}}(A,C,\delta_w,\mathcal{K})$ are not~PA. 

\section{PA of LTI Systems with Bounded-Noise} 
\label{sec:perfect}
\vspace{-6pt}

    


The notion of PA is introduced in~\cite{mo2010false,kwon2014analysis} for systems with a statistical ($\chi^2$) ID and a Kalman-filter implementing continuous (i.e., streamed) processing of sensor measurements. On the other hand, most existing RSEs for systems with bounded noise (e.g.,~\cite{pajic_tcns17,shoukry2017secure,sundaram_cdc10,sundaram2011distributed}) are based on batch-processing of sensor data -- i.e., processing a window of sensor measurements at each time step (mostly even without taking previous computations into account). Thus, the notion of PA needs to differentiate between PA at a single time point vs. PA over a time interval. In this section, we first define these two notions of PA 
for systems with one of the two IDs, before providing the necessary and sufficient conditions individually. 

\begin{definition}
\label{def:PA}
System $\Sigma_{\textit{\rom{1}}}(A,C,\delta_w,\mathcal{K})$ is perfectly attackable at a single time step if  for any $M>0$, there exists 
a stealthy sequence of attack signals $\mathbf{a}(t)$ over $N$ time~steps (i.e., satisfying~\eqref{eq:stealthiness}), 
for which the RSE estimation error satisfies $||\Delta{x}(t)||>M$. 
Such attack vector $\mathbf{a}(t)$ is called a perfect attack for the system $\Sigma_{\textit{\rom{1}}}(A,C,\delta_w,\mathcal{K})$. 
\end{definition}

Definition~\ref{def:PA} 
does not require that the attack is stealthy before or remains stealthy at time steps after $t$. Such notion of PA for the system $\Sigma_{\textit{\rom{2}}}(A,C,\delta_w,\mathcal{K})$ is not relevant because the \textit{ID$_{\textit{\rom{2}}}$}, from~\eqref{eq:ID2}, validates that the estimated states in every two consecutive steps do not violate plant dynamics. Thus, we characterize a more realistic requirements for stealthy attacks -- PA over a time-interval.

\begin{definition}
\label{def:PAtime}
System $\Sigma(A,C,\delta_w,\mathcal{K})$ is perfectly attackable over time if for all $M>0$ there exists 
a sequence of attack signals $\mathbf{a}(t), \mathbf{a}(t+1),...$ and  a time point $t'\geq t$ such that for all $k$, where $k\geq t'$, it holds that  $||\Delta{x}(k)||>M$, and for all time steps, the estimated attack vectors $\hat{\mathbf{a}}$ satisfies the corresponding stealthiness requirements in~\eqref{eq:stealthiness}.
\end{definition}

To simplify our presentation, 
instead of formally stating that the estimation error may be arbitrarily large, 
we may say that the estimation error is unbounded.

\begin{remark}
\label{rem:PA12}
If the system $\Sigma_{\textit{\rom{2}}}(A,C,\delta_w,\mathcal{K})$ is perfectly attackable over time, then the $\Sigma_{\textit{\rom{1}}}(A,C,\delta_w,\mathcal{K})$ is also perfectly attackable over time because the stealthiness condition in (\ref{eq:stealthiness}) for \textit{ID$_{\textit{\rom{2}}}$} also includes the condition for \textit{ID$_{\textit{\rom{1}}}$}. 
\end{remark}

Note that PA over time is a stronger notion than PA at a single time point because  $\mathcal{D}_\textit{\rom{1}}\big(\hat{\mathbf{a}}(t))\big)$ should be equal to zero  for all time steps.  
Therefore, the following~holds.

\begin{proposition}
If the system  $\Sigma_{\textit{\rom{1}}}(A,C,\delta_w,\mathcal{K})$ is PA over time, then it is also PA at a single time step. 
\end{proposition}

\vspace{-2pt}
\subsection{PA of  $\Sigma_{\textit{\rom{1}}}(A,C,\delta_w,\mathcal{K})$ System}
\vspace{-2pt}

We now capture 
conditions for PA at a single time.

\vspace{-4pt}
\begin{theorem}
\label{thm:PAt}
System $\Sigma_{\textit{\rom{1}}}(A,C,\delta_w,\mathcal{K})$ is PA at a singletime step if and only if pair  $(A,\mathcal{P}_{\mathcal{K}^\complement} C)$ is not observable. 
\end{theorem}   

\vspace{-12pt}
\begin{pf}
($\Rightarrow$) Let us assume that the pair  $(A,\mathcal{P}_{\mathcal{K}^\complement} C)$ is observable, while the system $\Sigma_{\textit{\rom{1}}}(A,C,\delta_w,\mathcal{K})$ is PA at a single time step, which we denote as $t$. Then, there exists a stealthy attack sequence ${\mathbf{a}}(t)$ for which the RSE estimated attack vector $\hat{\mathbf{a}}(t)=0$ and $\norm{\Delta{x}(t)}$ is unbounded.

Consider 
data from noncompromised sensors in~$\mathcal{K}^\complement$; i.e., 
\vspace{-10pt}
\begin{equation*}
\mathcal{P}_{\mathcal{K}^\complement}\mathbf{y}(t) \overset{\text{\small (i)}}{=}
\mathbf{O}_{\mathcal{K}^\complement}x(t)+\mathcal{P}_{\mathcal{K}^{\complement}}{\mathbf{w}}(t)
\overset{\text{\small (ii)}}{=}\mathbf{O}_{\mathcal{K}^\complement}\hat{x}(t)+\mathcal{P}_{\mathcal{K}^{\complement}}\hat{{\mathbf{w}}}(t),
\end{equation*}
where (i) holds from~\eqref{eq:outputs} as the sensors are noncompromised, 
whereas (ii) holds from~\eqref{eq:RSE} since the attack is stealthy (i.e., $\hat{\mathbf{a}}(t)=0$). 
Hence, it follows that
$
\mathbf{O}_{\mathcal{K}^\complement}\Delta{x}(t)=\mathcal{P}_{\mathcal{K}^{\complement}}\Delta{\mathbf{w}}(t), 
$
where $\Delta{\mathbf{w}}(t)={\mathbf{w}}(t)-\hat{{\mathbf{w}}}(t)$. Since the matrix $\mathbf{O}_{\mathcal{K}^\complement}$ is full rank,
$\Delta{x}(t)=\big(\mathbf{O}_{\mathcal{K}^\complement}\big)^\dagger\Big(\mathcal{P}_{\mathcal{K}^{\complement}}\Delta{\mathbf{w}}(t)\Big),$
and thus
\vspace{-6pt}
\begin{equation}
\begin{split}
\label{eq:t11}    
\norm{\Delta{x}(t)}
\leq& \norm{\big(\mathbf{O}_{\mathcal{K}^\complement}\big)^\dagger}\Big(\norm{\mathcal{P}_{\mathcal{K}^{\complement}}\Delta{\mathbf{w}}(t)}\Big).
\end{split}
\end{equation}
The matrix $\big(\mathbf{O}_{\mathcal{K}^\complement}\big)^\dagger$ has a bounded norm, $\mathbf{w}(t)$ and $\hat{\mathbf{w}}(t)$ are also bounded. 
Thus, the right side of~\eqref{eq:t11} is bounded, meaning that $\Delta x(t)$ is bounded, which is a contradiction. 

($\Leftarrow$) Suppose that the pair $(A,\mathcal{P}_{\mathcal{K}^\complement} C)$ is not observable; thus, there exists a nonzero vector ${z}$ such that $\mathbf{O}_{\mathcal{K}^\complement} {z}=0$. Let us assume that the system is in state $x(t)$ when attack ${\mathbf{a}}(t)=\begin{bmatrix}
(\mathcal{P}_{\mathcal{K}}{\mathbf{a}}(t))^T & (\mathcal{P}_{\mathcal{K}^\complement}{\mathbf{a}}(t))^T
\end{bmatrix}^T=\mathbf{O}{z}=\begin{bmatrix}
(\mathbf{O}_{\mathcal{K}}z)^T &0
\end{bmatrix}^T$
is applied. 
Then, 
from~\eqref{eq:outputs} we have that
%
\begin{equation}\label{eq{15}}
{\mathbf{y}}(t)\hspace{-1pt}=\mathbf{O}{x}(t)+{{\mathbf{w}}}(t)+{\mathbf{a}}(t)=\mathbf{O}\hat{x}(t)+\hat{{\mathbf{w}}}(t)+\hat{\mathbf{a}}(t).
\end{equation}  
Consider $\hat{{\mathbf{w}}}'(t)=\hat{{\mathbf{w}}}(t)$, $\hat{x}'(t)=\hat{x}(t)+z$  and $\hat{\mathbf{a}}'(t)=0$. Now,  $\left(\hat{{{x}}}'(t),\hat{{\mathbf{w}}}'(t),\hat{{\mathbf{a}}}'(t) \right)$ is a feasible point for the RSE optimization problem from~\eqref{eq:RSE} that also minimizes the objective to zero. Thus, the output of RSE $\left(\hat{x}(t),\hat{\mathbf{a}}(t)\right)$ also has to have the same value for the objective function -- i.e., $\hat{\mathbf{a}}=\mathbf{0}$, and the attack will not be detected. 

Since $(A,C)$ is observable, from~\eqref{eq{15}} and~\eqref{eq:DeltaX}, we have
$ \Delta x(t) = \mathbf{O}^\dagger\Delta{\mathbf{w}}(t) +\mathbf{O}^\dagger\mathbf{a}(t)=\mathbf{O}^\dagger\Delta{\mathbf{w}}(t) + z.$
As~$\Delta{\mathbf{w}}(t)$ is bounded, and ${z}$  is any nonzero vector in the null-space of $\mathbf{O}_{\mathcal{K}^\complement}$, it can be chosen with an arbitrarily large norm. Thus, $\Sigma_{\textit{\rom{1}}}(A,C,\delta_w,\mathcal{K})$ is PA at a single time~step.\end{pf}

As the plant $(A,C)$ is observable, 
the next result follows. 
\begin{corollary}
System $\Sigma_{\textit{\rom{1}}}(A,C,\delta_w,\mathcal{S})$ (i.e., all sensors compromised) is perfectly attackable at a single time step. 
\end{corollary}
\begin{corollary}
\label{cor:RSEerr}
If the attack to the system $\Sigma_{\textit{\rom{1}}}(A,C,\delta_w,\mathcal{K})$ has the form $\mathbf{a}(t)=\mathbf{O}z(t)$, for some $z\in\mathbb{R}^n$, then $\hat{\mathbf{a}}(t)=0$ and the RSE error 
satisfies $\Delta{x}(t)=\mathbf{O}^\dagger\Delta{{\mathbf{w}}}(t) + z$.
\end{corollary}
\begin{remark}
Although in this paper we consider $l_0$-based estimators, it is straightforward to show that the results of Theorem~\ref{thm:PAt} are valid for any batch processing estimators like $l_1$-based estimator or estimators from~\cite{shoukry2017secure,shoukry2016event}.
\end{remark}

\begin{example}\label{ex:1}
To illustrate PA at time point, consider~system $\Sigma_{\textit{\rom{1}}}(A,C,\delta_w,\mathcal{K})$ with $\delta_w=0$, $\mathcal{K}=\mathcal{S}=\{s_1\}$, $N=2$, 
$
A=\left[\begin{smallmatrix}
.3&1\\0&.5
\end{smallmatrix}\right],~~ C=\begin{bmatrix}
1&0
\end{bmatrix}.
$
%
The attack vector ${\mathbf{a}}(t)=\begin{bmatrix}a^T(t)&a^T(t+1) \end{bmatrix}^T=\mathbf{O}z$ results in estimation error $\Delta x(t)=z$ and $\hat{\mathbf{a}}(t)=0$, for $z$ being any arbitrary nonzero vector; thus, can generate a perfect attack vector at time~$t$.
\end{example}

We now provide a necessary and sufficient condition that the system $\Sigma_{\textit{\rom{1}}}(A,C,\delta_w,\mathcal{K})$ is PA over~time.

\begin{theorem}
\label{thm:PA}
Consider the system $\Sigma_{\textit{\rom{1}}}(A,C,\delta_w,\mathcal{K})$ and let us define the matrix $F(\mathcal{K},N)$ as 
\vspace{-12pt}
\begin{equation}
\label{eq:F}
{F(\mathcal{K},N)}=\begin{bmatrix}
\mathbf{O}_{\mathcal{K}^\complement}^T&(\mathcal{P}_{\mathcal{K}}C)^T&...&(\mathcal{P}_{\mathcal{K}}CA^{N-2})^T
\end{bmatrix}^T.
\end{equation}
%
a) Suppose $F(\mathcal{K},N)$ is not full rank. Then, the system $\Sigma_{\textit{\rom{1}}}(A,C,\delta_w,\mathcal{K})$ is perfectly attackable over time if and only if 
it is perfectly attackable at a single time step.

b) Suppose $F(\mathcal{K},N)$ is full rank. Then 
$\Sigma_{\textit{\rom{1}}}(A,C,\delta_w,\mathcal{K})$ is PA over time if and only if 
it is PA at a single time step, $A$ is unstable and at least one eigenvector $v_i$ corresponding to an unstable eigenvalue satisfies $v_i\in \mathcal{N}(\mathcal{O}_{\mathcal{K}^\complement})$. 
\end{theorem}

From Theorem~\ref{thm:PA} it holds that, unlike the notion of PA in systems with probabilistic noise and statistical IDs~\cite{mo2010false,kwon2014analysis,jovanov_tac19}, for systems with bounded noise and $l_0$-based RSEs, 
a system can be perfectly attackable over time even if the plant is not unstable. 
Before proving Theorem~\ref{thm:PA}, first we introduce the following lemmas used in the proof.

\begin{lemma}
\label{lem:BPA}
Consider attack on the system $\Sigma_{\textit{\rom{1}}}(A,C,\delta_w,\mathcal{K})$ in the form $\mathbf{a}(t)=\mathbf{O}z(t)$, where it also holds that if $\mathcal{N}(F(\mathcal{K},N))=0$ then $z(t)\not\in \mathcal{N}(A)$. If $z(t+1)=Az(t)+\alpha(t)$, where $\alpha(t)\in{\mathcal{N}(F(\mathcal{K},N))}$, then  $\mathbf{a}(t+1)=\mathbf{O}z(t+1)$ is also a stealthy attack vector for the system. 
\end{lemma}

\begin{pf}
For $\mathbf{a}(t+1)$ to be a feasible attack vector, we need to show that $\mathcal{P}_{\mathcal{K}^\complement}{\mathbf{a}}(t+1)=0$, which is  equivalent~to
\vspace{-4pt}
\begin{equation}
\label{eq:t2}
\mathbf{O}_{\mathcal{K}^\complement}z(t+1)=\mathbf{O}_{\mathcal{K}^\complement}(Az(t)+\alpha(t))=0.
\end{equation}
As $\alpha(t)\in{\mathcal{N}(F(\mathcal{K},N))}$, we have $\mathbf{O}_{\mathcal{K}^\complement}\alpha(t)=0$.~From Cayley-Hamilton theorem and assumption that $\mathbf{O}_{\mathcal{K}^\complement}z(t)=0$ (since 
$\mathcal{P}_{\mathcal{K}^\complement}\mathbf{a}(t)=0$ as sensors from~$\mathcal{K}^\complement$ are not compromised),
 we have $\mathbf{O}_{\mathcal{K}^\complement}Az(t)=0$;
 thus, \eqref{eq:t2} also~holds.

We now show time consistency for the attacks; i.e., that the corresponding elements of $\mathcal{P}_{\mathcal{K}}{a}(t+1)$,...,$\mathcal{P}_{\mathcal{K}}{a}(t+N-1)$ of  vectors $\mathcal{P}_{\mathcal{K}}{\mathbf{a}}(t)$ and $\mathcal{P}_{\mathcal{K}}{\mathbf{a}}(t+1)$ are equal. We have
\vspace{-10pt}
\begin{equation*}
\begin{split}
\mathcal{P}_{\mathcal{K}}{\mathbf{a}}&(t+1)=\begin{bmatrix}
(\mathcal{P}_{\mathcal{K}}C)^T&\dots&(\mathcal{P}_{\mathcal{K}}CA^{N-1})^T
\end{bmatrix}^T(Az(t)+\alpha(t))\\
&=\begin{bmatrix}
(\mathcal{P}_{\mathcal{K}}CA)^T&\dots&(\mathcal{P}_{\mathcal{K}}CA^{N})^T
\end{bmatrix}^Tz(t)+\mathbf{O}_{\mathcal{K}}\alpha(t)
\end{split}
\end{equation*}
\vspace{-6pt}
\begin{equation*}
\mathcal{P}_{\mathcal{K}}{\mathbf{a}}(t)=\begin{bmatrix}
(\mathcal{P}_{\mathcal{K}}C)^T&\dots&(\mathcal{P}_{\mathcal{K}}CA^{N-1})^T
\end{bmatrix}^Tz(t).
\end{equation*}
By comparing the shared elements of vectors $\mathcal{P}_{\mathcal{K}}{\mathbf{a}}(t)$ and $\mathcal{P}_{\mathcal{K}}{\mathbf{a}}(t+1)$ we have that they are equal, ending~the~proof.
\end{pf}

\begin{lemma}
\label{lem:EstimError}
Let the system $\Sigma_{\textit{\rom{1}}}(A,C,\delta_w,\mathcal{K})$, at two consecutive time steps $t$ and $t+1$  have estimation error $\Delta{x}(t)$ and $\Delta{x}(t+1)$, while $\mathcal{D}\big(\hat{\mathbf{a}}(t)\big)=\mathcal{D}\big(\hat{\mathbf{a}}(t+1)\big)=0$. Then $\Delta{x}(t+1)=A\Delta{x}(t)+\alpha(t)+p(t)$, where $\alpha(t)\in{\mathcal{N}(F(\mathcal{K},N))}$ and $p(t)$ is a bounded vector. 
\end{lemma}	

\begin{pf}
From $\mathcal{D}_{\textit{\rom{1}}}\big(\hat{\mathbf{a}}(t)\big)=\mathcal{D}_{\textit{\rom{1}}}\big(\hat{\mathbf{a}}(t+1)\big)=0$, as in~\eqref{eq{15}}
\begin{equation}
\begin{split}
{\mathbf{a}}(t)&=\mathbf{O}\Delta {x}(t)+\Delta{{\mathbf{w}}}(t) \\
{\mathbf{a}}(t+1)&=\mathbf{O}\Delta {x}(t+1)+\Delta{{\mathbf{w}}}(t+1). 
\end{split}
\end{equation}
Since the sensors from $\mathcal{K}^\complement$ are not compromised, 
$\mathbf{O}_{\mathcal{K}^\complement}\Delta {x}(t)=-\mathcal{P}_{\mathcal{K}^\complement}\Delta{{\mathbf{w}}}(t)$ holds. 
Thus, we have~that
\begin{equation*}
\begin{split}
&\mathbf{O}_{\mathcal{K}^\complement}A\Delta {x}(t)=\begin{bmatrix} 
(\mathcal{P}_{\mathcal{K}^\complement}CA)^T&\dots&(\mathcal{P}_{\mathcal{K}^\complement}CA^{N})^T
\end{bmatrix}\Delta{x}(t)=\\
&\begin{bmatrix} 
-(\mathcal{P}_{\mathcal{K}^\complement}\Delta{{{w}}}^t(t+1))^T&\dots&-(\mathcal{P}_{\mathcal{K}^\complement}\Delta{{{w}}}^t(t+N-1))^T&\beta^T(t)
\end{bmatrix}^T\\
&=h(t),   
\end{split}
\end{equation*}
where $\beta(t)=\mathcal{P}_{\mathcal{K}^\complement}CA^{N}\Delta{x}(t)$ and $\Delta{{{w}}}^t(k)=w(k)-\hat{w}^t(k)$ for $k=t,...,t+N-1$. 
Using Cayley-Hamilton theorem, it holds that $A^N=c_0I+...+c_{N-1}A^{N-1}$, for some $c_0,...,c_{N-1}\in\mathbb{R}$. Thus, we get $\beta(t)=\mathcal{P}_{\mathcal{K}^\complement}\big(c_0\Delta{{{w}}}^t(t)+...+c_{N-1}\Delta{{{w}}}^t(t+N-1)\big)$. 
On the other hand, we have
$\mathbf{O}_{\mathcal{K}^\complement}\Delta {x}(t+1)=-\mathcal{P}_{\mathcal{K}^\complement}\Delta{{{\mathbf{w}}}}(t+1)$. Hence, it follows that 
\begin{align} 
\mathbf{O}_{\mathcal{K}^\complement}(\Delta {x}(t+1)-A\Delta {x}(t))&=-\mathcal{P}_{\mathcal{K}^\complement}\Delta{{{\mathbf{w}}}}(t+1)-h(t)\nonumber\\
& \delequal r_1(t)\label{eq{17}}
\end{align}
The attack vectors for compromised sensors are
\begin{equation*}
\begin{split}
{\mathcal{P}_{\mathcal{K}}\mathbf{a}}(t)=
&\begin{bmatrix}
(\mathcal{P}_{\mathcal{K}}C)^T&(\mathcal{P}_{\mathcal{K}}CA)^T&\dots&(\mathcal{P}_{\mathcal{K}}CA^{N-1})^T
\end{bmatrix}^T\hspace{-6pt}\Delta{x}(t)\\
& +\mathcal{P}_{\mathcal{K}}\Delta{{\mathbf{w}}}(t)
\end{split}
\end{equation*}
\begin{equation*}
\begin{split}
{\mathcal{P}_{\mathcal{K}}\mathbf{a}}(t+1)
=&\begin{bmatrix}
(\mathcal{P}_{\mathcal{K}}C)^T&(\mathcal{P}_{\mathcal{K}}CA)^T&\dots&(\mathcal{P}_{\mathcal{K}}CA^{N-1})^T
\end{bmatrix}^T\\&\times\Delta{x}(t+1)
+\mathcal{P}_{\mathcal{K}}\Delta{{\mathbf{w}}}(t+1)
\end{split}
\end{equation*}
Let us define $\mathbf{O}_{\mathcal{K}}^{N-2}=\begin{bmatrix}
(\mathcal{P}_{\mathcal{K}}C)^T&\cdots&(\mathcal{P}_{\mathcal{K}}CA^{N-2})^T
\end{bmatrix}^T$.
Consistency in the overlapping terms of the vectors in above equations, implies that 
%
\begin{equation*}
\begin{split}
\mathbf{O}_{\mathcal{K}}^{N-2}
\Delta{x}(t+1)&+\begin{bmatrix} 
\mathcal{P}_{\mathcal{K}}\Delta{{{w}}}^{t+1}(t+1)\\\vdots\\\mathcal{P}_{\mathcal{K}}\Delta{{{w}}}^{t+1}(t+N-1)
\end{bmatrix}=\\
&\mathbf{O}_{\mathcal{K}}^{N-2}A\Delta{x}(t)+\begin{bmatrix} 
\mathcal{P}_{\mathcal{K}}\Delta{{{w}}}^{t}(t+1)\\\vdots\\\mathcal{P}_{\mathcal{K}}\Delta{{{w}}}^{t}(t+N-1)
\end{bmatrix}.
\end{split}   
\end{equation*}
Now, we define 
\vspace{-10pt}
$$r_2(t)\delequal \begin{bmatrix} 
\mathcal{P}_{\mathcal{K}}\Delta{{{w}}}^{t}(t+1)\\\vdots\\\mathcal{P}_{\mathcal{K}}\Delta{{{w}}}^{t}(t+N-1)
\end{bmatrix}-\begin{bmatrix} 
\mathcal{P}_{\mathcal{K}}\Delta{{{w}}}^{t+1}(t+1)\\\vdots\\\mathcal{P}_{\mathcal{K}}\Delta{{{w}}}^{t+1}(t+N-1)
\end{bmatrix}.$$
Combining the above equation with (\ref{eq{17}}) results in 
\vspace{-10pt}
\begin{equation}
\begin{split}
\label{eq:t1}
F(\mathcal{K},N)(\Delta{x}(t+1)-A\Delta{x}(t))&=\begin{bmatrix} r_1(t)\\r_2(t)\end{bmatrix}\delequal r(t),
\end{split}
\end{equation} 
where $r(t)$ is bounded since both $r_1(t)$ and $r_2(t)$ are bounded. 
Thus, the solution of~\eqref{eq:t1} can be captured~as
\vspace{-4pt}
\begin{equation}
\label{eq:alphap}
\Delta{x}(t+1)-A\Delta{x}(t)=\alpha(t)+p(t) 
\end{equation}
where $p(t)\in\mathbb{R}^n$ is any bounded vector that satisfies $F(\mathcal{K},N)p(t)=r(t)$ and $\alpha(t)\in{\mathcal{N}(F(\mathcal{K},N))}$. 
\end{pf}

\begin{lemma}
\label{lem:NoPA}
Suppose that  $\Sigma_{\textit{\rom{1}}}(A,C,\delta_w,\mathcal{K})$ is PA at a single time step and $\Delta{x}(t)$ is bounded while $\hat{\mathbf{a}}(t)=0$. If $F(\mathcal{K},N)$ is full rank, then there exists no attack vector ${\mathbf{a}}(t+1)$ such that $\Delta{x}(t+1)$ becomes arbitrarily large while $\hat{\mathbf{a}}(t+1)=0$. 
\end{lemma}

\vspace{-10pt}
\begin{pf}
Assume that we can find  attack  ${\mathbf{a}}(t+1)$ such that $\Delta{x}(t+1)$ becomes arbitrarily large while $\hat{\mathbf{a}}(t+1)=0$. Since $\Delta{x}(t)$ is bounded and $\hat{\mathbf{a}}(t)=0$, it means that ${\mathbf{a}}(t)$ is also bounded. Also, $\mathcal{P}_{\mathcal{K}^\complement}a(t+N)=0$. Let us define the augmented vectors as $\mathbf{a}_{\mathcal{F}}(t+1)=\begin{bmatrix}a^T(t+1)|...|a^T(t+N-1)|\mathcal{P}_{\mathcal{K}^\complement}a^T(t+N)\end{bmatrix}^T$;  similarly $\Delta{\mathbf{w}}_{\mathcal{F}}(t+1)$. Then, from the constraint of \eqref{eq:RSE} we have that
$\Delta{x}(t+1)=F^\dagger(\mathcal{K},N)\mathbf{a}_{\mathcal{F}}(t+1)-F^\dagger(\mathcal{K},N)\mathbf{w}_{\mathcal{F}}(t+1).
$
The right side of the equation is bounded whereas the left may be arbitrarily large, which is a contradiction.  
\end{pf}

\begin{corollary}
\label{cor:iniatialPA}
If $F(\mathcal{K},N)$ for the system $\Sigma_{\textit{\rom{1}}}(A,C,\delta_w,\mathcal{K})$ is full rank, then a stealthy cannot induce an unbounded estimation error in the initial step of the attack. 
\end{corollary}

\vspace{-10pt}
\begin{pf}
Before starting attack at time $t_0$, the estimation error is bounded. Now, based on Lemma~\ref{lem:NoPA}, if the matrix $F(\mathcal{K},N)$ is full rank, it will be impossible to have unbounded estimation error $\Delta{x}(t_0)$ while $\hat{\mathbf{a}}(t_0)=0$
\end{pf}
\begin{lemma}
\label{lem:ExistPA}
There exists a nonzero attack vector ${\mathbf{a}}(t)$ (i.e., $\epsilon<\norm{{\mathbf{a}}(t)}$ with $\epsilon>0$) such that $\mathcal{D}(\hat{\mathbf{a}}(i),\hat{x}(i))=0$ for any $t-(N-1)\leq i\leq t+(N-1)$.
\end{lemma}

\vspace{-8pt}
\begin{pf}
The claim should be proven for both $\mathcal{D}_{\textit{\rom{1}}}$ and $\mathcal{D}_{\textit{\rom{2}}}$. As the proof for $\mathcal{D}_{\textit{\rom{2}}}$ also covers  the case for $\mathcal{D}_{\textit{\rom{1}}}$ IDs, due to space constraint, we will focus on $\mathcal{D}_{\textit{\rom{2}}}$.

Based on the stealthiness condition $\mathcal{D}_{\textit{\rom{2}}}(\hat{\mathbf{a}}(i))=0$ for any $t-(N-1)\leq i\leq t+N-1$, it holds that
\begin{equation}
\label{eq:cons11}
{\mathbf{y}}(i)=\mathbf{O}{x}(i)+{{\mathbf{w}}}(i)+{\mathbf{a}}(i)=\mathbf{O}\hat{x}(i)+\hat{{\mathbf{w}}}(i). 
\end{equation}
$\hat{{\mathbf{w}}}(i)={{\mathbf{w}}}(i)+{\mathbf{a}}(i)$ and $\hat{x}(i)=x(i)$ are a feasible point for constraint~\eqref{eq:cons11}. In this case, $||\hat{x}(i)-A\hat{x}(i-1)||=||x(i)-Ax(i-1)||=0$ satisfies the second stealthiness condition of \text{ID$_{\textit{\rom{2}}}$} from~\eqref{eq:ID2}, for any $i$. 
Thus, we need to find a nonzero attack ${\mathbf{a}}(i)$ such that $||\hat{{\mathbf{w}}}(i)||\leq \sqrt{N}\delta_w$ is satisfied. 
If for any $t-(N-1)\leq i\leq t+N-1$ it holds that $||{{\mathbf{w}}}(i)||<\sqrt{N}\delta_w$, then any nonzero attack vector satisfying $||{\mathbf{a}}(i)||<\sqrt{N}\delta_w-||{\mathbf{w}}(i)||$ is stealthy -- note that no other constraint beyond the norm-bound is required. 
Similarly, if for some $i'$, $||{{\mathbf{w}}}(i')||=\sqrt{N}\delta_w$, then ${\mathbf{a}}(i)=\gamma {{\mathbf{w}}}(i)+{\mathbf{a}}'(i)$ with any  ${\mathbf{a}}'(i)$ satisfying $||{\mathbf{a}}'(i)||\leq (1-|\gamma+1|)\sqrt{N}\delta_w$ and $-2<\gamma<0$ is a stealthy nonzero attack vector. 
(again, $\mathbf{a}'(i')$ in only norm constrained). 
\end{pf}

\begin{remark}
\label{rem:findingPA}
In Definitions \ref{def:PA},~\ref{def:PAtime}, 
we only focus on whether there exists such a sequence of nonzero stealthy attack vectors that results in unbounded estimation errors, and thus, making  the system PA -- i.e., we do not consider how the attacker attempts to find it.
\end{remark}

\begin{pf} [Proof of Theorem~\ref{thm:PA}]
\\a) First, assume that the system $\Sigma_{\textit{\rom{1}}}(A,C,\delta_w,\mathcal{K})$ is PA over time. Based on Remark~\ref{rem:PA12}, it is also PA at a single time step. Inversely, assume that $\Sigma_{\textit{\rom{1}}}(A,C,\delta_w,\mathcal{K})$ is PA at a single time step. Suppose that the attack starts at time $t_0$. Thus, 
$\mathcal{D}_{\textit{\rom{1}}}(\hat{\mathbf{a}}(t))=0$ for any $t<t_0-(N-1)$. The augmented attack vector ${\mathbf{a}}(t_0-(N-1))$ will be
\vspace{-8pt}
\begin{equation}
\begin{split}
{\mathbf{a}}(t_0-(N-1))= 
\begin{bmatrix}
0^T &(\mathcal{P}_{\mathcal{K}}{\mathbf{a}}(t_0-(N-1)))^T
\end{bmatrix}^T,
\end{split}
\end{equation}
where ~~$
\mathcal{P}_{\mathcal{K}}{\mathbf{a}}(t_0-(N-1))=\begin{bmatrix}
0&\dots&0&(\mathcal{P}_{\mathcal{K}}a(t_0))^T
\end{bmatrix}^T.$

As $F(\mathcal{K},N)$ is not full rank, there exists a nonzero~vector $z(t_0-(N-1))$ where $F(\mathcal{K},N)z(t_0-(N-1))=0$ and
\vspace{-8pt}
\begin{equation}
\begin{split}
&{\mathbf{a}}(t_0-(N-1))=\begin{bmatrix}
0&\dots&0&(\mathcal{P}_{\mathcal{K}}a(t_0))^T
\end{bmatrix}^T\\
&=\begin{bmatrix}
\mathbf{O}_{\mathcal{K}^\complement}\\\mathbf{O}_{\mathcal{K}}
\end{bmatrix} z(t_0-(N-1))=\mathbf{O}z(t_0-(N-1)).
\end{split}
\end{equation}
Here, $z(t_0-(N-1))$ can be chosen arbitrarily large -- i.e., ${\mathbf{a}}(t_0-(N-1))$ is a perfect attack vector. 
Now, from Lemma~\ref{lem:BPA}, the consecutive perfect attack vectors can also be constructed using ${\mathbf{a}}(t)=\mathbf{O}z(t)$ with $z(t)=A^{t-t_0+(N-1)}z(t_0-(N-1))+\sum_{i=t_0}^{t-1}A^{t-i-1}\alpha(i)$ for any $t>t_0-(N-1)$, where $\alpha(i) \in {\mathcal{N}(F(\mathcal{K},N))}$. Since $\alpha$ can be arbitrarily large, the system will have arbitrarily large estimation error for $t\geq t_0-(N-1)$ while remaining stealthy from $\text{ID}_{\textit{\rom{1}}}$ -- i.e.,  $\Sigma_{\textit{\rom{1}}}(A,C,\delta_w,\mathcal{K})$ is PA over time.

\vspace{4pt}
b) ($\Leftarrow$) {Suppose that $A$} is unstable and the system is PA at a time step; thus, $\mathbf{O}_{{\mathcal{K}^\complement}}$ is not full rank. From {Lemma \ref{lem:ExistPA}} (and its proof), there exists a nonzero attack vector ${\mathbf{a}}(t_0)$ such that for any $t_0-(N-1)\leq i\leq t_0$, $\mathcal{D}_{\textit{\rom{1}}}(\hat{\mathbf{a}}(i))=0$, 
as well as ${\mathbf{a}}(t_0)=\mathbf{O}z(t_0)$  and  $\mathbf{O}_{{\mathcal{K}^\complement}}z(t_0)=0$ (this holds, from the proof of the lemma which only constraints ${\mathbf{a}}(t_0)$ to have a certain norm bound).



Based on {Lemma~\ref{lem:BPA}} if $z(t_0+1)=Az(t_0)+\mathcal{N}(F(\mathcal{K},N))$, it is possible to have $\mathbf{a}(t_0+1)=\mathbf{O}z(t_0+1)$ with $\mathcal{D}_{\textit{\rom{1}}}(\hat{\mathbf{a}}(t_0+1))=0$. Since $F(\mathcal{K},N)$ is full rank, $\mathcal{N}(F(\mathcal{K},N))=0$. By continuing inserting attack vector in the form of $\mathbf{a}(t)=\mathbf{O}z(t)$ for a period of time $[t_0,t]$, we can get $z(t)=A^{t-t_0}z(t_0)$. Now, we consider two cases:

\textit{\textbf{Case \rom{1}} --} The unstable eigenvalues of the matrix $A$ are diagonizable. Let us denote by $v_1,...,v_q$ eigenvectors that correspond to unstable eigenvalues of matrix $A$, which we sometimes refer to as `unstable' eigenvectors. From the theorem assumption, one of these eigenvectors $v_i\in \mathcal{N}(\mathcal{O}_{\mathcal{K}^\complement})$, $i\in\{1,...,q\}$. Now, if we consider $z(t_0)=cv_i\neq 0$, 
we get $\mathbf{O}_{{\mathcal{K}^\complement}}z(t_0)=0$ where $c$ is chosen so that $\norm{\mathbf{O}z(t_0)}= \epsilon$, for some $\epsilon>0$. Hence, we get   
$z(t)=A^{t-t_0}z(t_0)=c\lambda_i^{t-t_0}v_i$. Since $|\lambda_i|>1$, $||z(t)||$ will be unbounded if $t\to \infty$. Therefore, based on the Corollary~\ref{cor:RSEerr} and Definition~\ref{def:PAtime} the system is PA over time. \\
\textit{\textbf{Case \rom{2}} --} Unstable eigenvalues of $A$ are not diagonizable and we consider generalized eigenvectors. For each independent eigenvector $v_i$ associated with $|\lambda_i|\geq 1$, we index its generalized eigenvector chain with length of $q_i$ as $v_{i+1},...,v_{i+q_i}$, { where $v_i\in \mathcal{N}(\mathcal{O}_{\mathcal{K}^\complement})$}. Now, consider $z(t_0)=cv_{i+q_i}$.  
Similarly to the \textbf{Case I}, we get $z(t)=A^{t-t_0}z(t_0)=\sum \limits_{l=0}^{q_i}c {{t-t_0}\choose{l}}\lambda_i^{t-t_0-l}v_{i+q_i-l}$~\cite{golub2012matrix}.
Since $|\lambda_i|\geq 1$, $z(t)$ is unbounded when $t\to \infty$. Hence, 
the system will be PA over time.

($\Rightarrow{}$) Let us assume that the system is PA over time and $A$ is stable. From Definition \ref{def:PAtime}, for all $M>0$ there exists a time step $t'$ such that for any $t\geq t'$, $||\Delta{x}(t)||>M$. 

Since $F(\mathcal{K},N)$ is full rank, from Corollary~{\ref{cor:iniatialPA}},  the estimation error is bounded when attack starts at $t_0+N-1$, i.e., $||\Delta{x}(t_0)||\leq \delta$ for some $\delta >0$. Now, for the interval $t_0<t<t'$ from Lemma~\ref{lem:EstimError},  $\Delta{x}(t)=A^{t-t_0}\Delta{x}(t_0)+\sum_{i=t_0}^{t-1}A^{t-i-1}p(i)$. Since the eigenvectors of $A$ span $\mathbb{R}^n$ (here we assume $A$ is diagonizable, yet, the results can be easily extended to the undiagonizable case),~we~have
\vspace{-10pt}
\begin{equation*}\label{eq{25}}
\begin{split}
\norm{\Delta{x}(t')}&=\norm{A^{t'-t_0}\Delta{x}(t_0)+\sum_{i=t_0}^{t'-1}A^{t'-i-1}p(i)}\\
=&\norm{\sum_{j=1}^{n}d_j\lambda_j^{t'-t_0}v_j+\sum_{i=t_0}^{t'-1}\sum_{j=1}^{n}d'_{i,j}\lambda_j^{t'-i-1}v_j}\\
\leq &|\lambda_{max}|^{t'-t_0}\norm{\Delta{x}(t_0)}+\norm{\sum_{i=t_0}^{t'-1}|\lambda_{max}|^{t'-i-1}p(i)}\\
\leq & \delta+ \frac{1}{1-|\lambda_{max}|}p_{max},
\end{split}
\end{equation*}
where $\lambda_{max}$ is the largest-norm eigenvalue and $p_{max}=\max_{t_0\leq i\leq t'-1}||p(i)||$. As $|\lambda_{max}|<1$ ($A$ is stable),~for~all $t'>t_0$, we have $|\lambda_{max}|^{t'-t_0}<1$. Thus,~$\norm{\Delta{x}(t')}$~is~bounded for any $t'>t_0$, contradicting that the system is PA.

Now, assume that none of the unstable eigenvectors of $A$ belong to $\mathcal{N}(\mathcal{O}_{\mathcal{K}^\complement})$. 
Again, we have that $\Delta{x}(t')$ can be written as
$
\Delta{x}(t')=c_1(t')v_1+...+c_q(t')v_q+\sigma (t'),$
where $c_j(t')=(d_j\lambda_j^{t'-t_0}+\sum_{i=t_0}^{t'-1}d'_{i,j}\lambda_j^{t'-i-1})v_j$ for $j=1,...,q$ and $\sigma (t')$ is the expansion of $\Delta{x}(t')$ over stable eigenvalues (satisfying $\sigma (t')\rightarrow 0$ as $t'\rightarrow\infty$). As the system is PA 
over time,
at least one of the coefficients $c_j(t')$ should be arbitrarily large as $t'$ increases. Now, since $\hat{\mathbf{a}}(t)=0$, it follows that $\mathbf{O}_{\mathcal{K}^\complement}\Delta{x}(t')=\mathcal{P}_{\mathcal{K}^{\complement}}\Delta{\mathbf{w}}(t'),$ making $\mathbf{O}_{\mathcal{K}^\complement}\Delta{x}(t')$ bounded. 
Thus, $\mathbf{O}_{\mathcal{K}^\complement}c_j(t')v_j$ is bounded because $v_1,...,v_n$ span $\mathbb{R}^n$ (the results can be easily extended to the case where $A$ is not-diagonizable)
and the other unstable eigenvectors cannot be used to compensate for $c_j(t')v_j$. From
$\mathbf{O}_{\mathcal{K}^\complement}c_j(t')v_j$ being bounded while $c_j(t')$ is arbitrarily large, it holds that $v_j\in\mathcal{N}(\mathbf{O}_{\mathcal{K}^\complement})$, 
which is a contradiction -- i.e., there exists an unstable eigenvector that lies in $\mathcal{N}(\mathbf{O}_{\mathcal{K}^\complement})$. 
%
%
\end{pf}    

\begin{example}
Consider the system  $\Sigma_{\textit{\rom{1}}}(A,C,\delta_w,\mathcal{K})$ from Example~\ref{ex:1}; it holds that $F(\mathcal{K},N)=
C$ for $N=2$. If we assume that attack starts at time zero, then it suffices to have ${\mathbf{a}}(-1)=\begin{bmatrix}
a(-1)&a(0)
\end{bmatrix}^T=\mathbf{O}z(-1)$ with $a(-1)=0$. By solving this equation, we get $z(-1)=\begin{bmatrix}
0&\eta
\end{bmatrix}^T$ where $\eta$ can be chosen arbitrarily large to impose unbounded estimation error at time $-1$ (consider that although the attack starts at time $0$, delay of the RSE causes unbounded error even at time -1). By choosing $z(t)=Az(t-1)$ for $t\geq 0$, and using ${\mathbf{a}}(t)=\begin{bmatrix}
a(t-1)&a(t)
\end{bmatrix}^T=\mathbf{O}z(t)$, the attack vector can be  constructed over time. However, if we choose $N=3$, it is impossible to find the attack vector ${\mathbf{a}}(-2)=\begin{bmatrix}
a(-2)&a(-1)&a(0) \end{bmatrix}^T=\mathbf{O}z(-2)$ with $a(-2)=a(-1)=0$, and since matrix $A$ is stable, 
{it is impossible to perfectly attack the system over time.}
\end{example}

\subsection{Perfect Attackabilty for $\Sigma_{\textit{\rom{2}}}(A,C,\delta_w,\mathcal{K})$} 

As previously described, for the system $\Sigma_{\textit{\rom{2}}}(A,C,\delta_w,\mathcal{K})$ only PA over time should be considered; we now capture necessary and sufficient conditions. 

\begin{theorem}
System $\Sigma_{\textit{\rom{2}}}(A,C,\delta_w,\mathcal{K})$ is PA over time if and only if  $\Sigma_{\textit{\rom{1}}}(A,C,\delta_w,\mathcal{K})$ is PA at a single time step, $A$ is unstable {and least one eigenvector $v_i$ corresponding to an unstable eigenvalue satisfies $v_i\in \mathcal{N}(\mathcal{O}_{\mathcal{K}^\complement})$.} 
\end{theorem}

\begin{pf}
($\Rightarrow$) Assume that  $\Sigma_{\textit{\rom{2}}}(A,C,\delta_w,\mathcal{K})$ is PA over time. Then from Remark~\ref{rem:PA12},
$\Sigma_{\textit{\rom{1}}}(A,C,\delta_w,\mathcal{K})$ is PA at single time step. Hence, we  need to show that $A$ is unstable. So, let us assume that $A$ is stable while $\Sigma_{\textit{\rom{2}}}(A,C,\delta_w,\mathcal{K})$ is PA over time. 
From Definition \ref{def:PAtime}, $\forall M>0$ there exists a time point $t'$ such that for all $k\geq t'$, $||\Delta{x}(k)||>M$. 
Now, let us assume that the attack starts at $t_0+N-1$. Since $\Delta{x}(t_0-1)$ is bounded, from $\mathcal{D}_{\textit{\rom{2}}}\big(\hat{\mathbf{a}}(t_0),\hat{x}(t_0)\big)=0$ 
there exists $\delta>0$ such that $||\Delta{x}(t_0)||\leq \delta$.  Now, for the interval $t_0<t\leq t'$ we have $\Delta{x}(t)=A^{t-t_0}\Delta{x}(t_0)+\sum_{i=t_0}^{t-1}A^{t-i-1}(p(i)+\alpha(i))$ with $||\Delta{x}(t_0)||\leq \delta$. On the other hand, by combining the condition $\mathcal{D}_{\textit{\rom{2}}}\big(\hat{\mathbf{a}},\hat{x}\big)=0$ for all time steps $t\geq t_0$ with~\eqref{eq:plant_withoutB} ($x(t)=Ax(t-1)$) we~get 
\begin{equation}
\begin{split}
&||\hat{x}(t)-A\hat{x}(t-1)||\\
&=||\hat{x}(t)-A\hat{x}(t-1)-x(t)+Ax(t-1)||\\
&=||\Delta{x}(t)-A\Delta{x}(t-1)||=||p(t)+\alpha(t)||\leq d   
\end{split}
\end{equation}


Since the eigenvectors of $A$ span the space $\mathbb{R}^n$ (here we assume the matrix $A$ is diagonizable, however, the results can be easily extended to the undiagonizable case), it holds that $\Delta{x}(t_0)=\alpha_1v_1+...+\alpha_nv_n$,  $p(i)=\beta_{i,1}v_1+...+\beta_{i,n}v_n$ and $\alpha(i)=\gamma_{i,1}v_1+...+\gamma_{i,n}v_n$. Now, we have
\vspace{-6pt}
\begin{equation*}\label{eq{26}}
\begin{split}
||\Delta{x}(t')||&=\norm{A^{t'-t_0}\Delta{x}(t_0)+\sum_{i=t_0}^{t'-1}A^{t'-i-1}(p(i)+\alpha(i))}\\
=&\norm{\sum_{j=1}^{n}\alpha_j\lambda_j^{t'-t_0}v_j+\sum_{i=t_0}^{t'-1}\sum_{j=1}^{n}(\beta_{i,j}+\gamma_{i,j})\lambda_j^{t'-i-1}v_j}
\\
\leq  |\lambda_{max}|^{t'-t_0}&\norm{\Delta{x}(t_0)}+\norm{\sum_{i=t_0}^{t-1}|\lambda_{max}|^{t'-i-1}(p(i)+\alpha(i))}\\
&\leq  \delta+\frac{d}{1-|\lambda_{max}|},
\end{split}
\end{equation*}
where $\lambda_{max}$ is the eigenvalue with the largest absolute value. Based on our assumption, $|\lambda_{max}|<1$  and for $t>t_0$ we have also $|\lambda_{max}|^{t-t_0}<1$. Hence, $\norm{\Delta{x}(t')}$ will be bounded for any $t'>t_0$, contradicting our assumption that the system $\Sigma_{\textit{\rom{2}}}(A,C,\delta_w,\mathcal{K})$ is PA. Finally, proof that at least one unstable eigenvector belongs to  $\mathcal{N}(\mathcal{O}_{\mathcal{K}^\complement})$) directly follows the approach for Theorem~\ref{thm:PA}.

($\Leftarrow$) Suppose that matrix $A$ has at least one eigenvalue outside the unit circle. From {Lemma~\ref{lem:ExistPA}}, there exists a nonzero attack vector ${\mathbf{a}}(t_0)$ such that for any $t_0-(N-1)\leq i\leq t_0$, $\mathcal{D}_{\textit{\rom{2}}}(\hat{\mathbf{a}}(i))=0$. Thus, there exists $\epsilon>0$ such that $\norm{{\mathbf{a}}(t_0)}=\epsilon$, and  
similarly to the proof of Theorem~\ref{thm:PA}, we can consider ${\mathbf{a}}(t_0)=\mathbf{O}z(t_0)$. Since $\mathbf{O}$ is full rank and $\Sigma_{\textit{\rom{1}}}(A,C,\delta_w,\mathcal{K})$ is PA at a single time point, $z(t_0)$ can be any nonzero vector that satisfies $\norm{\mathbf{O}z(t_0)}= \epsilon$ with $\mathbf{O}_{\mathcal{K}^\complement}z({t_0})=0$; any such vector $z(t_0)$ may be chosen arbitrarily by the attacker. 

From Lemma~\ref{lem:BPA} if $z(t_0+1)=Az(t_0)$, then attack  $\mathbf{a}(t_0+1)=\mathbf{O}z(t_0+1)$ results in $\mathcal{D}_{\textit{\rom{1}}}(\hat{\mathbf{a}}(t_0+1))=0$. Now, we need to show $||\hat{x}(t_0+1)-A\hat{x}(t_0)||\leq d$. From {Corollary~\ref{cor:RSEerr}}
\begin{equation*}
\begin{split}
&||\hat{x}(t_0+1)-A\hat{x}(t_0)||=||\Delta{x}(t_0+1)-A\Delta{x}(t_0)|| \\
&=||\mathbf{O}^\dagger\Delta{{\mathbf{w}}}(t_0+1) + z(t_0+1)-A\mathbf{O}^\dagger\Delta{{\mathbf{w}}}(t_0)-A z(t_0)||\\
&=||\mathbf{O}^\dagger\Delta{{\mathbf{w}}}(t_0+1)-A\mathbf{O}^\dagger\Delta{{\mathbf{w}}}(t_0)||\leq d
\end{split}
\end{equation*}
By continuing with attacks in the form of $\mathbf{a}(t)=\mathbf{O}z(t)$ for a period of time $[t_0,t]$, we get $z(t)=A^{t-t_0}z(t_0)$ while remaining stealthy from \text{ID$_{\textit{\rom{2}}}$}. Now, consider two cases: 

\textit{\textbf{Case \rom{1}} --} The unstable eigenvalues of $A$ are diagonizable. Let us denote by $v_1,...,v_q$ eigenvectors that correspond to
unstable eigenvalues of matrix $A$. From our assumption, there exists $v_i\in \mathcal{N}(\mathcal{O}_{\mathcal{K}^\complement})$, $i\in\{1,...,q\}$. Now, if we consider $z(t_0)=cv_i\neq 0$, we get $\mathbf{O}_{{\mathcal{K}^\complement}}z(t_0)=0$, where $c$ is chosen such that $\norm{\mathbf{O}z(t_0)}\leq \epsilon$. Thus,    
$z(t)=A^{t-t_0}z(t_0)=c\lambda_1^{t-t_0}v_i$. Since $|\lambda_i|>1$, $||z(t)||$ will be unbounded if $t\to \infty$. Hence, from Corollary~\ref{cor:RSEerr} and Definition~\ref{def:PAtime}, the system will be PA over time. \\
\textit{\textbf{Case \rom{2}} --} The unstable eigenvalues of $A$ are not diagonizable and we consider generalized eigenvectors. For each independent eigenvector $v_i$ associated with $|\lambda_i|\geq 1$, we index its generalized eigenvector chain with length~$q_i$ as $v_{i+1},...,v_{i+q_i}$. Consider $z(t_0)=cv_{i+q_i}$, where $v_i\in \mathcal{N}(\mathcal{O}_{\mathcal{K}^\complement})$. Similarly to \textbf{Case~I}, $z(t)=A^{t-t_0}z(t_0)=\sum \limits_{l=0}^{q_i}c {{t-t_0}\choose{l}}\lambda_i^{t-t_0-l}v_{i+q_i-1}$~\cite{golub2012matrix}. Since $|\lambda_i|\geq 1$, $z(t)$ is unbounded as $t\to \infty$, and from Corollary~\ref{cor:RSEerr} 
the system is PA over time.
\end{pf}

\vspace{-2pt}
The condition of PA over time for $\Sigma_{\textit{\rom{2}}}(A,C,\delta_w,\mathcal{K})$ is the same as for $\Sigma_{\textit{\rom{1}}}(A,C,\delta_w,\mathcal{K})$ when $F(\mathcal{K},N)$ is full~rank. When $F(\mathcal{K},N)$ is rank deficient, we can use $N=n+1$ to make the matrix full rank and get the same PA condition as for $\Sigma_{\textit{\rom{2}}}(A,C,\delta_w,\mathcal{K})$. Yet, increasing $N$ would increase computational overhead at each time step, which may be a problem in resource-constrained systems. Instead, one can use  $\Sigma_{\textit{\rom{2}}}(A,C,\delta_w,\mathcal{K})$ (i.e., $ID_{\textit{\rom{2}}}$) that only requires additional comparison, from~\eqref{eq:ID2}, at each time~step. 

\vspace{-2pt}
\section{Estimation with Intermittent Authentication} \label{sec:intermittent}
\vspace{-4pt}

We now study the effects of intermittent data authentication (sometimes refered to as intermittent integrity enforcement~\cite{jovanov_tac19}) on estimation error of $\Sigma(A,C,\delta_w,\mathcal{K})$. 

\begin{definition}
The intermittent data authentication policy for $i$-th sensor ($s_i\in{\mathcal{S}}$), denoted by  $(\mu_i,L_i)$ where $\mu_i=\{t_k^i\}_{k=0}^\infty$ such that $t_{k}^i>t_{k-1}^i$ and $L_i=sup(t_k^i-t_{k-1}^i)$, ensures that
$
a_i(t_k^i)=0.$
\end{definition}

Intermittent data authentication for sensor $i$ guarantees that the attack injected through the $i$-th sensor is zero at some specific points ($t_k^i$), whereas the interval between each of consecutive points is at most $L_i$ time steps. A global intermittent authentication policy is defined if all sensors use same $(\mu_i,L_i)$. 
We now capture conditions that $\Sigma_{\textit{\rom{1}}}(A,C,\delta_w,\mathcal{K})$, satisfying Theorem~\ref{thm:PAt}, is not PA.

\begin{theorem}\label{thm:NotPA}
If ${\mathcal{I}}_{i}\subseteq{\mathcal{S}}$, $i\in\{1,...,N\}$, 
consider matrix 
\vspace{-2pt}
\begin{equation}
\mathbf{O}_{\mathcal{I},\mathcal{K}^\complement}=
\begin{bmatrix}
(\mathcal{P}_{{\mathcal{I}}_{1}\cup{\mathcal{K}}^\complement}C)^T&...(\mathcal{P}_{{\mathcal{I}}_{N}\cup{\mathcal{K}}^\complement}CA^{N-1})^T
\end{bmatrix}^T.
\end{equation}
If intermittent data authentication is used at time $t+i$ for each sensor set $\mathcal{I}_{i}$, $i\in\{0,...,N-1\}$, then $\Sigma_{\textit{\rom{1}}}(A,C,\delta_w,\mathcal{K})$ is not PA at time $t$  if and only if $\mathbf{O}_{\mathcal{I},\mathcal{K}^\complement}$ is full rank. 
\end{theorem}

\begin{pf}
($\Leftarrow$) Suppose 
$\Sigma_{\textit{\rom{1}}}(A,C,\delta_w,\mathcal{K})$ is PA at time $t$. Since for any $i\in{1,...,N}$ intermittent data authentication is used, $\mathcal{P}_{{\mathcal{I}}_{i}}{a}(t+i)=0$, and thus
\vspace{-10pt}
\begin{equation*}\label{eq{16}}
\begin{split}
\mathbf{O}_{\mathcal{I},\mathcal{K}^\complement}x(t)+&\begin{bmatrix}
\mathcal{P}_{{\mathcal{I}}_{1}\cup{\mathcal{K}}^\complement}{w}(t)\\\vdots\\\mathcal{P}_{{\mathcal{I}}_{N}\cup{\mathcal{K}}^\complement}{w}(t+N-1)
\end{bmatrix}=\\&\mathbf{O}_{\mathcal{I},\mathcal{K}^\complement}\hat{x}(t)
+\begin{bmatrix}
\mathcal{P}_{{\mathcal{I}}_{1}\cup{\mathcal{K}}^\complement}\hat{{w}}(t)\\\vdots\\\mathcal{P}_{{\mathcal{I}}_{N}\cup{\mathcal{K}}^\complement}\hat{{w}}(t+N-1)
\end{bmatrix} \Rightarrow
\end{split}
\end{equation*}
\vspace{-20pt}
\begin{equation}\label{eq{20}}
\begin{split}
&\mathbf{O}_{\mathcal{I},\mathcal{K}^\complement}\Delta{x}(t)=\\
&=\begin{bmatrix}
\mathcal{P}_{{\mathcal{I}}_{1}\cup{\mathcal{K}}^\complement}\Big({w}(t)-\hat{{w}}(t)\Big)\\\vdots\\\mathcal{P}_{{\mathcal{I}}_{N}\cup{\mathcal{K}}^\complement}\Big({w}(t+N-1)-\hat{{w}}(t+N-1)\Big)
\end{bmatrix}
\end{split}
\end{equation}
We denote the right side of~\eqref{eq{20}} as $f(t)$. Since $\mathbf{O}_{\mathcal{I},\mathcal{K}^\complement}$ is full rank, from~\eqref{eq{20}} we have
$\Delta{x}(t)=\mathbf{O}_{\mathcal{I},\mathcal{K}^\complement}^\dagger f(t);
$
i.e., 
$\norm{\Delta{x}(t)}=\norm{\mathbf{O}_{\mathcal{I},\mathcal{K}^\complement}^\dagger f(t)} \leq \norm{\mathbf{O}_{\mathcal{I},\mathcal{K}^\complement}^\dagger}\norm{f(t)}=\bar{c}.
$
As the actual and estimated noise are bounded, $\norm{f(t)}$ and $\bar{c}$ are also bounded, contradicting PA of the system.  

($\Rightarrow$) System $\Sigma_{\textit{\rom{1}}}(A,C,\delta_w,\mathcal{K})$ is not PA at time $t$, and 
assume that $\mathbf{O}_{\mathcal{I},\mathcal{K}^\complement}$ is not full rank. 
Then, exists a nonzero vector $z$ such that $\mathbf{O}_{\mathcal{I},\mathcal{K}^\complement}z=0$; thus,  $\mathbf{O}_{\mathcal{K}^\complement}z=0$ and the pair ($\mathcal{P}_{\mathcal{K}^\complement}C,A$) is not observable, From Theorem~\ref{thm:PAt},  $\Sigma_{\textit{\rom{1}}}(A,C,\delta_w,\mathcal{K})$ is PA at  $t$, which is a contradiction. 
\end{pf}

Theorem~\ref{thm:NotPA} provides an intermittent data authentication policy such that the system is not PA at a single time step. To derive conditions of not being PA for all time steps, the condition of Theorem~\ref{thm:NotPA} should be satisfied at each time. Our goal is to derive conditions that a system is not PA over time, and we start with the following. 

\begin{proposition}
\label{Proposition3}
Assume  $F(\mathcal{S},N)$ is not full rank. Then system
$\Sigma_{\textit{\rom{1}}}(A,C,\delta_w,\mathcal{K})$ is not PA over time for any compromised sensor set $\mathcal{K}$, if the intermittent data authentication policy is used with $L_i=1$, $\forall i\in \mathcal{F}$, where $\mathcal{F}$ is a sensor subset such that the pair $(A,\mathcal{P}_{\mathcal{F}}C)$ is observable. 
\end{proposition}

\vspace{-10pt}
\begin{pf}
As for any $t>0$, authentication is used in $\{t,...,t+N-1\}$, $\mathcal{P}_{\mathcal{F}}a(t)=\mathcal{P}_{\mathcal{F}}a(t+1)=...=\mathcal{P}_{\mathcal{F}}a(t+N-1)=0$. The corresponding matrix for the authentication policy is 
$
\mathbf{O}_{\mathcal{F}}=\begin{bmatrix}
\mathcal{P}_{\mathcal{F}}C^T&\mathcal{P}_{\mathcal{F}}(CA)^T&...&\mathcal{P}_{\mathcal{F}}(CA^{N-1})^T
\end{bmatrix}^T.
$
Since $\mathbf{O}_{\mathcal{F}}$ is full rank, from Theorem \ref{thm:NotPA}, system $\Sigma(A,C,\mathcal{K})$ is not PA at time $t$. As this holds for any time $t$, from Definition~\ref{def:PAtime} the system is not PA over~time.
\end{pf}

From Proposition \ref{Proposition3} it follows that if matrix $F(\mathcal{S},N)$ is not full rank, then we can avoid PA over time by using data authentication at each time step for some specific subset of sensors. Although this may seem conservative, but in the following example, we show that a perfect attack can be achieved by only compromising suitable sensors at a single time step.  

\begin{example}
Consider again the model from Example~\ref{ex:1}, and assume that the attack is only injected at time zero. There are two vectors $z(-1)$, $z(0)\in {\mathbb{R}}^2$ which can satisfy 
$
\mathbf{a}(-1)=\begin{bmatrix}
a(-1)\\a(0)
\end{bmatrix}=\mathbf{O}z(-1)=\begin{bmatrix}
1&0\\.3&1
\end{bmatrix}\begin{bmatrix}
z_1(-1)\\z_2(-1)
\end{bmatrix},$
$\mathbf{a}(0)=\begin{bmatrix}
a(0)\\a(1)
\end{bmatrix}=\mathbf{O}z(0)=\begin{bmatrix}
1&0\\.3&1
\end{bmatrix}\begin{bmatrix}
z_1(0)\\z_2(0)
\end{bmatrix},
$
with $a(-1)=a(1)=0$. Solving the above two equations gives $a(0)=z_2(-1)=z_1(0)=-\frac{z_2(0)}{.3}$ and $z_1(-1)=0$. Using Corollary~\ref{cor:RSEerr}, we get $\Delta{x}(-1)=\begin{bmatrix}
0&a(0)
\end{bmatrix}^T$ and $\Delta{x}(0)=\begin{bmatrix}
a(0)&-.3a(0)
\end{bmatrix}^T$ (which can be chosen arbitrarily large by controlling the scalar $a(0)$), whereas \text{ID}$_{\textit{\rom{1}}}$ will not trigger alarm in these two time steps. Consider that $a(0)$ is not included in other time steps; thus, by inserting attack vector only at time zero, the system $\Sigma_{\textit{\rom{1}}}(A,C,0,s_1)$ can have unbounded estimation error without triggering~alarm.
\end{example}

The above example shows that for $\Sigma_{\textit{\rom{1}}}(A,C,\delta_w,\mathcal{K})$ when $F(\mathcal{S},N)$ is not full rank, a stealthy attack can result in arbitrarily large estimation error, even 
by injecting false data only at one time step. 
Hence, it is essential to use data authentication at all time steps -- i.e., non-intermittently. However, as shown below, when  $F(\mathcal{S},N)$ is full rank, $\Sigma_{\textit{\rom{1}}}(A,C,\delta_w,\mathcal{K})$ cannot be PA over time even when only intermittent authentication is used; this holds for $\Sigma_{\textit{\rom{2}}}(A,C,\delta_w,\mathcal{K})$ independently of the $F(\mathcal{S},N)$~rank. 
\begin{theorem}
Consider two cases: 
a) $\Sigma_{\textit{\rom{1}}}(A,C,\delta_w,\mathcal{K})$ with full rank $F(\mathcal{S},N)$; 
b) $\Sigma_{\textit{\rom{2}}}(A,C,\delta_w,\mathcal{K})$.
Both (a) and (b) are not PA over time if the intermittent authentication policy is used with $L_i=\mathcal{T}$,  $\forall i\in \mathcal{F}$ for a bounded~$\mathcal{T}$, where $\mathcal{F}$ is any sensor set such that $(A,\mathcal{P}_{\mathcal{F}}C)$ is observable. 
\end{theorem}	
\begin{pf}
From Lemma~\ref{lem:EstimError} and~\eqref{eq{15}}, it follows that
\begin{equation}
\label{eq{51}}
\begin{split}
\Delta{x}(t+1)=&A\Delta{x}(t)+\alpha(t)+p(t)\\
{\mathbf{a}}(t)=&\mathbf{O}\Delta {x}(t)+\Delta{{\mathbf{w}}}(t)
\end{split}
\end{equation}
for any $t\geq t_0$ if the attacker initiates the attack at time $t_0+N-1$. 
For system (a), $\alpha(t)=0$ and since $p(t)$ is bounded at all time steps $t$, thus $p(t)+\alpha(t)$ is bounded. For system (b) the stealthiness condition $||\Delta{x}(t+1)-A\Delta{x}(t)||<d$ causes $||p(t)+\alpha(t)||<d$. Thus, for both cases $p(t)+\alpha(t)$ is bounded. Assume $t_{k_0}$ is the first time instant that authentication is used after $t_0$. Then $\forall i\in \mathcal{F}$ we have 
\begin{equation}
a_i(t_{k_0})=a_i(t_{k_0}+\mathcal{T})=...=0
\end{equation}
Hence, $\mathcal{P}_{\mathcal{F}}a(t_{k_0})=\mathcal{P}_{\mathcal{F}}a(t_{k_0}+\mathcal{T})=...=0$. On the other hand, from~\eqref{eq{51}} we get $\mathcal{P}_{\mathcal{F}}a(t)=\mathcal{P}_{\mathcal{F}}C\Delta{x}(t)+\mathcal{P}_{\mathcal{F}}\Delta{w}(t)$ for any $t\geq t_0+N-1$. Now, consider $\mathcal{P}_{\mathcal{F}}a(t_{k_0}+i\mathcal{T})$ for any $i\geq 0$. Then, for $j\in \{1,...,N-1\}$ we have
\begin{equation*}
\begin{split}
&\mathcal{P}_{\mathcal{F}}a(t_{k_0}+(i+j)\mathcal{T})=\mathcal{P}_{\mathcal{F}}CA^{j\mathcal{T}}\Delta{x}(t_{k_0}+i\mathcal{T})\\
&+\sum_{f=t_{k_0}+i\mathcal{T}}^{t_{k_0}+(i+j)\mathcal{T}-1}\mathcal{P}_{\mathcal{F}}CA^{t_{k_0}+(i+j)\mathcal{T}-1-f}(p(f)+\alpha(f))\\
&+\mathcal{P}_{\mathcal{F}}\Delta{w}(t_{k_0}+(i+j)\mathcal{T})=0;
\end{split}
\end{equation*}
for $j=0$, $\mathcal{P}_{\mathcal{F}}a(t_{k_0}+(i+j)\mathcal{T})=\mathcal{P}_{\mathcal{F}}CA^{j\mathcal{T}}\Delta{x}(t_{k_0}+i\mathcal{T})$. By augmenting $\mathcal{P}_{\mathcal{F}}a(t_{k_0}+(i+j)\mathcal{T})$, $\forall j\in \{0,...,N-1\}$, 
\begin{equation*}
\begin{split}
&\begin{bmatrix}
\mathcal{P}_{\mathcal{F}}a(t_{k_0}+i\mathcal{T})\\\vdots\\\mathcal{P}_{\mathcal{F}}a(t_{k_0}+(i+N-1)\mathcal{T})\end{bmatrix}
=0\Rightarrow
\\
&\begin{bmatrix}
(\mathcal{P}_{\mathcal{F}}C)^T&\dots&(\mathcal{P}_{\mathcal{F}}CA^{(N-1)\mathcal{T}})^T
\end{bmatrix}^T\Delta{x}(t_{k_0}+i\mathcal{T})=\\&\begin{bmatrix}
\sum_{f=t_{k_0}+i\mathcal{T}}^{t_{k_0}+(i)\mathcal{T}-1}\mathcal{P}_{\mathcal{F}}CA^{t_{k_0}+(i)\mathcal{T}-1-f}(p(f)+\alpha(f))\\\vdots\\\sum_{f=t_{k_0}+i\mathcal{T}}^{t_{k_0}+(i+N-1)\mathcal{T}-1}\mathcal{P}_{\mathcal{F}}CA^{t_{k_0}+(i+N-1)\mathcal{T}-1-f}(p(f)+\alpha(f))
\end{bmatrix}\\
&+\begin{bmatrix}
\mathcal{P}_{\mathcal{F}}\Delta{w}(t_{k_0}+(i)\mathcal{T})\\\vdots\\\mathcal{P}_{\mathcal{F}}\Delta{w}(t_{k_0}+(i+N-1)\mathcal{T})
\end{bmatrix}
\end{split}
\end{equation*}
Now, since  $\begin{bmatrix}
(\mathcal{P}_{\mathcal{F}}C)^T&(\mathcal{P}_{\mathcal{F}}CA^{\mathcal{T}})^T&\dots&(\mathcal{P}_{\mathcal{F}}CA^{(N-1)\mathcal{T}})^T
\end{bmatrix}^T$ is full rank and the right side of the above equation is bounded, we have $\Delta{x}(t_{k_0}+i\mathcal{T})$ is bounded for any $i\geq 0$. On the other hand, from~\eqref{eq{51}} and the fact that $\alpha(t)$ and $p(t)$ are bounded for $t\geq t_0$, we can conclude that $\Delta{x}(t)$ is bounded for any $i\mathcal{T}\leq t \leq (i+1)\mathcal{T}$ for any $i\geq 0$. 
\end{pf}

\section{Numerical Results}\label{sec:numerical}

We illustrate our results on a realistic case study -- Vehicle Trajectory Following (VTF). 
Specifically, we show how the attacker can perfectly attack the system when the necessary conditions are satisfied and how intermittent data authentication effectively prevents such attacks. 
%
We consider the model from~\cite{kerns2014unmanned}, discretized with sampling time .01~s; i.e.,
$
A=\left[\begin{smallmatrix}
1&.01\\0&1
\end{smallmatrix}\right], B=\left[\begin{smallmatrix} 
.0001\\.01
\end{smallmatrix}\right], C=\left[\begin{smallmatrix} 
1&0&0\\
0&1&1
\end{smallmatrix}\right] ^T.
$
We assume that all sensors are compromised -- i.e.,  $\mathcal{K}=\mathcal{S}=\{s_1,s_2,s_3\} $ . Therefore, the system is PA over time as $A$ is also unstable. For simulation, we also assume that each element of sensor and system noise comes from uniform distribution $v_P,v_M\sim U(-.05,.05)$. 
Moreover, $N=2$ and the maximum possible estimation error when the system is not under attack is obtained as $||\Delta{x}||\leq ||\mathbf{O}^\dagger||||\Delta{{\mathbf{w}}}||=.0789$ from~\eqref{eq:16}. 

\begin{figure*}[!t]
\begin{subfigure}{0.33\textwidth}
\includegraphics[width=1\linewidth, height=2.9cm]{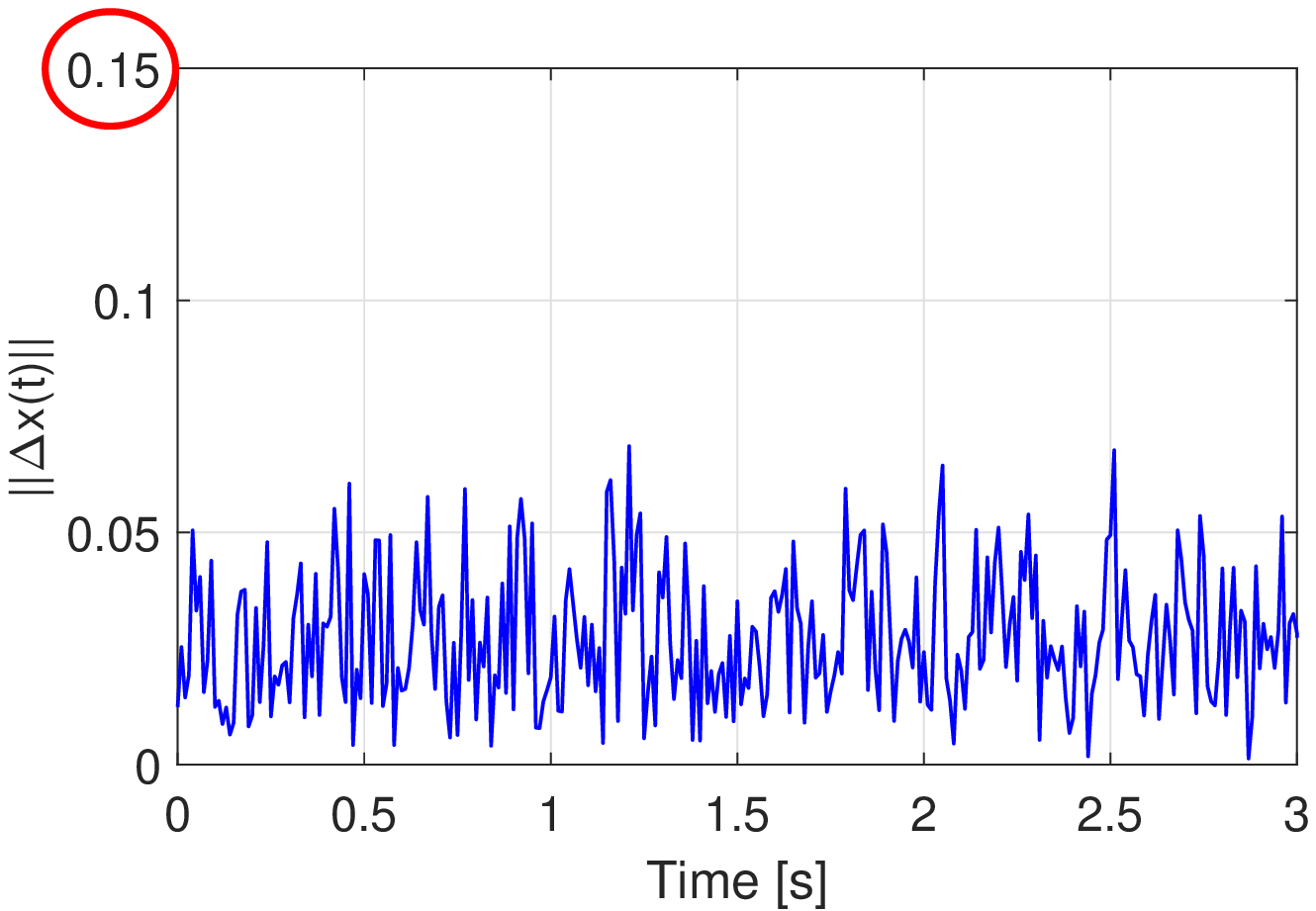} 
\caption{System without attack}
\label{fig:error_norm1}
\end{subfigure}
\begin{subfigure}{0.33\textwidth}
\includegraphics[width=1\linewidth, height=2.9cm]{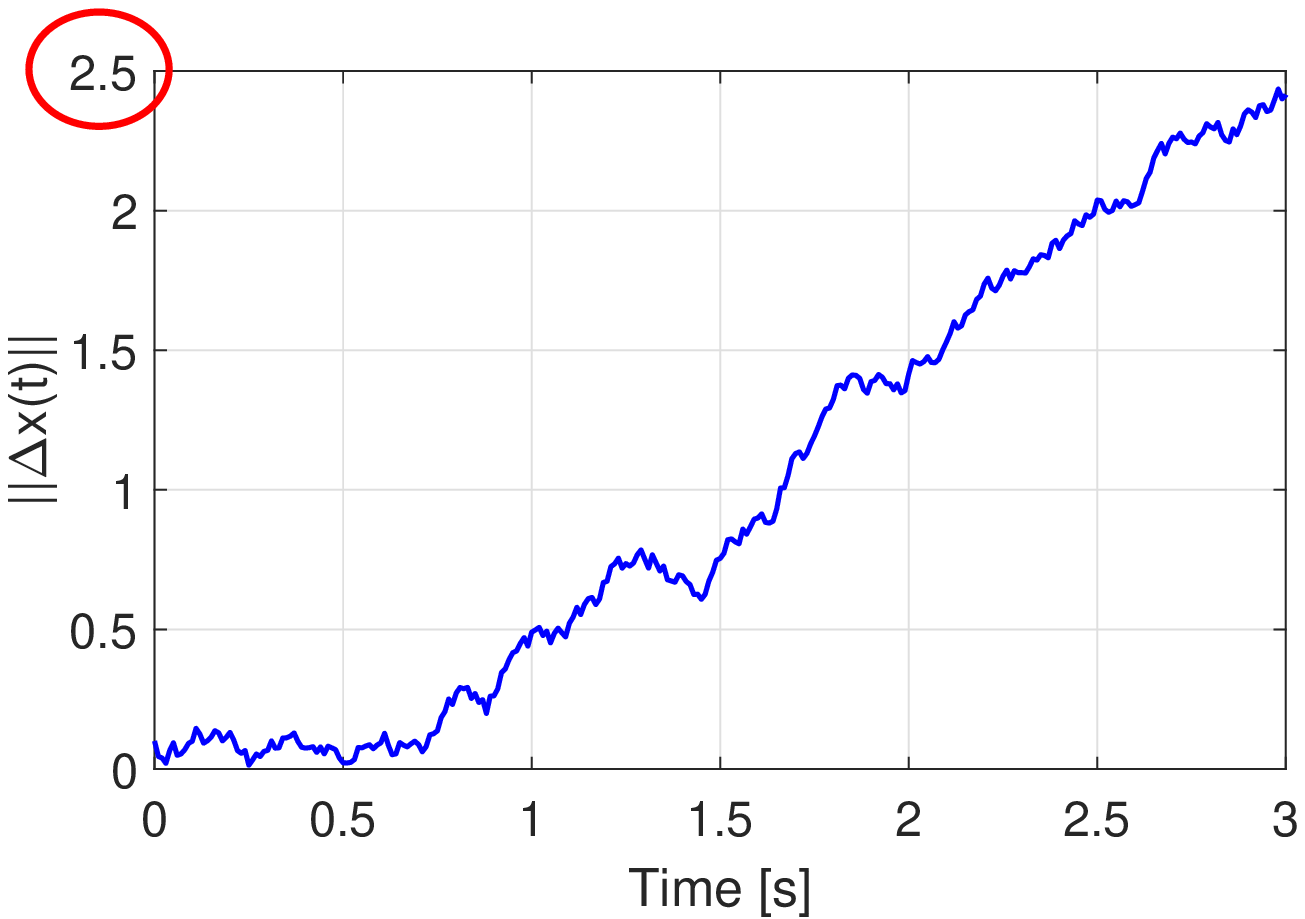}
\caption{Under a perfect attack, without authentication}
\label{fig:error_norm2}
\end{subfigure}
\begin{subfigure}{0.33\textwidth}
\includegraphics[width=1\linewidth, height=2.9cm]{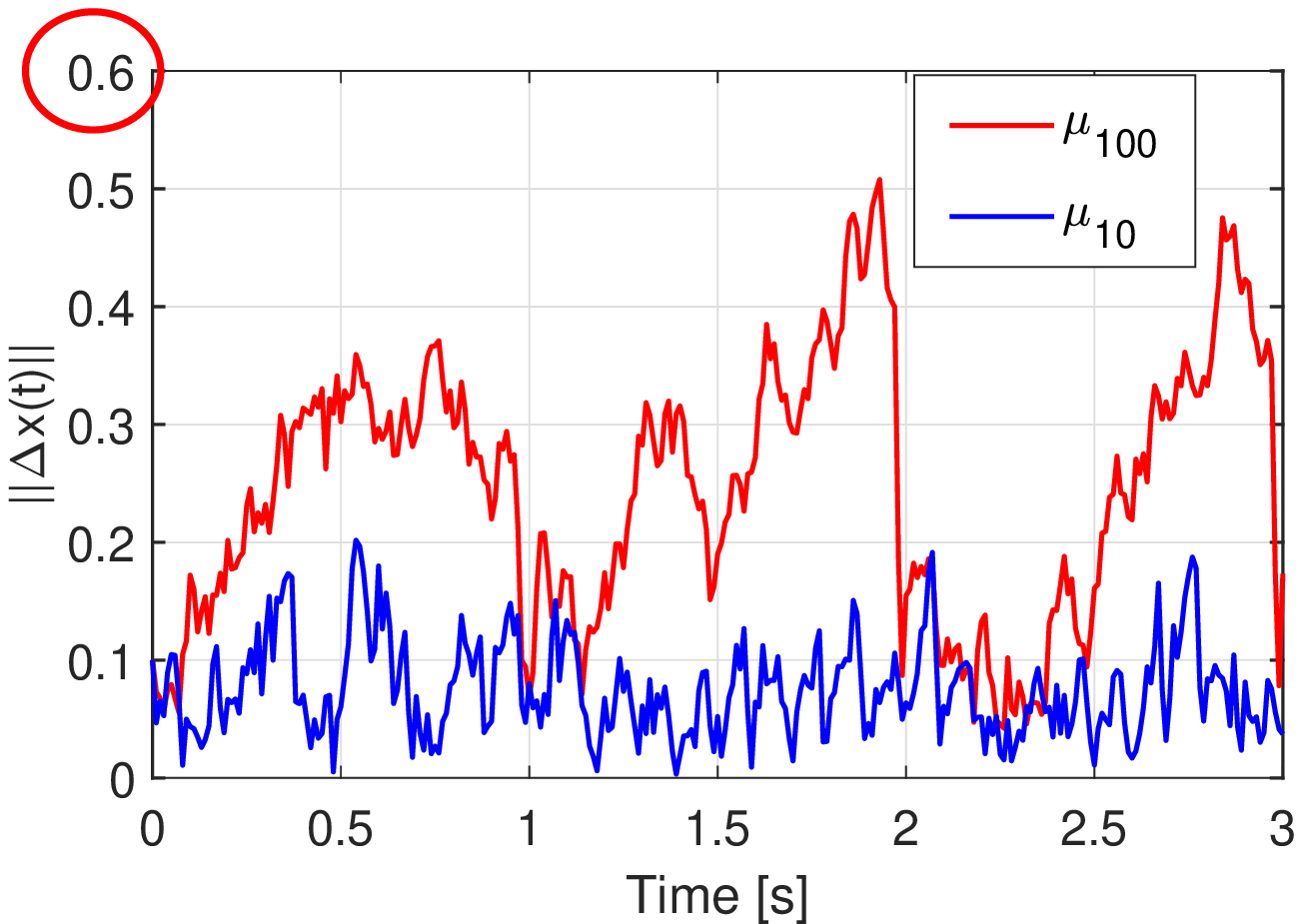}
\caption{With two different integrity enforcement policies, with periods $L=10,100$}
\label{fig:error_norm3}
\end{subfigure}
\vspace{-6pt}
\caption{Evolution of the estimation error norm for the VTF system; note the highlighted  (circled) different error ranges.
}
\label{fig:error_norm}
\end{figure*}

Fig. ~\ref{fig:error_norm} shows the evolution of the $l_2$ norm of estimation error in different scenarios. 
In Fig.~\ref{fig:error_norm1}, 
$||\Delta{x}(t)||$ is shown when the system is not under attack, whereas in Fig~\ref{fig:error_norm2} the system is under a perfect attack. 
In Fig~\ref{fig:error_norm3}, we considered two different data authentication policies $\mu_{10}$ and $\mu_{100}$; meaning $L=10$ and $L=100$, respectively, while the system is under by stealthy attack. As shown, when data authentication is used, the system is not PA over time and the estimation error remains bounded, and very low for less than 10\% of authenticated measurements; as the period of authentication increases, a stealthy attack can achieve higher maximum estimation~error. 

\begin{figure*}[!t]
\begin{subfigure}{0.33\textwidth}
\includegraphics[width=1\linewidth, height=2.9cm]{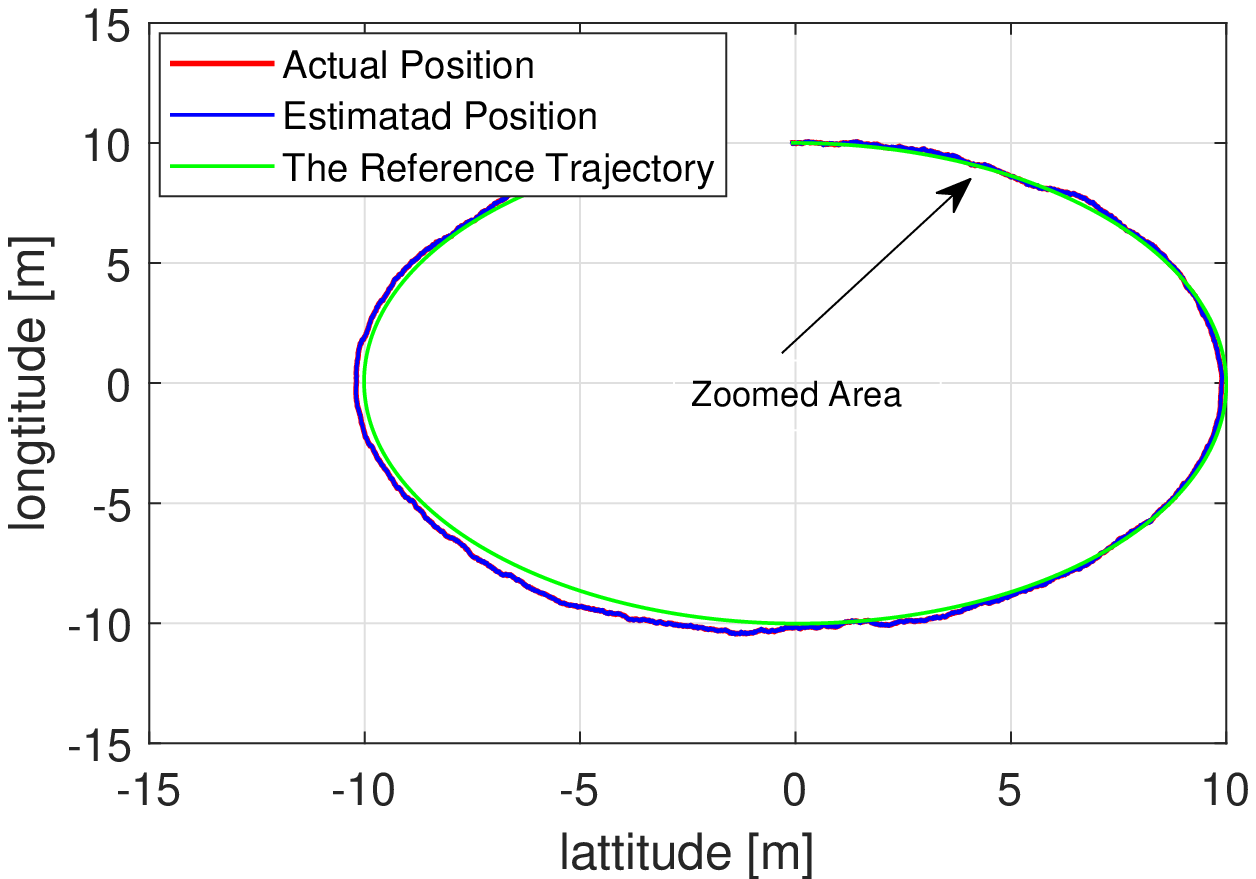} 
\caption{State estimates under stealthy attack with integrity enforcement policy $L=10$.}
\label{fig:tra1}
\end{subfigure}
\begin{subfigure}{0.33\textwidth}
\includegraphics[width=1\linewidth, height=2.9cm]{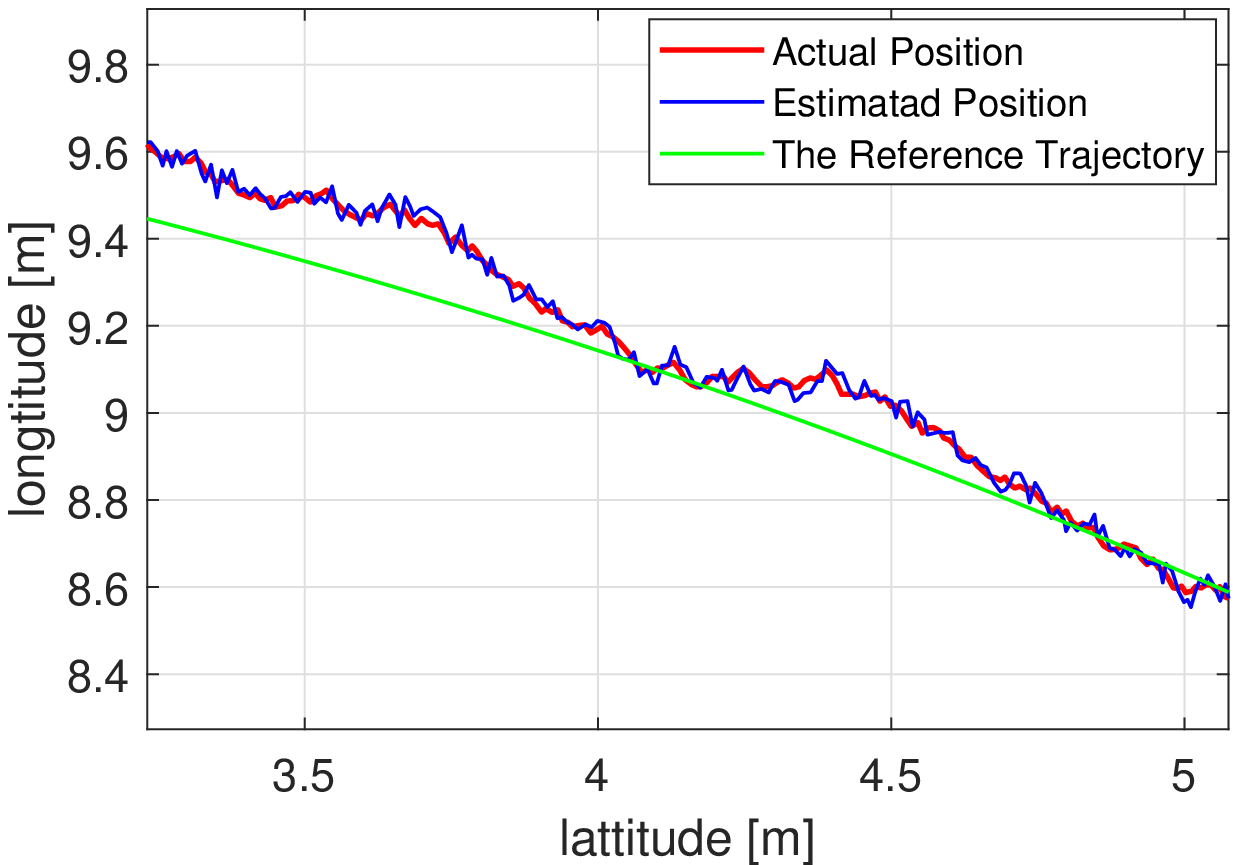}
\caption{Zoomed area of (a)}
\label{fig:tra2}
\end{subfigure}
\begin{subfigure}{0.33\textwidth}
\includegraphics[width=1\linewidth, height=2.9cm]{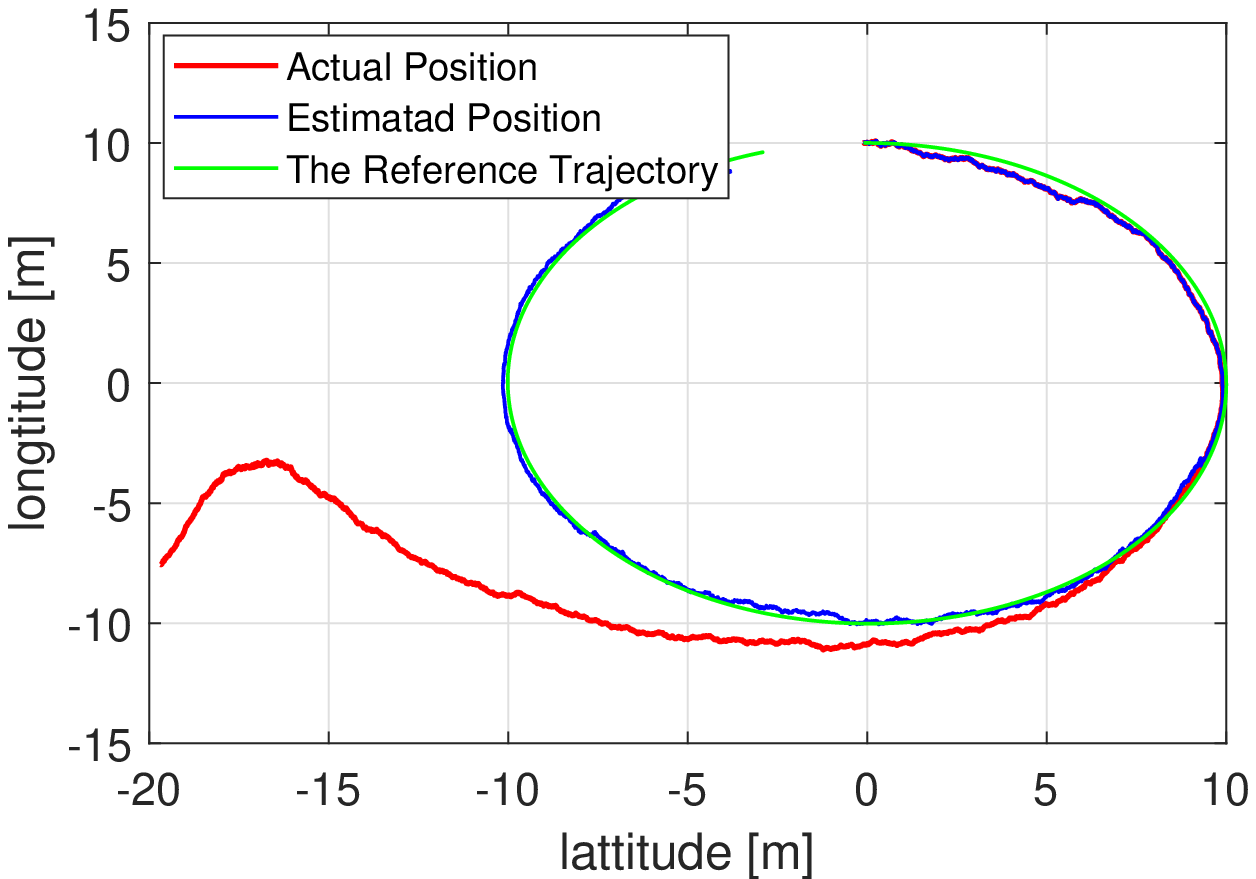}
\caption{State estimates under stealthy attack without integrity enforcement}
\label{fig:tra3}
\end{subfigure}
\caption{Simultaneous trajectory tracking and state estimation for the VTF system. At each time step, system state is estimated and used by the controller to track a circle trajectory. 
The duration of the simulation is $60~s$ and the attack starts at $t=20~s$. 
}
\label{fig:tra} 
\end{figure*}

Finally, we considered resilient state estimation within the VTF -- trajectory tracking; 
Fig~\ref{fig:tra} 
shows 60 seconds simulation. As shown, if a data authentication policy is used with $L=10$ (i.e., 10\% of authenticated messages), we obtain suitable control performance  even under stealthy attack. If  authentication is not used, a stealthy attack can force the system from the desired~path. 

\section{Conclusion}
\label{sec:concl}
\vspace{-10pt}
In this work, we considered the problem of resilient state estimation for LTI systems with bounded noise, when a subset of sensors are under attack. We defined two notions of perfect attackability (PA) -- at a time point and over time -- where stealthy attacks can cause an arbitrarily large estimation errors, and derived necessary and sufficient conditions for PA. We showed that,
unlike the Kalman filter-based observers,  batch processing-based resilient state estimators (RSE), such as $l_0$-based RSE, may be perfectly attackable 
even if the plant is not unstable. Furthermore, we studied the effects of intermittent data authentication on attack-induced estimation error. 
We showed that it is sufficient to even intermittently  use data authentication, once every bounded time period, to ensure that a system is not perfectly~attackable.


@article{Abl:45,
	author={B.C. Able},
	title={The examination of cell nuclei},
	journal={Birches. J.},
	year={1945},
	volume={35},
	pages={123--126},
}
@article{luo2019scalable,
  title={A Scalable and Optimal Graph-Search Method for Secure State Estimation},
  author={Luo, Xusheng and Pajic, Miroslav and Zavlanos, Michael M},
  journal={arXiv preprint arXiv:1903.10620},
  year={2019}
}
@article{Abl:56,
	author={B.C. Able},
	title={Nucleic acid content of microscope},
	journal={Nature},
	year={1956},
	volume={135},
	pages={7--9},
}

@incollection{AbTaRu:54,
	author={B.C. Able and R.A. Tagg and M. Rush},
	title={Enzyme-catalyzed cellular transanimations},
	pages={125--247},
	editor={A.F. Round},
	publisher={Academic  Press},
	year={1954},
	volume={2},
	booktitle={Advances in Enzymology},
	address={New York},
	edition={3rd},
}

@book{Bak:63a,
	author={R.C. Baker},
	title={Microscopic Staining Techniques},
	publisher={Butterworths},
	year={1963},
	address={London},
}

@article{Bak:63b,
	author={R.C. Baker},
	title={Methods of preparing thin-section slides},
	journal={J. Brit. Med. Assoc.},
	year={1963},
	volume={34},
	pages={184--186},
}

@book{Heritage:92,
      editor={A. H. Soukhanov},
	title="{The American Heritage. Dictionary of the American Language}",
	publisher={Houghton Mifflin Company},
	year={1992},

}

@article{ChaRou:66,
	author={F.H. Charlie and M.B. Routh},
	title={The Chemical Destination of Toxins},
	journal={J. Am. Chem. Soc.},
	year={1966},
	volume={66},
	pages={267--269},
}

@incollection{Dog:58,
	author={P.R. Dog},
	title={},
	booktitle={Chemical Carcinogenesis},
	publisher={Chapman \& Hall},
	year={1958},
	editor={R.W. Brown},
	chapter={7},
	volume={II},
	pages={56--98},
	address={London},
}

@book{Keo:58,
	author={R. Keohane},
	title={Power and Interdependence: World Politics in Transitions},
	publisher={Little, Brown \& Co.},
	year={1958},
	address={Boston},
}

@article{Pow:85,
	author={T. Powers},
	title={Is there a way out?},
	journal={Harpers},
	year={1985},
	pages={35--47},
	month=jun,
}


@misc{Sol:89,
	author={Solo},
	year={1989},
}






@article{chen2011lessons,
  title={Lessons from stuxnet},
  author={Chen, Thomas and Abu-Nimeh, Saeed},
  journal={Computer},
  volume={44},
  number={4},
  pages={91--93},
  year={2011}
}

@inproceedings{slay2007lessons,
  title={Lessons learned from the maroochy water breach},
  author={Slay, Jill and Miller, Michael},
  booktitle={International Conference on Critical Infrastructure Protection},
  pages={73--82},
  year={2007},
  organization={Springer}
}

@inproceedings{cardenas2008research,
  title={Research Challenges for the Security of Control Systems.},
  author={C{\'a}rdenas, Alvaro A and Amin, Saurabh and Sastry, Shankar},
  booktitle={HotSec},
  year={2008}
}

@article{teixeira2015secure,
  title={A secure control framework for resource-limited adversaries},
  author={Teixeira, Andr{\'e} and Shames, Iman and Sandberg, Henrik and Johansson, Karl Henrik},
  journal={Automatica},
  volume={51},
  pages={135--148},
  year={2015},
  publisher={Elsevier}
}

@inproceedings{mo2009secure,
  title={Secure control against replay attacks},
  author={Mo, Yilin and Sinopoli, Bruno},
  booktitle={47th Annual Allerton Conference on Communication, Control, and Computing},
  pages={911--918},
  year={2009},
  organization={IEEE}
}

@inproceedings{mo2010false,
  title={False data injection attacks in control systems},
  author={{Y. Mo and B. Sinopoli}},
  booktitle={First workshop on Secure Control Systems},
  pages={1--6},
  year={2010}
}

@article{smith2015covert,
  title={Covert misappropriation of networked control systems: Presenting a feedback structure},
  author={Smith, Roy S},
  journal={IEEE Control Systems Magazine},
  volume={35},
  number={1},
  pages={82--92},
  year={2015},
  publisher={IEEE}
}

@inproceedings{teixeira2012revealing,
  title={Revealing stealthy attacks in control systems},
  author={Teixeira, Andr{\'e} and Shames, Iman and Sandberg, Henrik and Johansson, Karl H},
  booktitle={2012 50th Annual Allerton Conference on Communication, Control, and Computing (Allerton)},
  pages={1806--1813},
  year={2012},
  organization={IEEE}
}



@article{fawzi2014secure,
  title={Secure estimation and control for cyber-physical systems under adversarial attacks},
  author={Fawzi, Hamza and Tabuada, Paulo and Diggavi, Suhas},
  journal={IEEE Transactions on Automatic control},
  volume={59},
  number={6},
  pages={1454--1467},
  year={2014},
  publisher={IEEE}
}




@article{zetter2016inside,
  title={Inside the cunning, unprecedented hack of Ukraine’s power grid},
  author={Zetter, Kim},
  journal={Wired},
  year={2016}
}


@article{langner2011stuxnet,
  title={Stuxnet: Dissecting a cyberwarfare weapon},
  author={Langner, Ralph},
  journal={IEEE Security \& Privacy},
  volume={9},
  number={3},
  pages={49--51},
  year={2011},
  publisher={IEEE}
}
@article{shoukry2017secure,
  title={Secure state estimation for cyber-physical systems under sensor attacks: A satisfiability modulo theory approach},
  author={Shoukry, Yasser and Nuzzo, Pierluigi and Puggelli, Alberto and Sangiovanni-Vincentelli, Alberto L and Seshia, Sanjit A and Tabuada, Paulo},
  journal={IEEE Transactions on Automatic Control},
  volume={62},
  number={10},
  pages={4917--4932},
  year={2017},
  publisher={IEEE}
}


@inproceedings{ncs_attack_model,
  title={Attack models and scenarios for networked control systems},
  author={Teixeira, Andr{\'e} and P{\'e}rez, Daniel and Sandberg, Henrik and Johansson, Karl Henrik},
  booktitle={1st ACM Int. Conf. on High Confidence Netw. Systems},
  pages={55--64},
  year={2012}
}





@article{shoukry2016event,
  title={Event-triggered state observers for sparse sensor noise/attacks},
  author={Shoukry, Yasser and Tabuada, Paulo},
  journal={IEEE Trans. on Aut. Control},
  volume={61},
  number={8},
  pages={2079--2091},
  year={2016},
  publisher={IEEE}
}

@article{kwon2014analysis,
  title={Analysis and design of stealthy cyber attacks on unmanned aerial systems},
  author={Kwon, Cheolhyeon and Liu, Weiyi and Hwang, Inseok},
  journal={Journal of Aerospace Information Systems},
  volume={11},
  number={8},
  pages={525--539},
  year={2014},
  publisher={American Institute of Aeronautics and Astronautics}
}

@article{sundaram2011distributed,
  title={Distributed function calculation via linear iterative strategies in the presence of malicious agents},
  author={Sundaram, Shreyas and Hadjicostis, Christoforos N},
  journal={IEEE Transactions on Automatic Control},
  volume={56},
  number={7},
  pages={1495--1508},
  year={2011},
  publisher={IEEE}
}
























@INPROCEEDINGS{dick_asilomar06,
author={C. Dick and F. Harris and M. Pajic and D. Vuletic},
booktitle={2006 Fortieth Asilomar Conference on Signals, Systems and Computers},
title={Real-Time QRD-Based Beamforming on an FPGA Platform},
year={2006},
pages={1200-1204},
keywords={array signal processing;digital signal processing chips;field programmable gate arrays;least squares approximations;linear algebra;logic design;real-time systems;FPGA resource utilization;Mathworks Simulink visual programming;Normal equations;System Generator;Xilinx Virtex;hardware verification tool;least-squares solution;linear algebra operations;model-based FPGA design flow;real-time QRD-based beamforming;Adaptive filters;Application software;Array signal processing;Baseband;Design engineering;Digital signal processing;Educational institutions;Field programmable gate arrays;Logic design;Systolic arrays},
doi={10.1109/ACSSC.2006.354945},
ISSN={1058-6393},
month={Oct},}


@article{dick_xcell07,
  title={Implementing a real-time beamformer on an FPGA platform},
  author={Dick, Chris and Harris, Fred and Pajic, Miroslav and Vuletic, Dragan},
  journal={Xcell journal},
  volume={86},
  year={2007}
}


@INPROCEEDINGS{jorgovanovic_miel08, 
author={Jorgovanovic, M. and Pajic, M. and Kvascev, G. and Popovic, J.}, 
booktitle={26th International Conference on Microelectronics (MIEL)}, 
title={FPGA design of arbitrary down-sampler}, 
year={2008}, 
month={May}, 
pages={391-394}, 
doi={10.1109/ICMEL.2008.4559303},}

@inproceedings {pajic_wess08,
title = {{WisperNet: Anti-Jamming for Wireless Sensor Networks}},
booktitle = {WESS 2008: 2nd Workshop on Embedded Systems Security - A Workshop of the IEEE/ACM EMSOFT'2008 and the Embedded Systems Week},
year = {2008},
author = {M. Pajic and R. Mangharam},
pages = {38-43},
}


@INPROCEEDINGS{evm_workshop09, 
author={Mangharam, R. and Pajic, M.}, 
booktitle={Distributed Computing Systems Workshops, 2009. ICDCS Workshops '09. 29th IEEE International Conference on}, 
title={Embedded Virtual Machines for Robust Wireless Control Systems}, 
year={2009}, 
month={June}, 
pages={38-43}, 
ISSN={1545-0678},}


@INPROCEEDINGS{pajic_ipsn09, 
author={Pajic, M. and Mangharam, R.}, 
booktitle={ International Conference on Information Processing in Sensor Networks (IPSN) }, 
title={Anti-jamming for embedded wireless networks}, 
year={2009}, 
month={April}, 
pages={301-312}, 
}

@article{pasqualetti2013attack,
  title={Attack detection and identification in cyber-physical systems},
  author={Pasqualetti, Fabio and D{\"o}rfler, Florian and Bullo, Francesco},
  journal={IEEE Transactions on Automatic Control},
  volume={58},
  number={11},
  pages={2715--2729},
  year={2013},
  publisher={IEEE}
}

@article{pajic_eurasip10,
  title={Spatio-temporal techniques for anti-jamming in embedded wireless networks},
  author={Pajic, Miroslav and Mangharam, Rahul},
  journal={EURASIP Journal on Wireless Communications and Networking},
  volume={2010},
  number={819318},
  year={2010},
  publisher={Springer International Publishing},
  doi = {10.1155/2010/819318}
}


@INPROCEEDINGS{pajic_rtas10, 
author={Pajic, M. and Mangharam, R.}, 
booktitle={16th IEEE Real-Time and Embedded Technology and Applications Symposium (RTAS)}, 
title={Embedded Virtual Machines for Robust Wireless Control and Actuation}, 
year={2010}, 
month={April}, 
pages={79-88}, 
doi={10.1109/RTAS.2010.43}, 
ISSN={1080-1812},}



@inproceedings{arney_iccps10,
 author = {Arney, David and Pajic, Miroslav and Goldman, Julian M. and Lee, Insup and Mangharam, Rahul and Sokolsky, Oleg},
 title = {Toward Patient Safety in Closed-loop Medical Device Systems},
 booktitle = {Proceedings of the 1st ACM/IEEE International Conference on Cyber-Physical Systems},
 series = {ICCPS '10},
 year = {2010},
 isbn = {978-1-4503-0066-7},
 location = {Stockholm, Sweden},
 pages = {139--148},
 numpages = {10},
 url = {http://doi.acm.org/10.1145/1795194.1795214},
 doi = {10.1145/1795194.1795214},
 acmid = {1795214},
 publisher = {ACM},
 address = {New York, NY, USA},
} 


@INPROCEEDINGS{jiang_ecrts10, 
author={Zhihao Jiang and Pajic, M. and Connolly, A and Dixit, S. and Mangharam, R.}, 
booktitle={22nd Euromicro Conference on Real-Time Systems (ECRTS)}, 
title={Real-Time Heart Model for Implantable Cardiac Device Validation and Verification}, 
year={2010}, 
month={July}, 
pages={239-248}, 
doi={10.1109/ECRTS.2010.36}, 
ISSN={1068-3070},}




@INPROCEEDINGS{pajic_cdc10, 
author={Pajic, M. and Sundaram, S. and Le Ny, Jerome and Pappas, G.J. and Mangharam, R.}, 
booktitle={49th IEEE Conference on Decision and Control (CDC)}, 
title={The Wireless Control Network: Synthesis and robustness}, 
year={2010}, 
month={Dec}, 
pages={7576-7581}, 
doi={10.1109/CDC.2010.5717159}, 
ISSN={0743-1546},}






@ARTICLE{pajic_tac11,
author={M. Pajic and S. Sundaram and G. J. Pappas and R. Mangharam},
journal={IEEE Transactions on Automatic Control},
title={The Wireless Control Network: A New Approach for Control Over Networks},
year={2011},
volume={56},
number={10},
pages={2305-2318},
doi={10.1109/TAC.2011.2163864},
ISSN={0018-9286},
month={Oct},}


@INPROCEEDINGS{pajic_asilomar11, 
author={Pajic, M. and Sundaram, S. and Pappas, G.J. and Mangharam, R.}, 
booktitle={Signals, Systems and Computers (ASILOMAR), 2011 Conference Record of the Forty Fifth Asilomar Conference on}, 
title={Network synthesis for dynamical system stabilization}, 
year={2011}, 
month={Nov}, 
pages={821-825},
doi={10.1109/ACSSC.2011.6190122}, 
ISSN={1058-6393},}


@INPROCEEDINGS{jiang_iccps11, 
author={Zhihao Jiang and Pajic, M. and Mangharam, R.}, 
booktitle={IEEE/ACM International Conference on Cyber-Physical Systems (ICCPS)}, 
title={Model-Based Closed-Loop Testing of Implantable Pacemakers}, 
year={2011}, 
month={April}, 
pages={131-140}, 
doi={10.1109/ICCPS.2011.28},}


@INPROCEEDINGS{pajic_cdc11,
author={M. Pajic and S. Sundaram and G. J. Pappas and R. Mangharam},
booktitle={50th IEEE Conference on Decision and Control and European Control Conference (CDC / ECC)},
title={Topological conditions for wireless control networks},
year={2011},
pages={2353-2360},
doi={10.1109/CDC.2011.6161347},
ISSN={0191-2216},
month={Dec},}


@article{hadziahmetovic_tvst12,
  title={The Oral Iron Chelator Deferiprone Protects Against Retinal Degeneration Induced through Diverse MechanismsHadziahmetovic et al.},
  author={Hadziahmetovic, Majda and Pajic, Miroslav and Grieco, Steven and Song, Ying and Song, Delu and Li, Yafeng and Cwanger, Alyssa and Iacovelli, Jared and Chu, Sally and Ying, Gui-shuang and others},
  journal={Translational vision science \& technology},
  volume={1},
  number={3},
  pages={2.1--2.12},
  year={2012},
  publisher={The Association for Research in Vision and Ophthalmology}
}


@ARTICLE{jiang_pieee12,
author={Z. Jiang and M. Pajic and R. Mangharam},
journal={Proceedings of the IEEE},
title={{Cyber-Physical Modeling of Implantable Cardiac Medical Devices}},
year={2012},
volume={100},
number={1},
pages={122-137},
doi={10.1109/JPROC.2011.2161241},
ISSN={0018-9219},
month={Jan},}


@article{pajic_tecs12,
 author = {Pajic, Miroslav and Chernoguzov, Alexander and Mangharam, Rahul},
 title = {Robust Architectures for Embedded Wireless Network Control and Actuation},
 journal = {ACM Transactions on Embedded Computing Systems},
 issue_date = {December 2012},
 volume = {11},
 number = {4},
 month = jan,
 year = {2013},
 issn = {1539-9087},
 pages = {82:1--82:24},
 articleno = {82},
 numpages = {24},
 url = {http://doi.acm.org/10.1145/2362336.2362349},
 doi = {10.1145/2362336.2362349},
 acmid = {2362349},
 publisher = {ACM},
 address = {New York, NY, USA},
 keywords = {Distributed systems, fault tolerance, wireless sensor networks},
} 


@article{hadziahmetovic_tvst12l,
  title={The Oral Iron Chelator Deferiprone Protects Against Retinal Degeneration Induced through Diverse MechanismsHadziahmetovic et al.},
  author={Hadziahmetovic, Majda and Pajic, Miroslav and Grieco, Steven and Song, Ying and Song, Delu and Li, Yafeng and Cwanger, Alyssa and Iacovelli, Jared and Chu, Sally and Ying, Gui-shuang and others},
  journal={Translational vision science \& technology},
  volume={1},
  number={3},
  pages={2--2},
  year={2012},
  publisher={The Association for Research in Vision and Ophthalmology}
}


@INPROCEEDINGS{pajic_rtas12, 
author={Pajic, M. and Zhihao Jiang and Insup Lee and Sokolsky, O. and Mangharam, R.}, 
booktitle={18th IEEE Real-Time and Embedded Technology and Applications Symposium (RTAS)}, 
title={From Verification to Implementation: A Model Translation Tool and a Pacemaker Case Study}, 
year={2012}, 
month={April}, 
pages={173-184}, 
doi={10.1109/RTAS.2012.25}, 
ISSN={1080-1812},}



@inproceedings{pajic_ipsn12,
 author = {Pajic, Miroslav and Sundaram, Shreyas and Le Ny, Jerome and Pappas, George J. and Mangharam, Rahul},
 title = {Closing the Loop: A Simple Distributed Method for Control over Wireless Networks},
 booktitle = {Proceedings of the 11th International Conference on Information Processing in Sensor Networks},
 series = {IPSN '12},
 year = {2012},
 isbn = {978-1-4503-1227-1},
 location = {Beijing, China},
 pages = {25--36},
 numpages = {12},
 url = {http://doi.acm.org/10.1145/2185677.2185681},
 doi = {10.1145/2185677.2185681},
 acmid = {2185681},
 publisher = {ACM},
 address = {New York, NY, USA},
 keywords = {cooperative control, decentralized control, networked control systems, wireless sensor networks},
} 


@inproceedings{jiang_tacas12,
 author = {Jiang, Zhihao and Pajic, Miroslav and Moarref, Salar and Alur, Rajeev and Mangharam, Rahul},
 title = {Modeling and Verification of a Dual Chamber Implantable Pacemaker},
 booktitle = {Proceedings of the 18th International Conference on Tools and Algorithms for the Construction and Analysis of Systems},
 series = {TACAS'12},
 year = {2012},
 isbn = {978-3-642-28755-8},
 location = {Tallinn, Estonia},
 pages = {188--203},
 numpages = {16},
 url = {http://dx.doi.org/10.1007/978-3-642-28756-5_14},
 doi = {10.1007/978-3-642-28756-5_14},
 acmid = {2260535},
 publisher = {Springer-Verlag},
 address = {Berlin, Heidelberg},
 keywords = {cyber-physical systems, implantable pacemaker, medical devices, software verification},
} 


@ARTICLE{pajic_jsac13,
author={M. Pajic and R. Mangharam and G. J. Pappas and S. Sundaram},
journal={IEEE Journal on Selected Areas in Communications},
title={Topological Conditions for In-Network Stabilization of Dynamical Systems},
year={2013},
volume={31},
number={4},
pages={794-807},
doi={10.1109/JSAC.2013.130415},
ISSN={0733-8716},
month={April},}



@ARTICLE{mangharam_iis13,
author={R. Mangharam and M. Pajic},
journal={Journal of the Indian Institute of Science},
title={Distributed Control for Cyber-Physical Systems},
year={2013},
volume={93},
number={3},
pages={358-387},
month={July-September}
}


@inproceedings{pajic_hicons13,
 author = {Pajic, Miroslav and Bezzo, Nicola and Weimer, James and Alur, Rajeev and Mangharam, Rahul and Michael, Nathan and Pappas, George J. and Sokolsky, Oleg and Tabuada, Paulo and Weirich, Stephanie and Lee, Insup},
 title = {Towards Synthesis of Platform-aware Attack-resilient Control Systems: Extended Abstract},
 booktitle = {Proceedings of the 2Nd ACM International Conference on High Confidence Networked Systems},
 series = {HiCoNS '13},
 year = {2013},
 isbn = {978-1-4503-1961-4},
 location = {Philadelphia, Pennsylvania, USA},
 pages = {75--76},
 numpages = {2},
 url = {http://doi.acm.org/10.1145/2461446.2461457},
 doi = {10.1145/2461446.2461457},
 acmid = {2461457},
 publisher = {ACM},
 address = {New York, NY, USA},
 keywords = {attack resilient control, cyber-physical system security},
} 


@INPROCEEDINGS{gatsis_cdc13, 
author={Gatsis, K. and Pajic, M. and Ribeiro, A and Pappas, G.J.}, 
booktitle={IEEE 52nd Annual Conference on Decision and Control (CDC)}, 
title={Power-aware communication for wireless sensor-actuator systems}, 
year={2013}, 
month={Dec}, 
pages={4006-4011}, 
doi={10.1109/CDC.2013.6760502}, 
ISSN={0743-1546},}


@INPROCEEDINGS{miao_acc13, 
author={Fei Miao and Pajic, M. and Mangharam, R. and Pappas, G.J.}, 
booktitle={American Control Conference (ACC)}, 
title={Networked realization of discrete-time controllers}, 
year={2013}, 
month={June}, 
pages={2996-3001}, 
doi={10.1109/ACC.2013.6580290}, 
ISSN={0743-1619},}


@INPROCEEDINGS{pajic_cdc13, 
author={Pajic, M. and Sundaram, S. and Pappas, G.J.}, 
booktitle={52nd IEEE Annual Conference on Decision and Control (CDC)},
title={Stabilizability over deterministic relay networks}, 
year={2013}, 
month={Dec}, 
pages={4018-4023},
doi={10.1109/CDC.2013.6760504}, 
ISSN={0743-1546},}


@INPROCEEDINGS{miao_cdc13, 
author={Fei Miao and Pajic, M. and Pappas, G.J.}, 
booktitle={IEEE 52nd Annual Conference on Decision and Control (CDC)}, 
title={Stochastic game approach for replay attack detection}, 
year={2013}, 
month={Dec}, 
pages={1854-1859}, 
doi={10.1109/CDC.2013.6760152}, 
ISSN={0743-1546}
}



@incollection{weimer_workshop13,
year={2013},
isbn={978-3-319-01158-5},
booktitle={Control of Cyber-Physical Systems},
volume={449},
series={Lecture Notes in Control and Information Sciences},
editor={Tarraf, Danielle C.},
doi={10.1007/978-3-319-01159-2_11},
title={Resilient Parameter-Invariant Control with Application to Vehicle Cruise Control},
url={http://dx.doi.org/10.1007/978-3-319-01159-2_11},
publisher={Springer International Publishing},
keywords={Secure Cyber-Physical Systems; Robust Control; Resilient Sensor Fusion},
author={Weimer, James and Bezzo, Nicola and Pajic, Miroslav and Pappas, GeorgeJ. and Sokolsky, Oleg and Lee, Insup},
pages={197-216},
language={English}
}


@article{pajic_tecs14,
 author = {Pajic, Miroslav and Jiang, Zhihao and Lee, Insup and Sokolsky, Oleg and Mangharam, Rahul},
 title = {Safety-critical Medical Device Development Using the UPP2SF Model Translation Tool},
 journal = {ACM Transactions on  Embedded Computing Systems},
 issue_date = {July 2014},
 volume = {13},
 number = {4s},
 month = apr,
 year = {2014},
 issn = {1539-9087},
 pages = {127:1--127:26},
 articleno = {127},
 numpages = {26},
 url = {http://doi.acm.org/10.1145/2584651},
 doi = {10.1145/2584651},
 acmid = {2584651},
 publisher = {ACM},
 address = {New York, NY, USA},
 keywords = {Model-based development, medical devices validation and verification, model translation, real-time embedded systems},
} 

@article{jakovljevic_jim14,
 author = {Jakovljevic, Zivana and Petrovic, Petar B. and Mikovic, Vladimir Dj. and Pajic, Miroslav},
 title = {Fuzzy Inference Mechanism for Recognition of Contact States in Intelligent Robotic Assembly},
 journal = {Journal of Intelligent Manufacturing},
 issue_date = {June 2014},
 volume = {25},
 number = {3},
 month = jun,
 year = {2014},
 issn = {0956-5515},
 pages = {571--587},
 numpages = {17},
 url = {http://dx.doi.org/10.1007/s10845-012-0706-x},
 doi = {10.1007/s10845-012-0706-x},
 acmid = {2629775},
 publisher = {Springer-Verlag New York, Inc.},
 address = {Secaucus, NJ, USA},
 keywords = {Contact states, Fuzzy inference mechanism, Part mating, Support vector machines},
} 


@ARTICLE{pajic_tii14,
author={M. Pajic and R. Mangharam and O. Sokolsky and D. Arney and J. Goldman and I. Lee},
journal={IEEE Transactions on Industrial Informatics},
title={Model-Driven Safety Analysis of Closed-Loop Medical Systems},
year={2014},
volume={10},
number={1},
pages={3-16},
doi={10.1109/TII.2012.2226594},
ISSN={1551-3203},
month={Feb},}


@article{jiang_sttt14,
  title={Closed-loop verification of medical devices with model abstraction and refinement},
  author={Jiang, Zhihao and Pajic, Miroslav and Alur, Rajeev and Mangharam, Rahul},
  journal={International Journal on Software Tools for Technology Transfer},
  volume={16},
  number={2},
  pages={191--213},
  year={2014},
  publisher={Springer}
}


@INPROCEEDINGS{ivanov_date14, 
author={Ivanov, R. and Pajic, M. and Insup Lee}, 
booktitle={Design, Automation and Test in Europe Conference and Exhibition (DATE)}, 
title={Attack-resilient sensor fusion}, 
year={2014}, 
month={March}, 
pages={1-6}, 
doi={10.7873/DATE.2014.067},}





@INPROCEEDINGS{miao_cdc14,
author={F. Miao and Q. Zhu and M. Pajic and G. J. Pappas},
booktitle={53rd IEEE Conference on Decision and Control},
title={Coding sensor outputs for injection attacks detection},
year={2014},
pages={5776-5781},
doi={10.1109/CDC.2014.7040293},
ISSN={0191-2216},
month={Dec},}

@INPROCEEDINGS{gatsis_cdc14,
author={K. Gatsis and M. Pajic and A. Ribeiro and G. J. Pappas},
booktitle={53rd IEEE Conference on Decision and Control},
title={Opportunistic sensor scheduling in wireless control systems},
year={2014},
pages={3777-3782},
doi={10.1109/CDC.2014.7039977},
ISSN={0191-2216},
month={Dec},}

@INPROCEEDINGS{sokolsky_cpsarch14, 
author={O. Sokolsky and M. Pajic and N. Bezzo and I. Lee}, 
booktitle={Workshop on Cyber-Physical System Architectures and Design Methodologies (CPSArch), part of Embedded Systems Week (ESWeek)}, 
title={Architecture-Centric Software Development for Cyber-Physical Systems}, 
year={2014},
numpages = {6}
}


@INPROCEEDINGS{gatsis_iccps14, 
author={Gatsis, K. and Pajic, M. and Ribeiro, A and Pappas, G.J.}, 
booktitle={ACM/IEEE International Conference on Cyber-Physical Systems (ICCPS)}, 
title={Opportunistic scheduling of control tasks over shared wireless channels}, 
year={2014}, 
month={April}, 
pages={48-59}, 
doi={10.1109/ICCPS.2014.6843710}
}


@INPROCEEDINGS{weimer_acc14, 
author={Weimer, J. and Bezzo, N. and Pajic, M. and Sokolsky, O. and Insup Lee}, 
booktitle={American Control Conference (ACC)}, 
title={Attack-resilient minimum mean-squared error estimation}, 
year={2014}, 
month={June}, 
pages={1114-1119},
doi={10.1109/ACC.2014.6859478}, 
ISSN={0743-1619},}


@inproceedings{ivanov_hicons14,
 author = {Ivanov, Radoslav and Pajic, Miroslav and Lee, Insup},
 title = {Resilient Multidimensional Sensor Fusion Using Measurement History},
 booktitle = {Proceedings of the 3rd International Conference on High Confidence Networked Systems},
 series = {HiCoNS '14},
 year = {2014},
 isbn = {978-1-4503-2652-0},
 location = {Berlin, Germany},
 pages = {1--10},
 numpages = {10},
 url = {http://doi.acm.org/10.1145/2566468.2566475},
 doi = {10.1145/2566468.2566475},
 acmid = {2566475},
 publisher = {ACM},
 address = {New York, NY, USA},
 keywords = {cps security, fault-tolerance, fault-tolerant algorithms, sensor fusion},
} 


@INPROCEEDINGS{bezzo_iros14,
author={N. Bezzo and J. Weimer and M. Pajic and O. Sokolsky and G. J. Pappas and I. Lee},
booktitle={2014 IEEE/RSJ International Conference on Intelligent Robots and Systems},
title={Attack resilient state estimation for autonomous robotic systems},
year={2014},
pages={3692-3698},
doi={10.1109/IROS.2014.6943080},
ISSN={2153-0858},
month={Sept},}


@inbook{ivanov14_book,
  title = {Resilient Sensor Fusion for Safety-Critical Cyber-Physical Systems},
  booktitle = {Multisensor Data Fusion: From Algorithm and Architecture Design to Applications},
  author = {Ivanov, R. and Pajic, M. and Lee, I.},
  year = {2014},
}



@ARTICLE{gatsis_tac15,
author={K. Gatsis and M. Pajic and A. Ribeiro and G. J. Pappas},
journal={IEEE Transactions on Automatic Control},
title={Opportunistic Control Over Shared Wireless Channels},
year={2015},
volume={60},
number={12},
pages={3140-3155},
doi={10.1109/TAC.2015.2416922},
ISSN={0018-9286},
month={Dec},}


@ARTICLE{jakovljevic_tii15,
author={Z. Jakovljevic and R. Puzovic and M. Pajic},
journal={IEEE Transactions on Industrial Informatics},
title={Recognition of Planar Segments in Point Cloud Based on Wavelet Transform},
year={2015},
volume={11},
number={2},
pages={342-352},
doi={10.1109/TII.2015.2389195},
ISSN={1551-3203},
month={April},}


@article{jakovljevic_aa15,
  title={Diagnosis of irregularities in the robotized part mating process based on contextual recognition of contact states transitions},
  author={Jakovljevic, Zivana and Petrovic, Petar B and Milkovic, Dragan and Pajic, Miroslav},
  journal={Assembly Automation},
  volume={35},
  number={2},
  pages={190--199},
  year={2015}
}


@inproceedings{pajic_emsoft15,
 author = {Pajic, Miroslav and Park, Junkil and Lee, Insup and Pappas, George J. and Sokolsky, Oleg},
 title = {Automatic Verification of Linear Controller Software},
 booktitle = {Proceedings of the 12th International Conference on Embedded Software},
 series = {EMSOFT '15},
 year = {2015},
 isbn = {978-1-4673-8079-9},
 location = {Amsterdam, The Netherlands},
 pages = {217--226},
 numpages = {10},
 url = {http://dl.acm.org/citation.cfm?id=2830865.2830889},
 acmid = {2830889},
 publisher = {IEEE Press},
 address = {Piscataway, NJ, USA},
} 


@inproceedings{junkil_iccps15,
 author = {Park, Junkil and Ivanov, Radoslav and Weimer, James and Pajic, Miroslav and Lee, Insup},
 title = {Sensor Attack Detection in the Presence of Transient Faults},
 booktitle = {Proceedings of the ACM/IEEE Sixth International Conference on Cyber-Physical Systems},
 series = {ICCPS '15},
 year = {2015},
 isbn = {978-1-4503-3455-6},
 location = {Seattle, Washington},
 pages = {1--10},
 numpages = {10},
 url = {http://doi.acm.org/10.1145/2735960.2735984},
 doi = {10.1145/2735960.2735984},
 acmid = {2735984},
 publisher = {ACM},
 address = {New York, NY, USA},
} 


@inproceedings{alfaruque_codes15,
 author = {Al Faruque, Mohammad and Regazzoni, Francesco and Pajic, Miroslav},
 title = {Design Methodologies for Securing Cyber-physical Systems},
 booktitle = {Proceedings of the 10th International Conference on Hardware/Software Codesign and System Synthesis},
 series = {CODES '15},
 year = {2015},
 isbn = {978-1-4673-8321-9},
 location = {Amsterdam, The Netherlands},
 pages = {30--36},
 numpages = {7},
 url = {http://dl.acm.org/citation.cfm?id=2830840.2830844},
 acmid = {2830844},
 publisher = {IEEE Press},
 address = {Piscataway, NJ, USA},
} 



@INPROCEEDINGS{pajic_cdc15, 
author={Pajic, M. and Tabuada, P. and Lee, I. and Pappas, G.J.}, 
booktitle={54th IEEE  Annual Conference on Decision and Control (CDC)}, 
title={Attack-Resilient State Estimation in the Presence of Noise}, 
location = {Osaka, Japan},
 year = {2015},
pages = {5827--5832},
numpages = {6},
month={Dec}
 }


@INPROCEEDINGS{ivanov_allerton15,
author={R. Ivanov and N. Atanasov and M. Pajic and G. Pappas and I. Lee},
booktitle={2015 53rd Annual Allerton Conference on Communication, Control, and Computing (Allerton)},
title={Robust estimation using context-aware filtering},
year={2015},
pages={590-597},
doi={10.1109/ALLERTON.2015.7447058},
month={Sept},}


@INPROCEEDINGS{ivanov_mvigro15, 
author={Ivanov, R. and Atanasov, N. and Pajic, M. and  Lee, I. and Pappas, G. J.}, 
booktitle={Workshop on Multi VIew Geometry in Robotics (MVIGRO), in conjunction with RSS}, 
title={Robust Localization Using Context-Aware Filtering}, 
year={2015}
}



@article{ivanov_tecs16,
 author = {Ivanov, Radoslav and Pajic, Miroslav and Lee, Insup},
 title = {Attack-Resilient Sensor Fusion for Safety-Critical Cyber-Physical Systems},
 journal = {ACM Transactions on Embedded Computing Systems},
 issue_date = {February 2016},
 volume = {15},
 number = {1},
 month = feb,
 year = {2016},
 issn = {1539-9087},
 pages = {21:1--21:24},
 articleno = {21},
 numpages = {24},
 url = {http://doi.acm.org/10.1145/2847418},
 doi = {10.1145/2847418},
 acmid = {2847418},
 publisher = {ACM},
} 




@incollection{junkil_tacas16,
  title={Scalable Verification of Linear Controller Software},
  author={Park, Junkil and Pajic, Miroslav and Lee, Insup and Sokolsky, Oleg},
  booktitle={Tools and Algorithms for the Construction and Analysis of Systems (TACAS)},
  pages={662--679},
  year={2016},
  publisher={Springer}
}



@INPROCEEDINGS{ivanov_iccps16,
author={R. Ivanov and N. Atanasov and J. Weimer and M. Pajic and A. Simpao and M. Rehman and G. Pappas and I. Lee},
booktitle={2016 ACM/IEEE 7th International Conference on Cyber-Physical Systems (ICCPS)},
title={Estimation of Blood Oxygen Content Using Context-Aware Filtering},
year={2016},
pages={1-10},
doi={10.1109/ICCPS.2016.7479102},
month={April},}


@INPROCEEDINGS{mangharam_comnets16,
author={R. Mangharam and H. Abbas and M. Behl and K. Jang and M. Pajic and Z. Jiang},
booktitle={2016 8th International Conference on Communication Systems and Networks (COMSNETS)},
title={Three challenges in cyber-physical systems},
year={2016},
pages={1-8},
doi={10.1109/COMSNETS.2016.7440015},
month={Jan},}









@inproceedings{bogdan_codes16,
 author = {Bogdan, Paul and Pajic, Miroslav and Pande, Partha Pratim and Raghunathan, Vijay},
 title = {Making the Internet-of-things a Reality: From Smart Models, Sensing and Actuation to Energy-efficient Architectures},
 booktitle = {Proceedings of the Eleventh IEEE/ACM/IFIP International Conference on Hardware/Software Codesign and System Synthesis},
 series = {CODES '16},
 year = {2016},
 isbn = {978-1-4503-4483-8},
 location = {Pittsburgh, Pennsylvania},
 pages = {25:1--25:10},
 articleno = {25},
 numpages = {10},
 url = {http://doi.acm.org/10.1145/2968456.2973272},
 doi = {10.1145/2968456.2973272},
 acmid = {2973272}
} 




@inproceedings{ibrahim_cases16,
 author = {Ibrahim, Mohamed and Boswell, Craig and Chakrabarty, Krishnendu and Scott, Kristin and Pajic, Miroslav},
 title = {A Real-time Digital-microfluidic Platform for Epigenetics},
 booktitle = {Proceedings of the International Conference on Compilers, Architectures and Synthesis for Embedded Systems},
 series = {CASES '16},
 year = {2016},
 pages = {10:1--10:10},
 doi = {10.1145/2968455.2968516}
} 




@inproceedings{lesi_etfa16,
 author = {V. Lesi and Z. Jakovljevic and M. Pajic},
 title = {{Towards Plug-n-Play Numerical Control for Reconfigurable Manufacturing Systems}},
 booktitle = {21st IEEE International Conference on Emerging Technologies and Factory Automation (ETFA)},
 year = {2016},
 pages = {1-8},
 doi={10.1109/ETFA.2016.7733524},
month={Sept},
 } 
 


@inproceedings{li_iccad16,
 author = {Li, Zipeng and Lai, Kelvin Yi-Tse and Yu, Po-Hsien and Chakrabarty, Krishnendu and Pajic, Miroslav and Ho, Tsung-Yi and Lee, Chen-Yi},
 title = {{Error Recovery in a Micro-electrode-dot-array Digital Microfluidic Biochip}},
 booktitle = {Proceedings of the 35th International Conference on Computer-Aided Design},
 series = {ICCAD '16},
 year = {2016},
 pages = {105:1--105:8},
 doi = {10.1145/2966986.2967035}
} 






@ARTICLE{miao_tcns17,
author={F. Miao and Q. Zhu and M. Pajic and G. J. Pappas},
journal={IEEE Transactions on Control of Network Systems},
title={Coding Schemes for Securing Cyber-Physical Systems Against Stealthy Data Injection Attacks},
year={2017},
volume={4},
number={1},
pages={106-117},
keywords={Algorithm design and analysis;Data models;Detectors;Encoding;Estimation;Kalman filters;Linear systems;Coding;detection;feasible coding matrix;state estimator;stealthy data injection attacks;time-varying coding},
doi={10.1109/TCNS.2016.2573039},
ISSN={2325-5870},
month={March}
}



@article{junkil_tcps17,
 author = {Park, Junkil and Ivanov, Radoslav and Weimer, James and Pajic, Miroslav and Son, Sang Hyuk and Lee, Insup},
 title = {Security of Cyber-Physical Systems in the Presence of Transient Sensor Faults},
 journal = {ACM Transactions on Cyber-Physical Systems},
 issue_date = {May 2017},
 volume = {1},
 number = {3},
 month = May,
 year = {2017},
 pages = {15:1--15:23}
} 


@incollection{junkil_tacas17,
author="Park, Junkil and Pajic, Miroslav and Sokolsky, Oleg and Lee, Insup",
title="Automatic Verification of Finite Precision Implementations of Linear Controllers",
bookTitle="Tools and Algorithms for the Construction and Analysis of Systems: 23rd International Conference, TACAS 2017",
year="2017",
publisher="Springer Berlin Heidelberg",
pages="153--169",
doi="10.1007/978-3-662-54577-5_9"
}


@INPROCEEDINGS{fricks_rams17, 
author={R. W. B. Fricks and H. H. Tseng and M. Pajic and K. S. Trivedi}, 
booktitle={2017 Annual Reliability and Maintainability Symposium (RAMS)}, 
title={Transient performance availability modeling in high volume outpatient clinics}, 
year={2017}, 
pages={1-6}, 
doi={10.1109/RAM.2017.7889777}, 
month={Jan}
}



@INPROCEEDINGS{jovanov_cdc17, 
author={I. Jovanov and M. Pajic}, 
booktitle={2017 IEEE 56th Annual Conference on Decision and Control (CDC)}, 
title={Sporadic data integrity for secure state estimation}, 
year={2017}, 
volume={}, 
number={}, 
pages={163-169}, 
doi={10.1109/CDC.2017.8263660}, 
ISSN={}, 
month={Dec},}





 @INPROCEEDINGS{lesi_rtss17, 
author={V. Lesi and I. Jovanov and M. Pajic}, 
booktitle={2017 IEEE Real-Time Systems Symposium (RTSS)}, 
title={Network Scheduling for Secure Cyber-Physical Systems}, 
year={2017}, 
volume={}, 
number={}, 
pages={45-55}, 
keywords={cryptography;cyber-physical systems;data integrity;integer programming;linear programming;message authentication;scheduling;QoC requirements;Quality-of-Control;cryptographic tools;cyber-physical systems attacks;cyber-physical systems security;data integrity;message authentication codes;mixed integer linear programming problem;network scheduling;opportunistic scheduling;real-time network messages;runtime bandwidth allocation;Authentication;Automotive engineering;Data integrity;Real-time systems;Sensors;Standards;CAN-bus;CPS-Security;Mixed-Integer-Linear-Programming;Quality-of-Control;Real-Time-Scheduling}, 
doi={10.1109/RTSS.2017.00012}, 
ISSN={}, 
month={Dec}
}



@article{elfar_tecs17,
 author = {Elfar, Mahmoud and Zhong, Zhanwei and Li, Zipeng and Chakrabarty, Krishnendu and Pajic, Miroslav},
 title = {Synthesis of Error-Recovery Protocols for Micro-Electrode-Dot-Array Digital Microfluidic Biochips},
 journal = {ACM Trans. Embed. Comput. Syst.},
 issue_date = {October 2017},
 volume = {16},
 number = {5s},
 month = sep,
 year = {2017},
 issn = {1539-9087},
 pages = {127:1--127:22},
 articleno = {127},
 numpages = {22},
 url = {http://doi.acm.org/10.1145/3126538},
 doi = {10.1145/3126538}
} 



@article{miao_automatica17,
title = "A hybrid stochastic game for secure control of cyber-physical systems",
journal = "Automatica",
volume = "93",
pages = "55 - 63",
year = "2018",
issn = "0005-1098",
doi = "https://doi.org/10.1016/j.automatica.2018.03.012",
url = "http://www.sciencedirect.com/science/article/pii/S0005109818300992",
author = "Fei Miao and Quanyan Zhu and Miroslav Pajic and George J. Pappas",
keywords = "Stochastic game, Secure control, Saddle-point equilibrium"
}


@ARTICLE{li_tcad17, 
author={Z. Li and K. Y. T. Lai and J. McCrone and P. H. Yu and K. Chakrabarty and M. Pajic and T. Y. Ho and C. Y. Lee}, 
journal={IEEE Transactions on Computer-Aided Design of Integrated Circuits and Systems}, 
title={Efficient and Adaptive Error Recovery in a Micro-Electrode-Dot-Array Digital Microfluidic Biochip}, 
year={2017}, 
volume={PP}, 
number={99}, 
pages={1-1}, 
doi={10.1109/TCAD.2017.2729347}, 
month={}
}

@Inbook{jakovljevic_newtech17a,
author="Jakovljevic, Zivana and Mitrovic, Stefan and Pajic, Miroslav",
title="Cyber Physical Production Systems---An IEC 61499 Perspective",
bookTitle="Proceedings of 5th International Conference on Advanced Manufacturing Engineering and Technologies: NEWTECH 2017",
year="2017",
publisher="Springer International Publishing",
pages="27--39",
doi="10.1007/978-3-319-56430-2_3"
}

@Inbook{jakovljevic_newtech17b,
author="Jakovljevic, Zivana and Majstorovic, Vidosav and Stojadinovic, Slavenko and Zivkovic, Srdjan and Gligorijevic, Nemanja and Pajic, Miroslav",
title="Cyber-Physical Manufacturing Systems (CPMS)",
bookTitle="Proceedings of 5th International Conference on Advanced Manufacturing Engineering and Technologies: NEWTECH 2017",
year="2017",
publisher="Springer International Publishing",
pages="199--214",
doi="10.1007/978-3-319-56430-2_14"
}


@article{jovanov_arxiv17,
  author    = {Jovanov, I. and Pajic, M.},
  title     = {Relaxing Integrity Requirements for Attack-resilient Cyber-Physical Systems},
  journal   = {CoRR},
  volume    = {abs/1707.02950},
  year      = {2017},
  url       = {https://arxiv.org/abs/1707.02950},
}


@article{ivanov_tac19,
  title={Continuous estimation using context-dependent discrete measurements},
  author={Ivanov, Radoslav and Atanasov, Nikolay and Pajic, Miroslav and Weimer, James and Pappas, George J and Lee, Insup},
  journal={IEEE Transactions on Automatic Control},
  volume={64},
  number={1},
  pages={235--250},
  year={2019},
  publisher={IEEE}
}

@inproceedings{elfar_iccps17,
  title={Platform for security-aware design of human-on-the-loop cyber-physical systems},
  author={Elfar, Mahmoud and Zhu, Haibei and Raghunathan, Adithya and Tay, Yi Y and Wubbenhorst, Jeffrey and Cummings, ML and Pajic, Miroslav},
  booktitle={Proc. of the 8th ACM ICCPS},
  year={2017}
}





@INPROCEEDINGS{jovanov_cdc18, 
author={I. {Jovanov} and M. {Pajic}}, 
booktitle={2018 IEEE Conference on Decision and Control (CDC)}, 
title={Secure State Estimation with Cumulative Message Authentication}, 
year={2018}, 
pages={2074-2079}, 
month={Dec}
}



@inproceedings{jovanov_iccps18,
  title={Platform for model-based design and testing for deep brain stimulation},
  author={Jovanov, Ilija and Naumann, Michael and Kumaravelu, Karthik and Grill, Warren M and Pajic, Miroslav},
  booktitle={Proceedings of the 9th ACM/IEEE International Conference on Cyber-Physical Systems (ICCPS)},
  pages={263--274},
  year={2018},
  organization={IEEE Press}
}

@article{zhu_thms19,
 author = {H. Zhu and M. Cummings and M. Elfar and Z. Wang and M. Pajic},
 title = {Operator Strategy Model Development in UAV Hacking Detection},
 journal = {IEEE Transactions on Human-Machine Systems},
 note = {to appear, available at \url{http://people.duke.edu/~mp275/publications.html}}
} 

@inproceedings{zhu_chii18,
  title={Human augmentation of uav cyber-attack detection},
  author={Zhu, Haibei and Elfar, Mahmoud and Pajic, Miroslav and Wang, Ziyao and Cummings, Mary L},
  booktitle={International Conference on Augmented Cognition},
  pages={154--167},
  year={2018},
  organization={Springer}
}


@inproceedings{elfar_icra19,
  title={Security-Aware Synthesis of Human-UAV Protocols},
  author={M. Elfar and H. Zhu and M. L. Cummings and M. Pajic},
  booktitle={2019 International Conference on Robotics and Automation (ICRA)},
  year={2019},
  note = {to appear, available at \url{http://people.duke.edu/~mp275/publications.html}}
}


@article{wang_cav19,
  author    = {Y. Wang and S. Nalluri  and B. Bonakdarpour and M. Pajic},
  title     = {Statistical Model Checking for Probabilistic Hyperproperties},
  journal   = {CoRR},
  volume    = {abs/1707.02950},
  year      = {2019},
  url       = {https://arxiv.org/abs/1707.02950},
}

@article{elfar_cav19,
  author    = {M. Elfar and Y. Wang and M. Pajic},
  title     = {Security-Aware Synthesis using Delayed Action Games},
  journal   = {CoRR},
  volume    = {abs/1902.04111},
  year      = {2019},
  url       = {https://arxiv.org/abs/1902.04111},
}


@inproceedings{checkoway2011comprehensive,
  title={Comprehensive experimental analyses of automotive attack surfaces.},
  author={Checkoway, Stephen and McCoy, Damon and Kantor, Brian and Anderson, Danny and Shacham, Hovav and Savage, Stefan and Koscher, Karl and Czeskis, Alexei and Roesner, Franziska and Kohno, Tadayoshi and others},
  booktitle={USENIX Security Symposium},
  volume={4},
  pages={447--462},
  year={2011},
  organization={San Francisco}
}



@inproceedings{khazraei2017new,
  title={A new watermarking approach for replay attack detection in lqg systems},
  author={Khazraei, Amir and Kebriaei, Hamed and Salmasi, Farzad Rajaei},
  booktitle={56th IEEE Annual Conf. on Decision and Control (CDC)},
  pages={5143--5148},
  year={2017},
}



@INPROCEEDINGS{dick_asilomar06,
author={C. Dick and F. Harris and M. Pajic and D. Vuletic},
booktitle={2006 Fortieth Asilomar Conference on Signals, Systems and Computers},
title={Real-Time QRD-Based Beamforming on an FPGA Platform},
year={2006},
pages={1200-1204},
keywords={array signal processing;digital signal processing chips;field programmable gate arrays;least squares approximations;linear algebra;logic design;real-time systems;FPGA resource utilization;Mathworks Simulink visual programming;Normal equations;System Generator;Xilinx Virtex;hardware verification tool;least-squares solution;linear algebra operations;model-based FPGA design flow;real-time QRD-based beamforming;Adaptive filters;Application software;Array signal processing;Baseband;Design engineering;Digital signal processing;Educational institutions;Field programmable gate arrays;Logic design;Systolic arrays},
doi={10.1109/ACSSC.2006.354945},
ISSN={1058-6393},
month={Oct},}


@article{dick_xcell07,
  title={Implementing a real-time beamformer on an FPGA platform},
  author={Dick, Chris and Harris, Fred and Pajic, Miroslav and Vuletic, Dragan},
  journal={Xcell journal},
  volume={86},
  year={2007}
}


@INPROCEEDINGS{jorgovanovic_miel08, 
author={Jorgovanovic, M. and Pajic, M. and Kvascev, G. and Popovic, J.}, 
booktitle={26th International Conference on Microelectronics (MIEL)}, 
title={FPGA design of arbitrary down-sampler}, 
year={2008}, 
month={May}, 
pages={391-394}, 
doi={10.1109/ICMEL.2008.4559303},}

@inproceedings {pajic_wess08,
title = {{WisperNet: Anti-Jamming for Wireless Sensor Networks}},
booktitle = {WESS 2008: 2nd Workshop on Embedded Systems Security - A Workshop of the IEEE/ACM EMSOFT'2008 and the Embedded Systems Week},
year = {2008},
author = {M. Pajic and R. Mangharam},
pages = {38-43},
}


@INPROCEEDINGS{evm_workshop09, 
author={Mangharam, R. and Pajic, M.}, 
booktitle={Distributed Computing Systems Workshops, 2009. ICDCS Workshops '09. 29th IEEE International Conference on}, 
title={Embedded Virtual Machines for Robust Wireless Control Systems}, 
year={2009}, 
month={June}, 
pages={38-43}, 
ISSN={1545-0678},}


@INPROCEEDINGS{pajic_ipsn09, 
author={Pajic, M. and Mangharam, R.}, 
booktitle={ International Conference on Information Processing in Sensor Networks (IPSN) }, 
title={Anti-jamming for embedded wireless networks}, 
year={2009}, 
month={April}, 
pages={301-312}, 
}



@article{pajic_eurasip10,
  title={Spatio-temporal techniques for anti-jamming in embedded wireless networks},
  author={Pajic, Miroslav and Mangharam, Rahul},
  journal={EURASIP Journal on Wireless Communications and Networking},
  volume={2010},
  number={819318},
  year={2010},
  publisher={Springer International Publishing},
  doi = {10.1155/2010/819318}
}


@INPROCEEDINGS{pajic_rtas10, 
author={Pajic, M. and Mangharam, R.}, 
booktitle={16th IEEE Real-Time and Embedded Technology and Applications Symposium (RTAS)}, 
title={Embedded Virtual Machines for Robust Wireless Control and Actuation}, 
year={2010}, 
month={April}, 
pages={79-88}, 
doi={10.1109/RTAS.2010.43}, 
ISSN={1080-1812},}



@inproceedings{arney_iccps10,
 author = {Arney, David and Pajic, Miroslav and Goldman, Julian M. and Lee, Insup and Mangharam, Rahul and Sokolsky, Oleg},
 title = {Toward Patient Safety in Closed-loop Medical Device Systems},
 booktitle = {Proceedings of the 1st ACM/IEEE International Conference on Cyber-Physical Systems},
 series = {ICCPS '10},
 year = {2010},
 isbn = {978-1-4503-0066-7},
 location = {Stockholm, Sweden},
 pages = {139--148},
 numpages = {10},
 url = {http://doi.acm.org/10.1145/1795194.1795214},
 doi = {10.1145/1795194.1795214},
 acmid = {1795214},
 publisher = {ACM},
 address = {New York, NY, USA},
} 


@INPROCEEDINGS{jiang_ecrts10, 
author={Zhihao Jiang and Pajic, M. and Connolly, A and Dixit, S. and Mangharam, R.}, 
booktitle={22nd Euromicro Conference on Real-Time Systems (ECRTS)}, 
title={Real-Time Heart Model for Implantable Cardiac Device Validation and Verification}, 
year={2010}, 
month={July}, 
pages={239-248}, 
doi={10.1109/ECRTS.2010.36}, 
ISSN={1068-3070},}




@INPROCEEDINGS{pajic_cdc10, 
author={Pajic, M. and Sundaram, S. and Le Ny, Jerome and Pappas, G.J. and Mangharam, R.}, 
booktitle={49th IEEE Conference on Decision and Control (CDC)}, 
title={The Wireless Control Network: Synthesis and robustness}, 
year={2010}, 
month={Dec}, 
pages={7576-7581}, 
doi={10.1109/CDC.2010.5717159}, 
ISSN={0743-1546},}





@ARTICLE{pajic_tac11,
author={M. Pajic and S. Sundaram and G. J. Pappas and R. Mangharam},
journal={IEEE Transactions on Automatic Control},
title={The Wireless Control Network: A New Approach for Control Over Networks},
year={2011},
volume={56},
number={10},
pages={2305-2318},
doi={10.1109/TAC.2011.2163864},
ISSN={0018-9286},
month={Oct},}


@INPROCEEDINGS{pajic_asilomar11, 
author={Pajic, M. and Sundaram, S. and Pappas, G.J. and Mangharam, R.}, 
booktitle={Signals, Systems and Computers (ASILOMAR), 2011 Conference Record of the Forty Fifth Asilomar Conference on}, 
title={Network synthesis for dynamical system stabilization}, 
year={2011}, 
month={Nov}, 
pages={821-825},
doi={10.1109/ACSSC.2011.6190122}, 
ISSN={1058-6393},}


@INPROCEEDINGS{jiang_iccps11, 
author={Zhihao Jiang and Pajic, M. and Mangharam, R.}, 
booktitle={IEEE/ACM International Conference on Cyber-Physical Systems (ICCPS)}, 
title={Model-Based Closed-Loop Testing of Implantable Pacemakers}, 
year={2011}, 
month={April}, 
pages={131-140}, 
doi={10.1109/ICCPS.2011.28},}


@INPROCEEDINGS{pajic_cdc11,
author={M. Pajic and S. Sundaram and G. J. Pappas and R. Mangharam},
booktitle={50th IEEE Conference on Decision and Control and European Control Conference (CDC / ECC)},
title={Topological conditions for wireless control networks},
year={2011},
pages={2353-2360},
doi={10.1109/CDC.2011.6161347},
ISSN={0191-2216},
month={Dec},}


@article{hadziahmetovic_tvst12,
  title={The Oral Iron Chelator Deferiprone Protects Against Retinal Degeneration Induced through Diverse MechanismsHadziahmetovic et al.},
  author={Hadziahmetovic, Majda and Pajic, Miroslav and Grieco, Steven and Song, Ying and Song, Delu and Li, Yafeng and Cwanger, Alyssa and Iacovelli, Jared and Chu, Sally and Ying, Gui-shuang and others},
  journal={Translational vision science \& technology},
  volume={1},
  number={3},
  pages={2.1--2.12},
  year={2012},
  publisher={The Association for Research in Vision and Ophthalmology}
}


@ARTICLE{jiang_pieee12,
author={Z. Jiang and M. Pajic and R. Mangharam},
journal={Proceedings of the IEEE},
title={{Cyber-Physical Modeling of Implantable Cardiac Medical Devices}},
year={2012},
volume={100},
number={1},
pages={122-137},
doi={10.1109/JPROC.2011.2161241},
ISSN={0018-9219},
month={Jan},}


@article{pajic_tecs12,
 author = {Pajic, Miroslav and Chernoguzov, Alexander and Mangharam, Rahul},
 title = {Robust Architectures for Embedded Wireless Network Control and Actuation},
 journal = {ACM Transactions on Embedded Computing Systems},
 issue_date = {December 2012},
 volume = {11},
 number = {4},
 month = jan,
 year = {2013},
 issn = {1539-9087},
 pages = {82:1--82:24},
 articleno = {82},
 numpages = {24},
 url = {http://doi.acm.org/10.1145/2362336.2362349},
 doi = {10.1145/2362336.2362349},
 acmid = {2362349},
 publisher = {ACM},
 address = {New York, NY, USA},
 keywords = {Distributed systems, fault tolerance, wireless sensor networks},
} 


@article{hadziahmetovic_tvst12l,
  title={The Oral Iron Chelator Deferiprone Protects Against Retinal Degeneration Induced through Diverse MechanismsHadziahmetovic et al.},
  author={Hadziahmetovic, Majda and Pajic, Miroslav and Grieco, Steven and Song, Ying and Song, Delu and Li, Yafeng and Cwanger, Alyssa and Iacovelli, Jared and Chu, Sally and Ying, Gui-shuang and others},
  journal={Translational vision science \& technology},
  volume={1},
  number={3},
  pages={2--2},
  year={2012},
  publisher={The Association for Research in Vision and Ophthalmology}
}


@INPROCEEDINGS{pajic_rtas12, 
author={Pajic, M. and Zhihao Jiang and Insup Lee and Sokolsky, O. and Mangharam, R.}, 
booktitle={18th IEEE Real-Time and Embedded Technology and Applications Symposium (RTAS)}, 
title={From Verification to Implementation: A Model Translation Tool and a Pacemaker Case Study}, 
year={2012}, 
month={April}, 
pages={173-184}, 
doi={10.1109/RTAS.2012.25}, 
ISSN={1080-1812},}



@inproceedings{pajic_ipsn12,
 author = {Pajic, Miroslav and Sundaram, Shreyas and Le Ny, Jerome and Pappas, George J. and Mangharam, Rahul},
 title = {Closing the Loop: A Simple Distributed Method for Control over Wireless Networks},
 booktitle = {Proceedings of the 11th International Conference on Information Processing in Sensor Networks},
 series = {IPSN '12},
 year = {2012},
 isbn = {978-1-4503-1227-1},
 location = {Beijing, China},
 pages = {25--36},
 numpages = {12},
 url = {http://doi.acm.org/10.1145/2185677.2185681},
 doi = {10.1145/2185677.2185681},
 acmid = {2185681},
 publisher = {ACM},
 address = {New York, NY, USA},
 keywords = {cooperative control, decentralized control, networked control systems, wireless sensor networks},
} 


@inproceedings{jiang_tacas12,
 author = {Jiang, Zhihao and Pajic, Miroslav and Moarref, Salar and Alur, Rajeev and Mangharam, Rahul},
 title = {Modeling and Verification of a Dual Chamber Implantable Pacemaker},
 booktitle = {Proceedings of the 18th International Conference on Tools and Algorithms for the Construction and Analysis of Systems},
 series = {TACAS'12},
 year = {2012},
 isbn = {978-3-642-28755-8},
 location = {Tallinn, Estonia},
 pages = {188--203},
 numpages = {16},
 url = {http://dx.doi.org/10.1007/978-3-642-28756-5_14},
 doi = {10.1007/978-3-642-28756-5_14},
 acmid = {2260535},
 publisher = {Springer-Verlag},
 address = {Berlin, Heidelberg},
 keywords = {cyber-physical systems, implantable pacemaker, medical devices, software verification},
} 


@ARTICLE{pajic_jsac13,
author={M. Pajic and R. Mangharam and G. J. Pappas and S. Sundaram},
journal={IEEE Journal on Selected Areas in Communications},
title={Topological Conditions for In-Network Stabilization of Dynamical Systems},
year={2013},
volume={31},
number={4},
pages={794-807},
doi={10.1109/JSAC.2013.130415},
ISSN={0733-8716},
month={April},}



@ARTICLE{mangharam_iis13,
author={R. Mangharam and M. Pajic},
journal={Journal of the Indian Institute of Science},
title={Distributed Control for Cyber-Physical Systems},
year={2013},
volume={93},
number={3},
pages={358-387},
month={July-September}
}


@inproceedings{pajic_hicons13,
 author = {Pajic, Miroslav and Bezzo, Nicola and Weimer, James and Alur, Rajeev and Mangharam, Rahul and Michael, Nathan and Pappas, George J. and Sokolsky, Oleg and Tabuada, Paulo and Weirich, Stephanie and Lee, Insup},
 title = {Towards Synthesis of Platform-aware Attack-resilient Control Systems: Extended Abstract},
 booktitle = {Proceedings of the 2Nd ACM International Conference on High Confidence Networked Systems},
 series = {HiCoNS '13},
 year = {2013},
 isbn = {978-1-4503-1961-4},
 location = {Philadelphia, Pennsylvania, USA},
 pages = {75--76},
 numpages = {2},
 url = {http://doi.acm.org/10.1145/2461446.2461457},
 doi = {10.1145/2461446.2461457},
 acmid = {2461457},
 publisher = {ACM},
 address = {New York, NY, USA},
 keywords = {attack resilient control, cyber-physical system security},
} 


@INPROCEEDINGS{gatsis_cdc13, 
author={Gatsis, K. and Pajic, M. and Ribeiro, A and Pappas, G.J.}, 
booktitle={IEEE 52nd Annual Conference on Decision and Control (CDC)}, 
title={Power-aware communication for wireless sensor-actuator systems}, 
year={2013}, 
month={Dec}, 
pages={4006-4011}, 
doi={10.1109/CDC.2013.6760502}, 
ISSN={0743-1546},}


@INPROCEEDINGS{miao_acc13, 
author={Fei Miao and Pajic, M. and Mangharam, R. and Pappas, G.J.}, 
booktitle={American Control Conference (ACC)}, 
title={Networked realization of discrete-time controllers}, 
year={2013}, 
month={June}, 
pages={2996-3001}, 
doi={10.1109/ACC.2013.6580290}, 
ISSN={0743-1619},}


@INPROCEEDINGS{pajic_cdc13, 
author={Pajic, M. and Sundaram, S. and Pappas, G.J.}, 
booktitle={52nd IEEE Annual Conference on Decision and Control (CDC)},
title={Stabilizability over deterministic relay networks}, 
year={2013}, 
month={Dec}, 
pages={4018-4023},
doi={10.1109/CDC.2013.6760504}, 
ISSN={0743-1546},}


@INPROCEEDINGS{miao_cdc13, 
author={Fei Miao and Pajic, M. and Pappas, G.J.}, 
booktitle={IEEE 52nd Annual Conference on Decision and Control (CDC)}, 
title={Stochastic game approach for replay attack detection}, 
year={2013}, 
month={Dec}, 
pages={1854-1859}, 
doi={10.1109/CDC.2013.6760152}, 
ISSN={0743-1546}
}



@incollection{weimer_workshop13,
year={2013},
isbn={978-3-319-01158-5},
booktitle={Control of Cyber-Physical Systems},
volume={449},
series={Lecture Notes in Control and Information Sciences},
editor={Tarraf, Danielle C.},
doi={10.1007/978-3-319-01159-2_11},
title={Resilient Parameter-Invariant Control with Application to Vehicle Cruise Control},
url={http://dx.doi.org/10.1007/978-3-319-01159-2_11},
publisher={Springer International Publishing},
keywords={Secure Cyber-Physical Systems; Robust Control; Resilient Sensor Fusion},
author={Weimer, James and Bezzo, Nicola and Pajic, Miroslav and Pappas, GeorgeJ. and Sokolsky, Oleg and Lee, Insup},
pages={197-216},
language={English}
}


@article{pajic_tecs14,
 author = {Pajic, Miroslav and Jiang, Zhihao and Lee, Insup and Sokolsky, Oleg and Mangharam, Rahul},
 title = {Safety-critical Medical Device Development Using the UPP2SF Model Translation Tool},
 journal = {ACM Transactions on  Embedded Computing Systems},
 issue_date = {July 2014},
 volume = {13},
 number = {4s},
 month = apr,
 year = {2014},
 issn = {1539-9087},
 pages = {127:1--127:26},
 articleno = {127},
 numpages = {26},
 url = {http://doi.acm.org/10.1145/2584651},
 doi = {10.1145/2584651},
 acmid = {2584651},
 publisher = {ACM},
 address = {New York, NY, USA},
 keywords = {Model-based development, medical devices validation and verification, model translation, real-time embedded systems},
} 

@article{jakovljevic_jim14,
 author = {Jakovljevic, Zivana and Petrovic, Petar B. and Mikovic, Vladimir Dj. and Pajic, Miroslav},
 title = {Fuzzy Inference Mechanism for Recognition of Contact States in Intelligent Robotic Assembly},
 journal = {Journal of Intelligent Manufacturing},
 issue_date = {June 2014},
 volume = {25},
 number = {3},
 month = jun,
 year = {2014},
 issn = {0956-5515},
 pages = {571--587},
 numpages = {17},
 url = {http://dx.doi.org/10.1007/s10845-012-0706-x},
 doi = {10.1007/s10845-012-0706-x},
 acmid = {2629775},
 publisher = {Springer-Verlag New York, Inc.},
 address = {Secaucus, NJ, USA},
 keywords = {Contact states, Fuzzy inference mechanism, Part mating, Support vector machines},
} 


@ARTICLE{pajic_tii14,
author={M. Pajic and R. Mangharam and O. Sokolsky and D. Arney and J. Goldman and I. Lee},
journal={IEEE Transactions on Industrial Informatics},
title={Model-Driven Safety Analysis of Closed-Loop Medical Systems},
year={2014},
volume={10},
number={1},
pages={3-16},
doi={10.1109/TII.2012.2226594},
ISSN={1551-3203},
month={Feb},}


@article{jiang_sttt14,
  title={Closed-loop verification of medical devices with model abstraction and refinement},
  author={Jiang, Zhihao and Pajic, Miroslav and Alur, Rajeev and Mangharam, Rahul},
  journal={International Journal on Software Tools for Technology Transfer},
  volume={16},
  number={2},
  pages={191--213},
  year={2014},
  publisher={Springer}
}


@INPROCEEDINGS{ivanov_date14, 
author={Ivanov, R. and Pajic, M. and Insup Lee}, 
booktitle={Design, Automation and Test in Europe Conference and Exhibition (DATE)}, 
title={Attack-resilient sensor fusion}, 
year={2014}, 
month={March}, 
pages={1-6}, 
doi={10.7873/DATE.2014.067},}





@INPROCEEDINGS{miao_cdc14,
author={F. Miao and Q. Zhu and M. Pajic and G. J. Pappas},
booktitle={53rd IEEE Conference on Decision and Control},
title={Coding sensor outputs for injection attacks detection},
year={2014},
pages={5776-5781},
doi={10.1109/CDC.2014.7040293},
ISSN={0191-2216},
month={Dec},}

@INPROCEEDINGS{gatsis_cdc14,
author={K. Gatsis and M. Pajic and A. Ribeiro and G. J. Pappas},
booktitle={53rd IEEE Conference on Decision and Control},
title={Opportunistic sensor scheduling in wireless control systems},
year={2014},
pages={3777-3782},
doi={10.1109/CDC.2014.7039977},
ISSN={0191-2216},
month={Dec},}

@INPROCEEDINGS{sokolsky_cpsarch14, 
author={O. Sokolsky and M. Pajic and N. Bezzo and I. Lee}, 
booktitle={Workshop on Cyber-Physical System Architectures and Design Methodologies (CPSArch), part of Embedded Systems Week (ESWeek)}, 
title={Architecture-Centric Software Development for Cyber-Physical Systems}, 
year={2014},
numpages = {6}
}


@INPROCEEDINGS{gatsis_iccps14, 
author={Gatsis, K. and Pajic, M. and Ribeiro, A and Pappas, G.J.}, 
booktitle={ACM/IEEE International Conference on Cyber-Physical Systems (ICCPS)}, 
title={Opportunistic scheduling of control tasks over shared wireless channels}, 
year={2014}, 
month={April}, 
pages={48-59}, 
doi={10.1109/ICCPS.2014.6843710}
}


@INPROCEEDINGS{weimer_acc14, 
author={Weimer, J. and Bezzo, N. and Pajic, M. and Sokolsky, O. and Insup Lee}, 
booktitle={American Control Conference (ACC)}, 
title={Attack-resilient minimum mean-squared error estimation}, 
year={2014}, 
month={June}, 
pages={1114-1119},
doi={10.1109/ACC.2014.6859478}, 
ISSN={0743-1619},}


@inproceedings{ivanov_hicons14,
 author = {Ivanov, Radoslav and Pajic, Miroslav and Lee, Insup},
 title = {Resilient Multidimensional Sensor Fusion Using Measurement History},
 booktitle = {Proceedings of the 3rd International Conference on High Confidence Networked Systems},
 series = {HiCoNS '14},
 year = {2014},
 isbn = {978-1-4503-2652-0},
 location = {Berlin, Germany},
 pages = {1--10},
 numpages = {10},
 url = {http://doi.acm.org/10.1145/2566468.2566475},
 doi = {10.1145/2566468.2566475},
 acmid = {2566475},
 publisher = {ACM},
 address = {New York, NY, USA},
 keywords = {cps security, fault-tolerance, fault-tolerant algorithms, sensor fusion},
} 


@INPROCEEDINGS{bezzo_iros14,
author={N. Bezzo and J. Weimer and M. Pajic and O. Sokolsky and G. J. Pappas and I. Lee},
booktitle={2014 IEEE/RSJ International Conference on Intelligent Robots and Systems},
title={Attack resilient state estimation for autonomous robotic systems},
year={2014},
pages={3692-3698},
doi={10.1109/IROS.2014.6943080},
ISSN={2153-0858},
month={Sept},}


@inbook{ivanov14_book,
  title = {Resilient Sensor Fusion for Safety-Critical Cyber-Physical Systems},
  booktitle = {Multisensor Data Fusion: From Algorithm and Architecture Design to Applications},
  author = {Ivanov, R. and Pajic, M. and Lee, I.},
  year = {2014},
}



@ARTICLE{gatsis_tac15,
author={K. Gatsis and M. Pajic and A. Ribeiro and G. J. Pappas},
journal={IEEE Transactions on Automatic Control},
title={Opportunistic Control Over Shared Wireless Channels},
year={2015},
volume={60},
number={12},
pages={3140-3155},
doi={10.1109/TAC.2015.2416922},
ISSN={0018-9286},
month={Dec},}


@ARTICLE{jakovljevic_tii15,
author={Z. Jakovljevic and R. Puzovic and M. Pajic},
journal={IEEE Transactions on Industrial Informatics},
title={Recognition of Planar Segments in Point Cloud Based on Wavelet Transform},
year={2015},
volume={11},
number={2},
pages={342-352},
doi={10.1109/TII.2015.2389195},
ISSN={1551-3203},
month={April},}


@article{jakovljevic_aa15,
  title={Diagnosis of irregularities in the robotized part mating process based on contextual recognition of contact states transitions},
  author={Jakovljevic, Zivana and Petrovic, Petar B and Milkovic, Dragan and Pajic, Miroslav},
  journal={Assembly Automation},
  volume={35},
  number={2},
  pages={190--199},
  year={2015}
}


@inproceedings{pajic_emsoft15,
 author = {Pajic, Miroslav and Park, Junkil and Lee, Insup and Pappas, George J. and Sokolsky, Oleg},
 title = {Automatic Verification of Linear Controller Software},
 booktitle = {Proceedings of the 12th International Conference on Embedded Software},
 series = {EMSOFT '15},
 year = {2015},
 isbn = {978-1-4673-8079-9},
 location = {Amsterdam, The Netherlands},
 pages = {217--226},
 numpages = {10},
 url = {http://dl.acm.org/citation.cfm?id=2830865.2830889},
 acmid = {2830889},
 publisher = {IEEE Press},
 address = {Piscataway, NJ, USA},
} 


@inproceedings{junkil_iccps15,
 author = {Park, Junkil and Ivanov, Radoslav and Weimer, James and Pajic, Miroslav and Lee, Insup},
 title = {Sensor Attack Detection in the Presence of Transient Faults},
 booktitle = {Proceedings of the ACM/IEEE Sixth International Conference on Cyber-Physical Systems},
 series = {ICCPS '15},
 year = {2015},
 isbn = {978-1-4503-3455-6},
 location = {Seattle, Washington},
 pages = {1--10},
 numpages = {10},
 url = {http://doi.acm.org/10.1145/2735960.2735984},
 doi = {10.1145/2735960.2735984},
 acmid = {2735984},
 publisher = {ACM},
 address = {New York, NY, USA},
} 


@inproceedings{alfaruque_codes15,
 author = {Al Faruque, Mohammad and Regazzoni, Francesco and Pajic, Miroslav},
 title = {Design Methodologies for Securing Cyber-physical Systems},
 booktitle = {Proceedings of the 10th International Conference on Hardware/Software Codesign and System Synthesis},
 series = {CODES '15},
 year = {2015},
 isbn = {978-1-4673-8321-9},
 location = {Amsterdam, The Netherlands},
 pages = {30--36},
 numpages = {7},
 url = {http://dl.acm.org/citation.cfm?id=2830840.2830844},
 acmid = {2830844},
 publisher = {IEEE Press},
 address = {Piscataway, NJ, USA},
} 



@INPROCEEDINGS{pajic_cdc15, 
author={Pajic, M. and Tabuada, P. and Lee, I. and Pappas, G.J.}, 
booktitle={54th IEEE  Annual Conference on Decision and Control (CDC)}, 
title={Attack-Resilient State Estimation in the Presence of Noise}, 
location = {Osaka, Japan},
 year = {2015},
pages = {5827--5832},
numpages = {6},
month={Dec}
 }


@INPROCEEDINGS{ivanov_allerton15,
author={R. Ivanov and N. Atanasov and M. Pajic and G. Pappas and I. Lee},
booktitle={2015 53rd Annual Allerton Conference on Communication, Control, and Computing (Allerton)},
title={Robust estimation using context-aware filtering},
year={2015},
pages={590-597},
doi={10.1109/ALLERTON.2015.7447058},
month={Sept},}


@INPROCEEDINGS{ivanov_mvigro15, 
author={Ivanov, R. and Atanasov, N. and Pajic, M. and  Lee, I. and Pappas, G. J.}, 
booktitle={Workshop on Multi VIew Geometry in Robotics (MVIGRO), in conjunction with RSS}, 
title={Robust Localization Using Context-Aware Filtering}, 
year={2015}
}



@article{ivanov_tecs16,
 author = {Ivanov, Radoslav and Pajic, Miroslav and Lee, Insup},
 title = {Attack-Resilient Sensor Fusion for Safety-Critical Cyber-Physical Systems},
 journal = {ACM Transactions on Embedded Computing Systems},
 issue_date = {February 2016},
 volume = {15},
 number = {1},
 month = feb,
 year = {2016},
 issn = {1539-9087},
 pages = {21:1--21:24},
 articleno = {21},
 numpages = {24},
 url = {http://doi.acm.org/10.1145/2847418},
 doi = {10.1145/2847418},
 acmid = {2847418},
 publisher = {ACM},
} 




@incollection{junkil_tacas16,
  title={Scalable Verification of Linear Controller Software},
  author={Park, Junkil and Pajic, Miroslav and Lee, Insup and Sokolsky, Oleg},
  booktitle={Tools and Algorithms for the Construction and Analysis of Systems (TACAS)},
  pages={662--679},
  year={2016},
  publisher={Springer}
}



@INPROCEEDINGS{ivanov_iccps16,
author={R. Ivanov and N. Atanasov and J. Weimer and M. Pajic and A. Simpao and M. Rehman and G. Pappas and I. Lee},
booktitle={2016 ACM/IEEE 7th International Conference on Cyber-Physical Systems (ICCPS)},
title={Estimation of Blood Oxygen Content Using Context-Aware Filtering},
year={2016},
pages={1-10},
doi={10.1109/ICCPS.2016.7479102},
month={April},}


@INPROCEEDINGS{mangharam_comnets16,
author={R. Mangharam and H. Abbas and M. Behl and K. Jang and M. Pajic and Z. Jiang},
booktitle={2016 8th International Conference on Communication Systems and Networks (COMSNETS)},
title={Three challenges in cyber-physical systems},
year={2016},
pages={1-8},
doi={10.1109/COMSNETS.2016.7440015},
month={Jan},}








@inproceedings{bogdan_codes16,
 author = {Bogdan, Paul and Pajic, Miroslav and Pande, Partha Pratim and Raghunathan, Vijay},
 title = {Making the Internet-of-things a Reality: From Smart Models, Sensing and Actuation to Energy-efficient Architectures},
 booktitle = {Proceedings of the Eleventh IEEE/ACM/IFIP International Conference on Hardware/Software Codesign and System Synthesis},
 series = {CODES '16},
 year = {2016},
 isbn = {978-1-4503-4483-8},
 location = {Pittsburgh, Pennsylvania},
 pages = {25:1--25:10},
 articleno = {25},
 numpages = {10},
 url = {http://doi.acm.org/10.1145/2968456.2973272},
 doi = {10.1145/2968456.2973272},
 acmid = {2973272}
} 




@inproceedings{ibrahim_cases16,
 author = {Ibrahim, Mohamed and Boswell, Craig and Chakrabarty, Krishnendu and Scott, Kristin and Pajic, Miroslav},
 title = {A Real-time Digital-microfluidic Platform for Epigenetics},
 booktitle = {Proceedings of the International Conference on Compilers, Architectures and Synthesis for Embedded Systems},
 series = {CASES '16},
 year = {2016},
 pages = {10:1--10:10},
 doi = {10.1145/2968455.2968516}
} 




@inproceedings{lesi_etfa16,
 author = {V. Lesi and Z. Jakovljevic and M. Pajic},
 title = {{Towards Plug-n-Play Numerical Control for Reconfigurable Manufacturing Systems}},
 booktitle = {21st IEEE International Conference on Emerging Technologies and Factory Automation (ETFA)},
 year = {2016},
 pages = {1-8},
 doi={10.1109/ETFA.2016.7733524},
month={Sept},
 } 
 


@inproceedings{li_iccad16,
 author = {Li, Zipeng and Lai, Kelvin Yi-Tse and Yu, Po-Hsien and Chakrabarty, Krishnendu and Pajic, Miroslav and Ho, Tsung-Yi and Lee, Chen-Yi},
 title = {{Error Recovery in a Micro-electrode-dot-array Digital Microfluidic Biochip}},
 booktitle = {Proceedings of the 35th International Conference on Computer-Aided Design},
 series = {ICCAD '16},
 year = {2016},
 pages = {105:1--105:8},
 doi = {10.1145/2966986.2967035}
} 






@ARTICLE{miao_tcns17,
author={F. Miao and Q. Zhu and M. Pajic and G. J. Pappas},
journal={IEEE Transactions on Control of Network Systems},
title={Coding Schemes for Securing Cyber-Physical Systems Against Stealthy Data Injection Attacks},
year={2017},
volume={4},
number={1},
pages={106-117},
keywords={Algorithm design and analysis;Data models;Detectors;Encoding;Estimation;Kalman filters;Linear systems;Coding;detection;feasible coding matrix;state estimator;stealthy data injection attacks;time-varying coding},
doi={10.1109/TCNS.2016.2573039},
ISSN={2325-5870},
month={March}
}



@article{junkil_tcps17,
 author = {Park, Junkil and Ivanov, Radoslav and Weimer, James and Pajic, Miroslav and Son, Sang Hyuk and Lee, Insup},
 title = {Security of Cyber-Physical Systems in the Presence of Transient Sensor Faults},
 journal = {ACM Transactions on Cyber-Physical Systems},
 issue_date = {May 2017},
 volume = {1},
 number = {3},
 month = May,
 year = {2017},
 pages = {15:1--15:23}
} 


@incollection{junkil_tacas17,
author="Park, Junkil and Pajic, Miroslav and Sokolsky, Oleg and Lee, Insup",
title="Automatic Verification of Finite Precision Implementations of Linear Controllers",
bookTitle="Tools and Algorithms for the Construction and Analysis of Systems: 23rd International Conference, TACAS 2017",
year="2017",
publisher="Springer Berlin Heidelberg",
pages="153--169",
doi="10.1007/978-3-662-54577-5_9"
}


@INPROCEEDINGS{fricks_rams17, 
author={R. W. B. Fricks and H. H. Tseng and M. Pajic and K. S. Trivedi}, 
booktitle={2017 Annual Reliability and Maintainability Symposium (RAMS)}, 
title={Transient performance availability modeling in high volume outpatient clinics}, 
year={2017}, 
pages={1-6}, 
doi={10.1109/RAM.2017.7889777}, 
month={Jan}
}



@INPROCEEDINGS{jovanov_cdc17, 
author={I. Jovanov and M. Pajic}, 
booktitle={2017 IEEE 56th Annual Conference on Decision and Control (CDC)}, 
title={Sporadic data integrity for secure state estimation}, 
year={2017}, 
volume={}, 
number={}, 
pages={163-169}, 
doi={10.1109/CDC.2017.8263660}, 
ISSN={}, 
month={Dec},}





 @INPROCEEDINGS{lesi_rtss17, 
author={V. Lesi and I. Jovanov and M. Pajic}, 
booktitle={2017 IEEE Real-Time Systems Symposium (RTSS)}, 
title={Network Scheduling for Secure Cyber-Physical Systems}, 
year={2017}, 
volume={}, 
number={}, 
pages={45-55}, 
keywords={cryptography;cyber-physical systems;data integrity;integer programming;linear programming;message authentication;scheduling;QoC requirements;Quality-of-Control;cryptographic tools;cyber-physical systems attacks;cyber-physical systems security;data integrity;message authentication codes;mixed integer linear programming problem;network scheduling;opportunistic scheduling;real-time network messages;runtime bandwidth allocation;Authentication;Automotive engineering;Data integrity;Real-time systems;Sensors;Standards;CAN-bus;CPS-Security;Mixed-Integer-Linear-Programming;Quality-of-Control;Real-Time-Scheduling}, 
doi={10.1109/RTSS.2017.00012}, 
ISSN={}, 
month={Dec}
}



@article{elfar_tecs17,
 author = {Elfar, Mahmoud and Zhong, Zhanwei and Li, Zipeng and Chakrabarty, Krishnendu and Pajic, Miroslav},
 title = {Synthesis of Error-Recovery Protocols for Micro-Electrode-Dot-Array Digital Microfluidic Biochips},
 journal = {ACM Trans. Embed. Comput. Syst.},
 issue_date = {October 2017},
 volume = {16},
 number = {5s},
 month = sep,
 year = {2017},
 issn = {1539-9087},
 pages = {127:1--127:22},
 articleno = {127},
 numpages = {22},
 url = {http://doi.acm.org/10.1145/3126538},
 doi = {10.1145/3126538}
} 



@article{miao_automatica17,
title = "A hybrid stochastic game for secure control of cyber-physical systems",
journal = "Automatica",
volume = "93",
pages = "55 - 63",
year = "2018",
issn = "0005-1098",
doi = "https://doi.org/10.1016/j.automatica.2018.03.012",
url = "http://www.sciencedirect.com/science/article/pii/S0005109818300992",
author = "Fei Miao and Quanyan Zhu and Miroslav Pajic and George J. Pappas",
keywords = "Stochastic game, Secure control, Saddle-point equilibrium"
}


@ARTICLE{li_tcad17, 
author={Z. Li and K. Y. T. Lai and J. McCrone and P. H. Yu and K. Chakrabarty and M. Pajic and T. Y. Ho and C. Y. Lee}, 
journal={IEEE Transactions on Computer-Aided Design of Integrated Circuits and Systems}, 
title={Efficient and Adaptive Error Recovery in a Micro-Electrode-Dot-Array Digital Microfluidic Biochip}, 
year={2017}, 
volume={PP}, 
number={99}, 
pages={1-1}, 
doi={10.1109/TCAD.2017.2729347}, 
month={}
}

@Inbook{jakovljevic_newtech17a,
author="Jakovljevic, Zivana and Mitrovic, Stefan and Pajic, Miroslav",
title="Cyber Physical Production Systems---An IEC 61499 Perspective",
bookTitle="Proceedings of 5th International Conference on Advanced Manufacturing Engineering and Technologies: NEWTECH 2017",
year="2017",
publisher="Springer International Publishing",
pages="27--39",
doi="10.1007/978-3-319-56430-2_3"
}

@Inbook{jakovljevic_newtech17b,
author="Jakovljevic, Zivana and Majstorovic, Vidosav and Stojadinovic, Slavenko and Zivkovic, Srdjan and Gligorijevic, Nemanja and Pajic, Miroslav",
title="Cyber-Physical Manufacturing Systems (CPMS)",
bookTitle="Proceedings of 5th International Conference on Advanced Manufacturing Engineering and Technologies: NEWTECH 2017",
year="2017",
publisher="Springer International Publishing",
pages="199--214",
doi="10.1007/978-3-319-56430-2_14"
}



@article{ivanov_tac19,
  title={Continuous estimation using context-dependent discrete measurements},
  author={Ivanov, Radoslav and Atanasov, Nikolay and Pajic, Miroslav and Weimer, James and Pappas, George J and Lee, Insup},
  journal={IEEE Transactions on Automatic Control},
  volume={64},
  number={1},
  pages={235--250},
  year={2019},
  publisher={IEEE}
}

@inproceedings{elfar_iccps17,
  title={Platform for security-aware design of human-on-the-loop cyber-physical systems},
  author={Elfar, Mahmoud and Zhu, Haibei and Raghunathan, Adithya and Tay, Yi Y and Wubbenhorst, Jeffrey and Cummings, ML and Pajic, Miroslav},
  booktitle={Proc. of the 8th ACM ICCPS},
  year={2017}
}





@INPROCEEDINGS{jovanov_cdc18, 
author={I. {Jovanov} and M. {Pajic}}, 
booktitle={2018 IEEE Conference on Decision and Control (CDC)}, 
title={Secure State Estimation with Cumulative Message Authentication}, 
year={2018}, 
pages={2074-2079}, 
month={Dec}
}



@inproceedings{jovanov_iccps18,
  title={Platform for model-based design and testing for deep brain stimulation},
  author={Jovanov, Ilija and Naumann, Michael and Kumaravelu, Karthik and Grill, Warren M and Pajic, Miroslav},
  booktitle={Proceedings of the 9th ACM/IEEE International Conference on Cyber-Physical Systems (ICCPS)},
  pages={263--274},
  year={2018},
  organization={IEEE Press}
}

@article{zhu_thms19,
 author = {H. Zhu and M. Cummings and M. Elfar and Z. Wang and M. Pajic},
 title = {Operator Strategy Model Development in UAV Hacking Detection},
 journal = {IEEE Transactions on Human-Machine Systems},
 note = {to appear, available at \url{http://people.duke.edu/~mp275/publications.html}}
} 

@inproceedings{zhu_chii18,
  title={Human augmentation of uav cyber-attack detection},
  author={Zhu, Haibei and Elfar, Mahmoud and Pajic, Miroslav and Wang, Ziyao and Cummings, Mary L},
  booktitle={International Conference on Augmented Cognition},
  pages={154--167},
  year={2018},
  organization={Springer}
}


@inproceedings{elfar_icra19,
  title={Security-Aware Synthesis of Human-UAV Protocols},
  author={M. Elfar and H. Zhu and M. L. Cummings and M. Pajic},
  booktitle={2019 International Conference on Robotics and Automation (ICRA)},
  year={2019},
  note = {to appear, available at \url{http://people.duke.edu/~mp275/publications.html}}
}


@article{wang_cav19,
  author    = {Y. Wang and S. Nalluri  and B. Bonakdarpour and M. Pajic},
  title     = {Statistical Model Checking for Probabilistic Hyperproperties},
  journal   = {CoRR},
  volume    = {abs/1707.02950},
  year      = {2019},
  url       = {https://arxiv.org/abs/1707.02950},
}

@article{elfar_cav19,
  author    = {M. Elfar and Y. Wang and M. Pajic},
  title     = {Security-Aware Synthesis using Delayed Action Games},
  journal   = {CoRR},
  volume    = {abs/1902.04111},
  year      = {2019},
  url       = {https://arxiv.org/abs/1902.04111},
}



@inproceedings{nateghi2018cyber,
  title={Cyber Attack Reconstruction of Nonlinear Systems via Higher-Order Sliding-Mode Observer and Sparse Recovery Algorithm},
  author={Nateghi, Shamila and Shtessel, Yuri and Barbot, Jean-Pierre and Edwards, Christopher},
  booktitle={IEEE Conf. on Decision and Control (CDC)},
  pages={5963--5968},
  year={2018},
  organization={}
 }
 
 @article{hu2017secure,
  title={Secure state estimation and control for cyber security of the nonlinear power systems},
  author={Hu, Qie and Fooladivanda, Dariush and Chang, Young Hwan and Tomlin, Claire J},
  journal={IEEE Transactions on Control of Network Systems},
  volume={5},
  number={3},
  pages={1310--1321},
  year={2017},
  publisher={IEEE}
}

@article{kerns2014unmanned,
  title={Unmanned aircraft capture and control via GPS spoofing},
  author={Kerns, Andrew J and Shepard, Daniel P and Bhatti, Jahshan A and Humphreys, Todd E},
  journal={Journal of Field Robotics},
  volume={31},
  number={4},
  pages={617--636},
  year={2014},
  publisher={Wiley Online Library}
}

@inproceedings{khazraei2017replay,
  title={Replay attack detection in a multi agent system using stability analysis and loss effective watermarking},
  author={Khazraei, Amir and Kebriaei, Hamed and Salmasi, Farzad Rajaei},
  booktitle={2017 American Control Conference (ACC)},
  pages={4778--4783},
  year={2017},
  organization={IEEE}
}



@book{golub2012matrix,
  title={Matrix computations},
  author={Golub, Gene H and Van Loan, Charles F},
  volume={3},
  year={2012},
  publisher={JHU press}
}








@INPROCEEDINGS{dick_asilomar06,
author={C. Dick and F. Harris and M. Pajic and D. Vuletic},
booktitle={2006 Fortieth Asilomar Conference on Signals, Systems and Computers},
title={Real-Time QRD-Based Beamforming on an FPGA Platform},
year={2006},
pages={1200-1204},
keywords={array signal processing;digital signal processing chips;field programmable gate arrays;least squares approximations;linear algebra;logic design;real-time systems;FPGA resource utilization;Mathworks Simulink visual programming;Normal equations;System Generator;Xilinx Virtex;hardware verification tool;least-squares solution;linear algebra operations;model-based FPGA design flow;real-time QRD-based beamforming;Adaptive filters;Application software;Array signal processing;Baseband;Design engineering;Digital signal processing;Educational institutions;Field programmable gate arrays;Logic design;Systolic arrays},
doi={10.1109/ACSSC.2006.354945},
ISSN={1058-6393},
month={Oct},}


@article{dick_xcell07,
  title={Implementing a real-time beamformer on an FPGA platform},
  author={Dick, Chris and Harris, Fred and Pajic, Miroslav and Vuletic, Dragan},
  journal={Xcell journal},
  volume={86},
  year={2007}
}


@INPROCEEDINGS{jorgovanovic_miel08, 
author={Jorgovanovic, M. and Pajic, M. and Kvascev, G. and Popovic, J.}, 
booktitle={26th International Conference on Microelectronics (MIEL)}, 
title={FPGA design of arbitrary down-sampler}, 
year={2008}, 
month={May}, 
pages={391-394}, 
doi={10.1109/ICMEL.2008.4559303},}

@inproceedings {pajic_wess08,
title = {{WisperNet: Anti-Jamming for Wireless Sensor Networks}},
booktitle = {WESS 2008: 2nd Workshop on Embedded Systems Security - A Workshop of the IEEE/ACM EMSOFT'2008 and the Embedded Systems Week},
year = {2008},
author = {M. Pajic and R. Mangharam},
pages = {38-43},
}


@INPROCEEDINGS{evm_workshop09, 
author={Mangharam, R. and Pajic, M.}, 
booktitle={Distributed Computing Systems Workshops, 2009. ICDCS Workshops '09. 29th IEEE International Conference on}, 
title={Embedded Virtual Machines for Robust Wireless Control Systems}, 
year={2009}, 
month={June}, 
pages={38-43}, 
ISSN={1545-0678},}


@INPROCEEDINGS{pajic_ipsn09, 
author={Pajic, M. and Mangharam, R.}, 
booktitle={ International Conference on Information Processing in Sensor Networks (IPSN) }, 
title={Anti-jamming for embedded wireless networks}, 
year={2009}, 
month={April}, 
pages={301-312}, 
}


@article{pajic_eurasip10,
  title={Spatio-temporal techniques for anti-jamming in embedded wireless networks},
  author={Pajic, Miroslav and Mangharam, Rahul},
  journal={EURASIP Journal on Wireless Communications and Networking},
  volume={2010},
  number={819318},
  year={2010},
  publisher={Springer International Publishing},
  doi = {10.1155/2010/819318}
}


@INPROCEEDINGS{pajic_rtas10, 
author={Pajic, M. and Mangharam, R.}, 
booktitle={16th IEEE Real-Time and Embedded Technology and Applications Symposium (RTAS)}, 
title={Embedded Virtual Machines for Robust Wireless Control and Actuation}, 
year={2010}, 
month={April}, 
pages={79-88}, 
doi={10.1109/RTAS.2010.43}, 
ISSN={1080-1812},}



@inproceedings{arney_iccps10,
 author = {Arney, David and Pajic, Miroslav and Goldman, Julian M. and Lee, Insup and Mangharam, Rahul and Sokolsky, Oleg},
 title = {Toward Patient Safety in Closed-loop Medical Device Systems},
 booktitle = {Proceedings of the 1st ACM/IEEE International Conference on Cyber-Physical Systems},
 series = {ICCPS '10},
 year = {2010},
 isbn = {978-1-4503-0066-7},
 location = {Stockholm, Sweden},
 pages = {139--148},
 numpages = {10},
 url = {http://doi.acm.org/10.1145/1795194.1795214},
 doi = {10.1145/1795194.1795214},
 acmid = {1795214},
 publisher = {ACM},
 address = {New York, NY, USA},
} 


@INPROCEEDINGS{jiang_ecrts10, 
author={Zhihao Jiang and Pajic, M. and Connolly, A and Dixit, S. and Mangharam, R.}, 
booktitle={22nd Euromicro Conference on Real-Time Systems (ECRTS)}, 
title={Real-Time Heart Model for Implantable Cardiac Device Validation and Verification}, 
year={2010}, 
month={July}, 
pages={239-248}, 
doi={10.1109/ECRTS.2010.36}, 
ISSN={1068-3070},}




@INPROCEEDINGS{pajic_cdc10, 
author={Pajic, M. and Sundaram, S. and Le Ny, Jerome and Pappas, G.J. and Mangharam, R.}, 
booktitle={49th IEEE Conference on Decision and Control (CDC)}, 
title={The Wireless Control Network: Synthesis and robustness}, 
year={2010}, 
month={Dec}, 
pages={7576-7581}, 
ISSN={0743-1546},}


@INPROCEEDINGS{sundaram_cdc10, 
author={Sundaram, S. and Pajic, M. and Hadjicostis, C.N. and Mangharam, R. and Pappas, G.J.}, 
booktitle={49th IEEE Conference on Decision and Control (CDC)}, 
title={{The Wireless Control Network: Monitoring for Malicious Behavior}}, 
year={2010}, 
month={Dec}, 
pages={5979-5984}, 
ISSN={0743-1546},}




@ARTICLE{pajic_tac11,
author={M. Pajic and S. Sundaram and G. J. Pappas and R. Mangharam},
journal={IEEE Transactions on Automatic Control},
title={The Wireless Control Network: A New Approach for Control Over Networks},
year={2011},
volume={56},
number={10},
pages={2305-2318},
doi={10.1109/TAC.2011.2163864},
ISSN={0018-9286},
month={Oct},}


@INPROCEEDINGS{pajic_asilomar11, 
author={Pajic, M. and Sundaram, S. and Pappas, G.J. and Mangharam, R.}, 
booktitle={Signals, Systems and Computers (ASILOMAR), 2011 Conference Record of the Forty Fifth Asilomar Conference on}, 
title={Network synthesis for dynamical system stabilization}, 
year={2011}, 
month={Nov}, 
pages={821-825},
doi={10.1109/ACSSC.2011.6190122}, 
ISSN={1058-6393},}


@INPROCEEDINGS{jiang_iccps11, 
author={Zhihao Jiang and Pajic, M. and Mangharam, R.}, 
booktitle={IEEE/ACM International Conference on Cyber-Physical Systems (ICCPS)}, 
title={Model-Based Closed-Loop Testing of Implantable Pacemakers}, 
year={2011}, 
month={April}, 
pages={131-140}, 
doi={10.1109/ICCPS.2011.28},}


@INPROCEEDINGS{pajic_cdc11,
author={M. Pajic and S. Sundaram and G. J. Pappas and R. Mangharam},
booktitle={50th IEEE Conference on Decision and Control and European Control Conference (CDC / ECC)},
title={Topological conditions for wireless control networks},
year={2011},
pages={2353-2360},
doi={10.1109/CDC.2011.6161347},
ISSN={0191-2216},
month={Dec},}


@article{hadziahmetovic_tvst12,
  title={The Oral Iron Chelator Deferiprone Protects Against Retinal Degeneration Induced through Diverse MechanismsHadziahmetovic et al.},
  author={Hadziahmetovic, Majda and Pajic, Miroslav and Grieco, Steven and Song, Ying and Song, Delu and Li, Yafeng and Cwanger, Alyssa and Iacovelli, Jared and Chu, Sally and Ying, Gui-shuang and others},
  journal={Translational vision science \& technology},
  volume={1},
  number={3},
  pages={2.1--2.12},
  year={2012},
  publisher={The Association for Research in Vision and Ophthalmology}
}


@ARTICLE{jiang_pieee12,
author={Z. Jiang and M. Pajic and R. Mangharam},
journal={Proceedings of the IEEE},
title={{Cyber-Physical Modeling of Implantable Cardiac Medical Devices}},
year={2012},
volume={100},
number={1},
pages={122-137},
doi={10.1109/JPROC.2011.2161241},
ISSN={0018-9219},
month={Jan},}


@article{pajic_tecs13,
 author = {Pajic, Miroslav and Chernoguzov, Alexander and Mangharam, Rahul},
 title = {Robust Architectures for Embedded Wireless Network Control and Actuation},
 journal = {ACM Transactions on Embedded Computing Systems},
 issue_date = {December 2012},
 volume = {11},
 number = {4},
 month = jan,
 year = {2013},
 issn = {1539-9087},
 pages = {82:1--82:24},
 articleno = {82},
 numpages = {24},
 url = {http://doi.acm.org/10.1145/2362336.2362349},
 doi = {10.1145/2362336.2362349},
 acmid = {2362349},
 publisher = {ACM},
 address = {New York, NY, USA},
 keywords = {Distributed systems, fault tolerance, wireless sensor networks},
} 


@article{hadziahmetovic_tvst12,
  title={The Oral Iron Chelator Deferiprone Protects Against Retinal Degeneration Induced through Diverse MechanismsHadziahmetovic et al.},
  author={Hadziahmetovic, Majda and Pajic, Miroslav and Grieco, Steven and Song, Ying and Song, Delu and Li, Yafeng and Cwanger, Alyssa and Iacovelli, Jared and Chu, Sally and Ying, Gui-shuang and others},
  journal={Translational vision science \& technology},
  volume={1},
  number={3},
  pages={2--2},
  year={2012},
  publisher={The Association for Research in Vision and Ophthalmology}
}


@INPROCEEDINGS{pajic_rtas12, 
author={Pajic, M. and Zhihao Jiang and Insup Lee and Sokolsky, O. and Mangharam, R.}, 
booktitle={18th IEEE Real-Time and Embedded Technology and Applications Symposium (RTAS)}, 
title={From Verification to Implementation: A Model Translation Tool and a Pacemaker Case Study}, 
year={2012}, 
month={April}, 
pages={173-184}, 
doi={10.1109/RTAS.2012.25}, 
ISSN={1080-1812},}



@inproceedings{pajic_ipsn12,
 author = {Pajic, Miroslav and Sundaram, Shreyas and Le Ny, Jerome and Pappas, George J. and Mangharam, Rahul},
 title = {Closing the Loop: A Simple Distributed Method for Control over Wireless Networks},
 booktitle = {Proceedings of the 11th International Conference on Information Processing in Sensor Networks},
 series = {IPSN '12},
 year = {2012},
 isbn = {978-1-4503-1227-1},
 location = {Beijing, China},
 pages = {25--36},
 numpages = {12},
 url = {http://doi.acm.org/10.1145/2185677.2185681},
 doi = {10.1145/2185677.2185681},
 acmid = {2185681},
 publisher = {ACM},
 address = {New York, NY, USA},
 keywords = {cooperative control, decentralized control, networked control systems, wireless sensor networks},
} 


@inproceedings{jiang_tacas12,
 author = {Jiang, Zhihao and Pajic, Miroslav and Moarref, Salar and Alur, Rajeev and Mangharam, Rahul},
 title = {Modeling and Verification of a Dual Chamber Implantable Pacemaker},
 booktitle = {Proceedings of the 18th International Conference on Tools and Algorithms for the Construction and Analysis of Systems},
 series = {TACAS'12},
 year = {2012},
 isbn = {978-3-642-28755-8},
 location = {Tallinn, Estonia},
 pages = {188--203},
 numpages = {16},
 url = {http://dx.doi.org/10.1007/978-3-642-28756-5_14},
 doi = {10.1007/978-3-642-28756-5_14},
 acmid = {2260535},
 publisher = {Springer-Verlag},
 address = {Berlin, Heidelberg},
 keywords = {cyber-physical systems, implantable pacemaker, medical devices, software verification},
} 


@ARTICLE{pajic_jsac13,
author={M. Pajic and R. Mangharam and G. J. Pappas and S. Sundaram},
journal={IEEE Journal on Selected Areas in Communications},
title={Topological Conditions for In-Network Stabilization of Dynamical Systems},
year={2013},
volume={31},
number={4},
pages={794-807},
doi={10.1109/JSAC.2013.130415},
ISSN={0733-8716},
month={April},}



@ARTICLE{mangharam_iis13,
author={R. Mangharam and M. Pajic},
journal={Journal of the Indian Institute of Science},
title={Distributed Control for Cyber-Physical Systems},
year={2013},
volume={93},
number={3},
pages={358-387},
month={July-September}
}


@inproceedings{pajic_hicons13,
 author = {Pajic, Miroslav and Bezzo, Nicola and Weimer, James and Alur, Rajeev and Mangharam, Rahul and Michael, Nathan and Pappas, George J. and Sokolsky, Oleg and Tabuada, Paulo and Weirich, Stephanie and Lee, Insup},
 title = {Towards Synthesis of Platform-aware Attack-resilient Control Systems: Extended Abstract},
 booktitle = {Proceedings of the 2Nd ACM International Conference on High Confidence Networked Systems},
 series = {HiCoNS '13},
 year = {2013},
 isbn = {978-1-4503-1961-4},
 location = {Philadelphia, Pennsylvania, USA},
 pages = {75--76},
 numpages = {2},
 url = {http://doi.acm.org/10.1145/2461446.2461457},
 doi = {10.1145/2461446.2461457},
 acmid = {2461457},
 publisher = {ACM},
 address = {New York, NY, USA},
 keywords = {attack resilient control, cyber-physical system security},
} 


@INPROCEEDINGS{gatsis_cdc13, 
author={Gatsis, K. and Pajic, M. and Ribeiro, A and Pappas, G.J.}, 
booktitle={IEEE 52nd Annual Conference on Decision and Control (CDC)}, 
title={Power-aware communication for wireless sensor-actuator systems}, 
year={2013}, 
month={Dec}, 
pages={4006-4011}, 
doi={10.1109/CDC.2013.6760502}, 
ISSN={0743-1546},}


@INPROCEEDINGS{miao_acc13, 
author={Fei Miao and Pajic, M. and Mangharam, R. and Pappas, G.J.}, 
booktitle={American Control Conference (ACC)}, 
title={Networked realization of discrete-time controllers}, 
year={2013}, 
month={June}, 
pages={2996-3001}, 
doi={10.1109/ACC.2013.6580290}, 
ISSN={0743-1619},}


@INPROCEEDINGS{pajic_cdc13, 
author={Pajic, M. and Sundaram, S. and Pappas, G.J.}, 
booktitle={52nd IEEE Annual Conference on Decision and Control (CDC)},
title={Stabilizability over deterministic relay networks}, 
year={2013}, 
month={Dec}, 
pages={4018-4023},
doi={10.1109/CDC.2013.6760504}, 
ISSN={0743-1546},}


@INPROCEEDINGS{miao_cdc13, 
author={Fei Miao and Pajic, M. and Pappas, G.J.}, 
booktitle={IEEE 52nd Annual Conference on Decision and Control (CDC)}, 
title={Stochastic game approach for replay attack detection}, 
year={2013}, 
month={Dec}, 
pages={1854-1859}, 
doi={10.1109/CDC.2013.6760152}, 
ISSN={0743-1546}
}



@incollection{weimer_workshop13,
year={2013},
isbn={978-3-319-01158-5},
booktitle={Control of Cyber-Physical Systems},
volume={449},
series={Lecture Notes in Control and Information Sciences},
editor={Tarraf, Danielle C.},
doi={10.1007/978-3-319-01159-2_11},
title={Resilient Parameter-Invariant Control with Application to Vehicle Cruise Control},
url={http://dx.doi.org/10.1007/978-3-319-01159-2_11},
publisher={Springer International Publishing},
keywords={Secure Cyber-Physical Systems; Robust Control; Resilient Sensor Fusion},
author={Weimer, James and Bezzo, Nicola and Pajic, Miroslav and Pappas, GeorgeJ. and Sokolsky, Oleg and Lee, Insup},
pages={197-216},
language={English}
}


@article{pajic_tecs14,
 author = {Pajic, Miroslav and Jiang, Zhihao and Lee, Insup and Sokolsky, Oleg and Mangharam, Rahul},
 title = {Safety-critical Medical Device Development Using the UPP2SF Model Translation Tool},
 journal = {ACM Transactions on  Embedded Computing Systems},
 issue_date = {July 2014},
 volume = {13},
 number = {4s},
 month = apr,
 year = {2014},
 issn = {1539-9087},
 pages = {127:1--127:26},
 articleno = {127},
 numpages = {26},
 url = {http://doi.acm.org/10.1145/2584651},
 doi = {10.1145/2584651},
 acmid = {2584651},
 publisher = {ACM},
 address = {New York, NY, USA},
 keywords = {Model-based development, medical devices validation and verification, model translation, real-time embedded systems},
} 

@article{jakovljevic_jim14,
 author = {Jakovljevic, Zivana and Petrovic, Petar B. and Mikovic, Vladimir Dj. and Pajic, Miroslav},
 title = {Fuzzy Inference Mechanism for Recognition of Contact States in Intelligent Robotic Assembly},
 journal = {Journal of Intelligent Manufacturing},
 issue_date = {June 2014},
 volume = {25},
 number = {3},
 month = jun,
 year = {2014},
 issn = {0956-5515},
 pages = {571--587},
 numpages = {17},
 url = {http://dx.doi.org/10.1007/s10845-012-0706-x},
 doi = {10.1007/s10845-012-0706-x},
 acmid = {2629775},
 publisher = {Springer-Verlag New York, Inc.},
 address = {Secaucus, NJ, USA},
 keywords = {Contact states, Fuzzy inference mechanism, Part mating, Support vector machines},
} 


@ARTICLE{pajic_tii14,
author={M. Pajic and R. Mangharam and O. Sokolsky and D. Arney and J. Goldman and I. Lee},
journal={IEEE Transactions on Industrial Informatics},
title={Model-Driven Safety Analysis of Closed-Loop Medical Systems},
year={2014},
volume={10},
number={1},
pages={3-16},
doi={10.1109/TII.2012.2226594},
ISSN={1551-3203},
month={Feb},}


@article{jiang_sttt14,
  title={Closed-loop verification of medical devices with model abstraction and refinement},
  author={Jiang, Zhihao and Pajic, Miroslav and Alur, Rajeev and Mangharam, Rahul},
  journal={International Journal on Software Tools for Technology Transfer},
  volume={16},
  number={2},
  pages={191--213},
  year={2014},
  publisher={Springer}
}


@INPROCEEDINGS{ivanov_date14, 
author={Ivanov, R. and Pajic, M. and Insup Lee}, 
booktitle={Design, Automation and Test in Europe Conference and Exhibition (DATE)}, 
title={Attack-resilient sensor fusion}, 
year={2014}, 
month={March}, 
pages={1-6}, 
doi={10.7873/DATE.2014.067},}


@INPROCEEDINGS{pajic_iccps14, 
author={Pajic, M. and Weimer, J. and Bezzo, N. and Tabuada, P. and Sokolsky, O. and Insup Lee and Pappas, G.J.}, 
booktitle={ACM/IEEE International Conference on Cyber-Physical Systems (ICCPS)}, 
title={Robustness of attack-resilient state estimators}, 
year={2014}, 
month={April}, 
pages={163-174},}


@INPROCEEDINGS{miao_cdc14,
author={F. Miao and Q. Zhu and M. Pajic and G. J. Pappas},
booktitle={53rd IEEE Conference on Decision and Control},
title={Coding sensor outputs for injection attacks detection},
year={2014},
pages={5776-5781},
doi={10.1109/CDC.2014.7040293},
ISSN={0191-2216},
month={Dec},}

@INPROCEEDINGS{gatsis_cdc14,
author={K. Gatsis and M. Pajic and A. Ribeiro and G. J. Pappas},
booktitle={53rd IEEE Conference on Decision and Control},
title={Opportunistic sensor scheduling in wireless control systems},
year={2014},
pages={3777-3782},
doi={10.1109/CDC.2014.7039977},
ISSN={0191-2216},
month={Dec},}

@INPROCEEDINGS{sokolsky_cpsarch14, 
author={O. Sokolsky and M. Pajic and N. Bezzo and I. Lee}, 
booktitle={Workshop on Cyber-Physical System Architectures and Design Methodologies (CPSArch), part of Embedded Systems Week (ESWeek)}, 
title={Architecture-Centric Software Development for Cyber-Physical Systems}, 
year={2014},
numpages = {6}
}


@INPROCEEDINGS{gatsis_iccps14, 
author={Gatsis, K. and Pajic, M. and Ribeiro, A and Pappas, G.J.}, 
booktitle={ACM/IEEE International Conference on Cyber-Physical Systems (ICCPS)}, 
title={Opportunistic scheduling of control tasks over shared wireless channels}, 
year={2014}, 
month={April}, 
pages={48-59}, 
doi={10.1109/ICCPS.2014.6843710}
}


@INPROCEEDINGS{weimer_acc14, 
author={Weimer, J. and Bezzo, N. and Pajic, M. and Sokolsky, O. and Insup Lee}, 
booktitle={American Control Conference (ACC)}, 
title={Attack-resilient minimum mean-squared error estimation}, 
year={2014}, 
month={June}, 
pages={1114-1119},
doi={10.1109/ACC.2014.6859478}, 
ISSN={0743-1619},}


@inproceedings{ivanov_hicons14,
 author = {Ivanov, Radoslav and Pajic, Miroslav and Lee, Insup},
 title = {Resilient Multidimensional Sensor Fusion Using Measurement History},
 booktitle = {Proceedings of the 3rd International Conference on High Confidence Networked Systems},
 series = {HiCoNS '14},
 year = {2014},
 isbn = {978-1-4503-2652-0},
 location = {Berlin, Germany},
 pages = {1--10},
 numpages = {10},
 url = {http://doi.acm.org/10.1145/2566468.2566475},
 doi = {10.1145/2566468.2566475},
 acmid = {2566475},
 publisher = {ACM},
 address = {New York, NY, USA},
 keywords = {cps security, fault-tolerance, fault-tolerant algorithms, sensor fusion},
} 


@INPROCEEDINGS{bezzo_iros14,
author={N. Bezzo and J. Weimer and M. Pajic and O. Sokolsky and G. J. Pappas and I. Lee},
booktitle={2014 IEEE/RSJ International Conference on Intelligent Robots and Systems},
title={Attack resilient state estimation for autonomous robotic systems},
year={2014},
pages={3692-3698},
doi={10.1109/IROS.2014.6943080},
ISSN={2153-0858},
month={Sept},}


@inbook{ivanov14_book,
  title = {Resilient Sensor Fusion for Safety-Critical Cyber-Physical Systems},
  booktitle = {Multisensor Data Fusion: From Algorithm and Architecture Design to Applications},
  author = {Ivanov, R. and Pajic, M. and Lee, I.},
  year = {2014},
}



@ARTICLE{gatsis_tac15,
author={K. Gatsis and M. Pajic and A. Ribeiro and G. J. Pappas},
journal={IEEE Transactions on Automatic Control},
title={Opportunistic Control Over Shared Wireless Channels},
year={2015},
volume={60},
number={12},
pages={3140-3155},
doi={10.1109/TAC.2015.2416922},
ISSN={0018-9286},
month={Dec},}


@ARTICLE{jakovljevic_tii15,
author={Z. Jakovljevic and R. Puzovic and M. Pajic},
journal={IEEE Transactions on Industrial Informatics},
title={Recognition of Planar Segments in Point Cloud Based on Wavelet Transform},
year={2015},
volume={11},
number={2},
pages={342-352},
doi={10.1109/TII.2015.2389195},
ISSN={1551-3203},
month={April},}


@article{jakovljevic_aa15,
  title={Diagnosis of irregularities in the robotized part mating process based on contextual recognition of contact states transitions},
  author={Jakovljevic, Zivana and Petrovic, Petar B and Milkovic, Dragan and Pajic, Miroslav},
  journal={Assembly Automation},
  volume={35},
  number={2},
  pages={190--199},
  year={2015}
}


@inproceedings{pajic_emsoft15,
 author = {Pajic, Miroslav and Park, Junkil and Lee, Insup and Pappas, George J. and Sokolsky, Oleg},
 title = {Automatic Verification of Linear Controller Software},
 booktitle = {Proceedings of the 12th International Conference on Embedded Software},
 series = {EMSOFT '15},
 year = {2015},
 isbn = {978-1-4673-8079-9},
 location = {Amsterdam, The Netherlands},
 pages = {217--226},
 numpages = {10},
 url = {http://dl.acm.org/citation.cfm?id=2830865.2830889},
 acmid = {2830889},
 publisher = {IEEE Press},
 address = {Piscataway, NJ, USA},
} 


@inproceedings{junkil_iccps15,
 author = {Park, Junkil and Ivanov, Radoslav and Weimer, James and Pajic, Miroslav and Lee, Insup},
 title = {Sensor Attack Detection in the Presence of Transient Faults},
 booktitle = {Proceedings of the ACM/IEEE Sixth International Conference on Cyber-Physical Systems},
 series = {ICCPS '15},
 year = {2015},
 isbn = {978-1-4503-3455-6},
 location = {Seattle, Washington},
 pages = {1--10},
 numpages = {10},
 url = {http://doi.acm.org/10.1145/2735960.2735984},
 doi = {10.1145/2735960.2735984},
 acmid = {2735984},
 publisher = {ACM},
 address = {New York, NY, USA},
} 


@inproceedings{alfaruque_codes15,
 author = {Al Faruque, Mohammad and Regazzoni, Francesco and Pajic, Miroslav},
 title = {Design Methodologies for Securing Cyber-physical Systems},
 booktitle = {Proceedings of the 10th International Conference on Hardware/Software Codesign and System Synthesis},
 series = {CODES '15},
 year = {2015},
 isbn = {978-1-4673-8321-9},
 location = {Amsterdam, The Netherlands},
 pages = {30--36},
 numpages = {7},
 url = {http://dl.acm.org/citation.cfm?id=2830840.2830844},
 acmid = {2830844},
 publisher = {IEEE Press},
 address = {Piscataway, NJ, USA},
} 


@INPROCEEDINGS{pajic_cdc15, 
author={M. {Pajic} and P. {Tabuada} and I. {Lee} and G. J. {Pappas}}, 
booktitle={2015 54th IEEE Conference on Decision and Control (CDC)}, 
title={Attack-resilient state estimation in the presence of noise}, 
year={2015}, 
volume={}, 
number={}, 
pages={5827-5832}, 
keywords={convex programming;integer programming;linear programming;linear systems;state estimation;attack vector;sound attack identification;sound attack detection;unbounded state-estimation error;rigorous analytic bound;l1-based state estimator;mixed-integer linear program;l0-based state estimator;sensor attack;attack-resilient state estimation;State estimation;Noise measurement;Symmetric matrices;Optimization;Linear systems;Size measurement;Security}, 
doi={10.1109/CDC.2015.7403135}, 
ISSN={null}, 
month={Dec},}



@INPROCEEDINGS{ivanov_allerton15,
author={R. Ivanov and N. Atanasov and M. Pajic and G. Pappas and I. Lee},
booktitle={2015 53rd Annual Allerton Conference on Communication, Control, and Computing (Allerton)},
title={Robust estimation using context-aware filtering},
year={2015},
pages={590-597},
doi={10.1109/ALLERTON.2015.7447058},
month={Sept},}


@INPROCEEDINGS{ivanov_mvigro15, 
author={Ivanov, R. and Atanasov, N. and Pajic, M. and  Lee, I. and Pappas, G. J.}, 
booktitle={Workshop on Multi VIew Geometry in Robotics (MVIGRO), in conjunction with RSS}, 
title={Robust Localization Using Context-Aware Filtering}, 
year={2015}
}



@article{ivanov_tecs16,
 author = {Ivanov, Radoslav and Pajic, Miroslav and Lee, Insup},
 title = {Attack-Resilient Sensor Fusion for Safety-Critical Cyber-Physical Systems},
 journal = {ACM Transactions on Embedded Computing Systems},
 issue_date = {February 2016},
 volume = {15},
 number = {1},
 month = feb,
 year = {2016},
 issn = {1539-9087},
 pages = {21:1--21:24},
 articleno = {21},
 numpages = {24},
 url = {http://doi.acm.org/10.1145/2847418},
 doi = {10.1145/2847418},
 acmid = {2847418},
 publisher = {ACM},
} 




@incollection{junkil_tacas16,
  title={Scalable Verification of Linear Controller Software},
  author={Park, Junkil and Pajic, Miroslav and Lee, Insup and Sokolsky, Oleg},
  booktitle={Tools and Algorithms for the Construction and Analysis of Systems (TACAS)},
  pages={662--679},
  year={2016},
  publisher={Springer}
}


@INPROCEEDINGS{ivanov_iccps16, 
author={R. {Ivanov} and N. {Atanasov} and J. {Weimer} and M. {Pajic} and A. {Simpao} and M. {Rehman} and G. {Pappas} and I. {Lee}}, 
booktitle={2016 ACM/IEEE 7th International Conference on Cyber-Physical Systems (ICCPS)}, 
title={Estimation of Blood Oxygen Content Using Context-Aware Filtering}, 
year={2016}, 
volume={}, 
number={}, 
pages={1-10}, 
keywords={biomedical measurement;blood;oxygen;paediatrics;surgery;blood oxygen content;context-aware filtering;blood oxygen concentration;noninvasive measurements;binary information;continuous measurements;discrete detection events;physiological variables;Blood;Current measurement;Estimation;Monitoring;Atmospheric modeling;Semiconductor device measurement;Biomedical monitoring}, 
doi={10.1109/ICCPS.2016.7479102}, 
ISSN={null}, 
month={April},}


@INPROCEEDINGS{mangharam_comnets16,
author={R. Mangharam and H. Abbas and M. Behl and K. Jang and M. Pajic and Z. Jiang},
booktitle={2016 8th International Conference on Communication Systems and Networks (COMSNETS)},
title={Three challenges in cyber-physical systems},
year={2016},
pages={1-8},
doi={10.1109/COMSNETS.2016.7440015},
month={Jan},}




@ARTICLE{pajic_tcns17, 
author={M. {Pajic} and I. {Lee} and G. J. {Pappas}}, 
journal={IEEE Transactions on Control of Network Systems}, 
title={Attack-Resilient State Estimation for Noisy Dynamical Systems}, 
year={2017}, 
volume={4}, 
number={1}, 
pages={82-92}, 
keywords={computer network security;cyber-physical systems;integer programming;linear programming;state estimation;attack-resilient state estimation;noisy dynamical systems;cyberphysical systems;bounded-size noise;sensor attacks;mixed-integer linear program;convex relaxation;sound attack detection;sound attack identification;State estimation;Security;Symmetric matrices;Control systems;Noise measurement;Standards;Cyber-physical systems;Attack-resilient state estimation;robustness of state estimators;cyberphysical systems security;linear systems},}


@ARTICLE{pajic_csm17, 
author={M. {Pajic} and J. {Weimer} and N. {Bezzo} and O. {Sokolsky} and G. J. {Pappas} and I. {Lee}}, 
journal={IEEE Control Systems Magazine}, 
title={Design and Implementation of Attack-Resilient Cyberphysical Systems: With a Focus on Attack-Resilient State Estimators}, 
year={2017}, 
volume={37}, 
number={2}, 
pages={66-81}, 
keywords={cyber-physical systems;safety-critical software;security of data;state estimation;attack-resilient cyberphysical systems;attack-resilient state estimators;security;high-profile attacks;critical infrastructure;Maroochy Water breach;industrial systems;StuxNet virus attack;industrial supervisory control and data acquisition system;German steel mill cyberattack;high-assurance military systems;attack vulnerability;RQ-170 Sentinel US drone;CPS;safety-critical control systems;networked control systems;embedded control systems;Security;Actuators;State estimation;Linear systems;Resilience;Real-time systems}, }



@inproceedings{bogdan_codes16,
 author = {Bogdan, Paul and Pajic, Miroslav and Pande, Partha Pratim and Raghunathan, Vijay},
 title = {Making the Internet-of-things a Reality: From Smart Models, Sensing and Actuation to Energy-efficient Architectures},
 booktitle = {Proceedings of the Eleventh IEEE/ACM/IFIP International Conference on Hardware/Software Codesign and System Synthesis},
 series = {CODES '16},
 year = {2016},
 isbn = {978-1-4503-4483-8},
 location = {Pittsburgh, Pennsylvania},
 pages = {25:1--25:10},
 articleno = {25},
 numpages = {10},
 url = {http://doi.acm.org/10.1145/2968456.2973272},
 doi = {10.1145/2968456.2973272},
 acmid = {2973272}
} 


@INPROCEEDINGS{ibrahim_cases16, 
author={M. {Ibrahim} and C. {Boswell} and K. {Chakrabarty} and K. {Scott} and M. {Pajic}}, 
booktitle={2016 International Conference on Compliers, Architectures, and Sythesis of Embedded Systems (CASES)}, 
title={A real-time digital-microfluidic platform for epigenetics}, 
year={2016}, 
volume={}, 
number={}, 
pages={1-10}, 
keywords={biological techniques;bioMEMS;computational complexity;embedded systems;genetics;lab-on-a-chip;microcontrollers;microfluidics;processor scheduling;real-time digital-microfluidic platform;digital-microfluidic biochips;biomolecular assays;fluidic task assignment;dynamic decision-making;quantitative epigenetics;chromatin structure;gene function regulation;biochip design specifications;real-time multiprocessor scheduling;NP-hard problem;resource-limited biochip;embedded microcontroller;Protocols;Real-time systems;Dynamic scheduling;DNA;Gene expression;Algorithm design and analysis}, 
doi={10.1145/2968455.2968516}, 
ISSN={null}, 
month={Oct}
}


 
@INPROCEEDINGS{lesi_etfa16, 
author={V. {Lesi} and Z. {Jakovljevic} and M. {Pajic}}, 
booktitle={2016 IEEE 21st International Conference on Emerging Technologies and Factory Automation (ETFA)}, 
title={Towards Plug-n-Play numerical control for Reconfigurable Manufacturing Systems}, 
year={2016}, 
volume={}, 
number={}, 
pages={1-8}, 
keywords={computerised numerical control;control system synthesis;decentralised control;ISO standards;machine control;machine tools;manufacturing systems;motion control;plug-n-play numerical control;reconfigurable manufacturing system;market condition fluctuation;product diversification;RMS utilization;modular equipment;dynamic controller architecture;control system design approach;reconfigurable machine tool;modularized CNC control;decentralized CNC control;plug-n-play automation system;fully decentralized motion control architecture;individual axis control module;machine configuration;machine control system interoperability;machine control system modularity;machine control system reconfigurability;real-time operating system;wireless networking;low-cost ARM Cortex-M3 MCU;ISO 10791-7 standard;software-in-the-loop testbed;Computer numerical control;Machining;Computer architecture;Interpolation;Motion control;Real-time systems}, 
doi={10.1109/ETFA.2016.7733524}, 
ISSN={null}, 
month={Sep.},}
 


@INPROCEEDINGS{li_iccad16, 
author={Z. {Li} and K. Y. {Lai} and P. {Yu} and K. {Chakrabarty} and M. {Pajic} and T. {Ho} and C. {Lee}}, 
booktitle={2016 IEEE/ACM International Conference on Computer-Aided Design (ICCAD)}, 
title={Error recovery in a micro-electrode-dot-array digital microfluidic biochip}, 
year={2016}, 
volume={}, 
number={}, 
pages={1-8}, 
keywords={bioMEMS;drops;lab-on-a-chip;microelectrodes;microfluidics;attractive technology platform;automating laboratory procedures;biochemistry;droplet volume variation;integrated sensors;micro-electrode-dot-array;MEDA digital microfluidic biochip;analytical chemistry benchmarks;PRISM model checker;MEDA chip fabrication;bioassay;probabilistic-timed-automata;local recovery strategies;on-chip sensors;error-recovery technique;biochemical experiments;droplet manipulation;Sensors;Microelectrodes;Sequential analysis;Real-time systems;Computer architecture;Biological system modeling}, 
doi={10.1145/2966986.2967035}, 
ISSN={1558-2434}, 
month={Nov},}





@ARTICLE{miao_tcns17, 
author={F. {Miao} and Q. {Zhu} and M. {Pajic} and G. J. {Pappas}}, 
journal={IEEE Transactions on Control of Network Systems}, 
title={Coding Schemes for Securing Cyber-Physical Systems Against Stealthy Data Injection Attacks}, 
year={2017}, 
volume={4}, 
number={1}, 
pages={106-117}, 
keywords={actuators;cryptography;cyber-physical systems;encoding;matrix algebra;sensors;state estimation;coding schemes;cyber-physical system;stealthy data injection attacks;sensors;actuators;statistical fault detector;state estimation errors;encryption schemes;communication networks;coding matrix;Encoding;Data models;Estimation;Linear systems;Algorithm design and analysis;Detectors;Kalman filters;Coding;detection;feasible coding matrix;state estimator;stealthy data injection attacks;time-varying coding}, 
doi={10.1109/TCNS.2016.2573039}, 
ISSN={2372-2533}, 
month={March},}



@article{junkil_tcps17,
 author = {Park, Junkil and Ivanov, Radoslav and Weimer, James and Pajic, Miroslav and Son, Sang Hyuk and Lee, Insup},
 title = {Security of Cyber-Physical Systems in the Presence of Transient Sensor Faults},
 year = {2017},
 issue_date = {July 2017},
 publisher = {Association for Computing Machinery},
 address = {New York, NY, USA},
 volume = {1},
 number = {3},
 issn = {2378-962X},
 url = {https://doi.org/10.1145/3064809},
 doi = {10.1145/3064809},
 journal = {ACM Trans. Cyber-Phys. Syst.},
 month = may,
 articleno = {Article 15},
 numpages = {23},
 keywords = {Cyber-physical systems security, fault-tolerance, fault-tolerant algorithms, sensor fusion}
}



@incollection{junkil_tacas17,
author="Park, Junkil and Pajic, Miroslav and Sokolsky, Oleg and Lee, Insup",
title="Automatic Verification of Finite Precision Implementations of Linear Controllers",
bookTitle="Tools and Algorithms for the Construction and Analysis of Systems: 23rd International Conference, TACAS 2017",
year="2017",
publisher="Springer Berlin Heidelberg",
pages="153--169",
doi="10.1007/978-3-662-54577-5_9"
}


@INPROCEEDINGS{fricks_rams17, 
author={R. W. B. {Fricks} and H. H. {Tseng} and M. {Pajic} and K. S. {Trivedi}}, 
booktitle={2017 Annual Reliability and Maintainability Symposium (RAMS)}, 
title={Transient performance availability modeling in high volume outpatient clinics}, 
year={2017}, 
volume={}, 
number={}, 
pages={1-6}, 
keywords={discrete event simulation;electronic health records;health care;maximum likelihood estimation;patient treatment;stochastic processes;transient performance;availability modeling;outpatient clinics;modeling tools;health care performance evaluation;discrete event simulation solutions;non-Markovian stochastic reward nets;academic medical center;glaucoma;clinic process flow diagrams;electronic health record data;maximum likelihood estimation;treatment phases;data fitting;Medical services;Testing;Stochastic processes;Data models;IEEE Fellows;Petri nets;Delays;R&M applications in health care;risk analysis and management;stochastic Petri nets;discrete event simulation}, 
doi={10.1109/RAM.2017.7889777}, 
ISSN={null}, 
month={Jan},}


@INPROCEEDINGS{elfar_iccps17demo, 
author={M. {Elfar} and H. {Zhu} and A. {Raghunathan} and Y. Y. {Tay} and J. {Wubbenhorst} and M. L. {Cummings} and M. {Pajic}}, 
booktitle={2017 ACM/IEEE 8th International Conference on Cyber-Physical Systems (ICCPS)}, 
title={WiP Abstract: Platform for Security-Aware Design of Human-on-the-Loop Cyber-Physical Systems}, 
year={2017}, 
volume={}, 
number={}, 
pages={93-94}, 
keywords={cyber-physical systems;mobile robots;safety;security of data;telerobotics;security-aware design;human-on-the-loop cyber-physical systems;RESCHU-SA;HOL;inductive reasoning;supervisory control of heterogeneous unmanned vehicles;unmanned aerial vehicles;UAV missions;interface usability;human-CPS;supervisory control systems;Security;Feeds;Cyber-physical systems;Privacy;Graphical user interfaces;Supervisory control;Unmanned aerial vehicles;CPS Security;Human-on-the-Loop CPS;Unmanned Aerial Vehicles}, 
doi={}, 
ISSN={null}, 
month={April},}


@INPROCEEDINGS{jovanov_cdc17, 
author={I. {Jovanov} and M. {Pajic}}, 
booktitle={2017 IEEE 56th Annual Conference on Decision and Control (CDC)}, 
title={Sporadic data integrity for secure state estimation}, 
year={2017}, 
volume={}, 
number={}, 
pages={163-169}, 
keywords={authorisation;data integrity;message authentication;state estimation;secure state estimation;network-based attacks;Man-in-the-Middle attacks;standard state estimators;strict integrity requirements;sporadic data integrity enforcement;message authentication;stealthy attacks;unbounded state estimation error;State estimation;Sensor systems;Scheduling;Detectors;Covariance matrices}, 
doi={10.1109/CDC.2017.8263660}, 
ISSN={null}, 
month={Dec},}


@article{lesi_tecs17,
 author = {Lesi, Vuk and Jovanov, Ilija and Pajic, Miroslav},
 title = {Security-Aware Scheduling of Embedded Control Tasks},
 journal = {ACM Trans. Embed. Comput. Syst.},
 issue_date = {October 2017},
 volume = {16},
 number = {5s},
 year = {2017},
 pages = {188:1--188:21},
 articleno = {188},
 numpages = {21}
} 


@INPROCEEDINGS{lesi_rtss17, 
author={V. {Lesi} and I. {Jovanov} and M. {Pajic}}, 
booktitle={2017 IEEE Real-Time Systems Symposium (RTSS)}, 
title={Network Scheduling for Secure Cyber-Physical Systems}, 
year={2017}, 
volume={}, 
number={}, 
pages={45-55}, 
keywords={cryptography;cyber-physical systems;data integrity;integer programming;linear programming;message authentication;scheduling;cyber-physical systems security;message authentication codes;cyber-physical systems attacks;cryptographic tools;runtime bandwidth allocation;opportunistic scheduling;QoC requirements;mixed integer linear programming problem;Quality-of-Control;real-time network messages;data integrity;network scheduling;Sensors;Real-time systems;Authentication;Standards;Data integrity;Automotive engineering;Real-Time-Scheduling;Mixed-Integer-Linear-Programming;CPS-Security;CAN-bus;Quality-of-Control}, 
doi={10.1109/RTSS.2017.00012}, 
ISSN={2576-3172}, 
month={Dec},}



@article{elfar_tecs17,
 author = {Elfar, Mahmoud and Zhong, Zhanwei and Li, Zipeng and Chakrabarty, Krishnendu and Pajic, Miroslav},
 title = {Synthesis of Error-Recovery Protocols for Micro-Electrode-Dot-Array Digital Microfluidic Biochips},
 journal = {ACM Trans. Embed. Comput. Syst.},
 issue_date = {October 2017},
 volume = {16},
 number = {5s},
 year = {2017},
 pages = {127:1--127:22},
 articleno = {127},
 numpages = {22},
} 



@Inbook{jakovljevic_newtech17a,
author="Jakovljevic, Zivana and Mitrovic, Stefan and Pajic, Miroslav",
title="Cyber Physical Production Systems---An IEC 61499 Perspective",
bookTitle="Proceedings of 5th International Conference on Advanced Manufacturing Engineering and Technologies: NEWTECH 2017",
year="2017",
publisher="Springer International Publishing",
pages="27--39",
doi="10.1007/978-3-319-56430-2_3"
}

@Inbook{jakovljevic_newtech17b,
author="Jakovljevic, Zivana and Majstorovic, Vidosav and Stojadinovic, Slavenko and Zivkovic, Srdjan and Gligorijevic, Nemanja and Pajic, Miroslav",
title="Cyber-Physical Manufacturing Systems (CPMS)",
bookTitle="Proceedings of 5th International Conference on Advanced Manufacturing Engineering and Technologies: NEWTECH 2017",
year="2017",
publisher="Springer International Publishing",
pages="199--214",
doi="10.1007/978-3-319-56430-2_14"
}



@ARTICLE{ivanov_tac19, 
author={R. {Ivanov} and N. {Atanasov} and M. {Pajic} and J. {Weimer} and G. J. {Pappas} and I. {Lee}}, 
journal={IEEE Transactions on Automatic Control}, 
title={Continuous Estimation Using Context-Dependent Discrete Measurements}, 
year={2019}, 
volume={64}, 
number={1}, 
pages={238-253}, 
keywords={Bayes methods;convergence;covariance matrices;eigenvalues and eigenfunctions;filtering theory;Gaussian distribution;patient monitoring;probability;sensors;state estimation;statistical analysis;context-dependent discrete measurements;continuous state estimation;discrete context-based measurements;context measurements;state information;standard continuous measurements;threshold-based measurements;recursive context-aware filter;Atmospheric measurements;Particle measurements;Estimation;Context modeling;Robot sensing systems;Context-aware state estimation;discrete context measurements;estimation of blood oxygen content;probit measurement model}, 
doi={10.1109/TAC.2018.2797839}, 
ISSN={2334-3303}, 
month={Jan},}




@ARTICLE{jovanov_tac19, 
author={I. Jovanov and M. Pajic}, 
journal={IEEE Transactions on Automatic Control}, 
title={Relaxing Integrity Requirements for Attack-Resilient Cyber-Physical Systems}, 
year={2019}, 
volume={64}, 
number={12}, 
pages={4843-4858}, 
keywords={State estimation;Detectors;Data integrity;Estimation error;Kalman filters;Cyber-physical systems;Schedules;Attack detection;attack-resilient state estimation;cyber-physical systems (CPS) security;Kalman filtering;linear systems}, 
ISSN={2334-3303}, }



@INPROCEEDINGS{jovanov_cdc18, 
author={I. {Jovanov} and M. {Pajic}}, 
booktitle={2018 IEEE Conference on Decision and Control (CDC)}, 
title={Secure State Estimation with Cumulative Message Authentication}, 
year={2018}, 
volume={}, 
number={}, 
pages={2074-2079}, 
keywords={computer network security;cryptography;data integrity;Kalman filters;message authentication;probability;state estimation;secure state estimation;cumulative message authentication;network-based attacks;Man-in-the-Middle attacks;false data;closed-loop system;intermittently integrity;delivered sensor measurements;standard cryptographic techniques;Message Authentication Codes;computation overhead;cumulative MACs;network overhead;Kalman filter-based state estimators;sequential probability ratio test intrusion detectors;MitM attacks;single cumulative MAC;design-time methodology;reachable state-estimation errors;stealthy attacks;cumulative enforcement policies;data integrity;sensor measurements;cumulative integrity enforcement policy;Detectors;State estimation;Data integrity;Kalman filters;Message authentication;Estimation error}, 
doi={10.1109/CDC.2018.8619250}, 
ISSN={0743-1546}, 
month={Dec},}


@INPROCEEDINGS{jovanov_iccps18demo, 
author={I. {Jovanov} and M. {Nauman} and K. {Kumaravelu} and V. {Lesi} and A. {Zutshi} and W. M. {Grill} and M. {Pajic}}, 
booktitle={2018 ACM/IEEE 9th International Conference on Cyber-Physical Systems (ICCPS)}, 
title={Learning-Based Control Design for Deep Brain Stimulation}, 
year={2018}, 
volume={}, 
number={}, 
pages={349-350}, 
keywords={bioelectric phenomena;biomedical electrodes;brain;diseases;learning (artificial intelligence);neuromuscular stimulation;patient treatment;basal ganglia model;control design;quality-of-control;deep brain stimulation devices;low-voltage electrical stimulation;outperform conventional DBS controllers;design-state exploration;physiologically relevant BG modeling;considered control;temporal pattern;control configuration;design-space exploration;optimization based methods;developed framework exploits machine learning;variable temporal patterns;electrical pulses;suitable control policies;design-time framework;BGM platform;hardware implementation;DBS devices;Parkinson's disease;neurological disorders;Satellite broadcasting;Brain modeling;Real-time systems;Field programmable gate arrays;Indexes;Basal ganglia;Physiology;Deep brain stimulation;model based design;demonstration;medical devices and systems;cyber physical system}, 
doi={10.1109/ICCPS.2018.00048}, 
ISSN={null}, 
month={April},}


@INPROCEEDINGS{jovanov_iccps18, 
author={I. {Jovanov} and M. {Naumann} and K. {Kumaravelu} and W. M. {Grill} and M. {Pajic}}, 
booktitle={2018 ACM/IEEE 9th International Conference on Cyber-Physical Systems (ICCPS)}, 
title={Platform for Model-Based Design and Testing for Deep Brain Stimulation}, 
year={2018}, 
volume={}, 
number={}, 
pages={263-274}, 
keywords={automata theory;biomedical electrodes;brain;diseases;feedback;medical disorders;neural nets;neurophysiology;patient treatment;BGM platform;physiologically relevant responses;DBS effectiveness;neurological disorders;Parkinson's disease;safety-critical nature;integrated design;deep brain stimulation;device testing;enabling real-time model simulation;hybrid-automata representation;neural activation events;nonlinear hybrid automata;physiologically relevant basal-ganglia model;DBS controllers;model-based design framework;Satellite broadcasting;Neurons;Testing;Monitoring;Basal ganglia;Medical treatment;Brain modeling;deep brain stimulation;model based design;medical cyber physical systems}, 
doi={10.1109/ICCPS.2018.00033}, 
ISSN={null}, 
month={April},}


@InProceedings{zhu_chii18,
author="Zhu, Haibei
and Elfar, Mahmoud
and Pajic, Miroslav
and Wang, Ziyao
and Cummings, Mary L.",
editor="Schmorrow, Dylan D.
and Fidopiastis, Cali M.",
title="Human Augmentation of UAV Cyber-Attack Detection",
booktitle="Augmented Cognition: Users and Contexts",
year="2018",
publisher="Springer International Publishing",
address="Cham",
pages="154--167",
abstract="Unmanned aerial vehicles (UAVs) have extensive applications in both civilian and military applications. Nevertheless, the continued development of UAVs has been accompanied by security concerns. UAV navigation systems are potentially vulnerable to malicious attacks that target their Global Positioning System (GPS). Thus, efficient GPS hacking detection with high success rate is paramount. Significant effort has been put into developing autonomous hacking detection techniques. However, little research has considered how a human operator can contribute to the security of such systems. In this paper, we propose a human-autonomy collaborative approach for a single operator of multiple-UAV supervisory control systems, where human geo-location is used to help detect possible UAV cyber-attacks. An experiment was designed and conducted using the RESCHU-SA experiment platform to evaluate this approach. The primary results show that 65{\%} of all experiment sessions reached over 80{\%} success rate in UAV hacking detection, while only 17{\%} of participants lost one or more UAVs because of incorrect hacking detections. These results suggest that such an approach could help achieve better security guarantees for human-in-the-loop autonomous UAV systems that are prone to cyber-attacks.",
isbn="978-3-319-91467-1"
}





@article{miao_automatica18,
title = "A hybrid stochastic game for secure control of cyber-physical systems",
journal = "Automatica",
volume = "93",
pages = "55 - 63",
year = "2018",
issn = "0005-1098",
doi = "https://doi.org/10.1016/j.automatica.2018.03.012",
author = "Fei Miao and Quanyan Zhu and Miroslav Pajic and George J. Pappas",
keywords = "Stochastic game, Secure control, Saddle-point equilibrium"
}

@ARTICLE{li_tcad18, 
author={Z. {Li} and K. Y. {Lai} and J. {McCrone} and P. {Yu} and K. {Chakrabarty} and M. {Pajic} and T. {Ho} and C. {Lee}}, 
journal={IEEE Transactions on Computer-Aided Design of Integrated Circuits and Systems}, 
title={Efficient and Adaptive Error Recovery in a Micro-Electrode-Dot-Array Digital Microfluidic Biochip}, 
year={2018}, 
volume={37}, 
number={3}, 
pages={601-614}, 
keywords={biochemistry;biological techniques;bioMEMS;lab-on-a-chip;linear programming;microfluidics;attractive technology platform;automating laboratory procedures;MEDA biochips;droplet manipulation;real-time sensing ability;chip degradation;biochemical experiments;efficient error-recovery strategy;droplet sensing;error-recovery technique;real-time data;on-chip sensors;local recovery strategies;probabilistic-timed-automata;online synthesis technique;global error recovery;local-recovery time;laboratory experiments;fabricated MEDA chip;microelectrode-dot-array digital microfluidic biochip;adaptive error recovery;DMFB;control flow;integer linear programming-based method;PRISM model checker;Sensors;Microelectrodes;Computer architecture;Real-time systems;Probabilistic logic;Biological system modeling;Digital microfluidics;error recovery;micro-electrode-dot-array (MEDA);online synthesis;optimization}, 
doi={10.1109/TCAD.2017.2729347}, 
ISSN={1937-4151}, 
month={March},}


@ARTICLE{zhu_thms19, 
author={H. {Zhu} and M. L. {Cummings} and M. {Elfar} and Z. {Wang} and M. {Pajic}}, 
journal={IEEE Transactions on Human-Machine Systems}, 
title={Operator Strategy Model Development in UAV Hacking Detection}, 
year={2019}, 
volume={49}, 
number={6}, 
pages={540-549}, 
keywords={Global Positioning System;Hidden Markov models;Computer crime;Task analysis;Unmanned aerial vehicles;Supervisory control;Cyberattack;Cyber-attack detection;hidden Markov model (HMM);human geo-location;human supervisory control;strategy classification;unmanned aerial vehicle (UAV)}, 
doi={10.1109/THMS.2018.2888578}, 
ISSN={2168-2305}, 
month={Dec},}



@INPROCEEDINGS{elfar_icra19, 
author={M. {Elfar} and H. {Zhu} and M. L. {Cummings} and M. {Pajic}}, 
booktitle={2019 International Conference on Robotics and Automation (ICRA)}, 
title={Security-Aware Synthesis of Human-UAV Protocols}, 
year={2019}, 
volume={}, 
number={}, 
pages={8011-8017}, 
keywords={autonomous aerial vehicles;command and control systems;control engineering computing;formal verification;learning (artificial intelligence);military aircraft;stochastic games;security-aware synthesis;human-UAV protocols;collaboration protocols;human-unmanned aerial vehicle;geolocation task;stochastic game-based model;stealthy false-data injection attacks;collected experimental data;human-UAV coalition;H-UAV protocol synthesis;human operators;UAV hidden-information constraint;RESCHU-SA testbed;geolocation strategies;model checkers;command and control systems;Games;Task analysis;Protocols;Geology;Stochastic processes;Security;Global Positioning System}, 
doi={10.1109/ICRA.2019.8794385}, 
ISSN={1050-4729}, 
month={May},}




##############################


@inproceedings{lesi_iotdi19,
 author = {Lesi, Vuk and Jakovljevic, Zivana and Pajic, Miroslav},
 title = {Reliable Industrial IoT-based Distributed Automation},
 booktitle = {Proceedings of the International Conference on Internet of Things Design and Implementation},
 series = {IoTDI '19},
 year = {2019},
 isbn = {978-1-4503-6283-2},
 location = {Montreal, Quebec, Canada},
 pages = {94--105},
 numpages = {12},
 url = {http://doi.acm.org/10.1145/3302505.3310072},
 doi = {10.1145/3302505.3310072},
 acmid = {3310072},
 publisher = {ACM},
} 


@InProceedings{borzoo_isola18,
author="Bonakdarpour, Borzoo
and Deshmukh, Jyotirmoy V.
and Pajic, Miroslav",
editor="Margaria, Tiziana
and Steffen, Bernhard",
title="Opportunities and Challenges in Monitoring Cyber-Physical Systems Security",
booktitle="Leveraging Applications of Formal Methods, Verification and Validation. Industrial Practice",
year="2018",
publisher="Springer International Publishing",
address="Cham",
pages="9--18",
abstract="Technological advances in distributed cyber-physical systems (CPS) will fundamentally alter the way present and future human societies lead their lives. From a security or privacy perspective, a (multi-agent) cyber-physical system is a network of sensors, actuators, and computation nodes, i.e., a system with multiple attack surfaces and latent exploits that originate both through software attacks and physical attacks. In this paper, we argue that we are in pressing need to bring about a paradigm shift in software development for multi-agent CPS. To this end, security and privacy policies should be made a critical ingredient of agent interfaces with a goal of ensuring both localized safety and privacy for each agent, as well as guaranteeing global system safety and security. We present our vision on new theory, algorithms, and tools to foster a culture of secure-by-design multi-agent CPS.",
isbn="978-3-030-03427-6"
}



@InProceedings{elfar_cav19,
author="Elfar, Mahmoud
and Wang, Yu
and Pajic, Miroslav",
title="Security-Aware Synthesis Using Delayed-Action Games",
booktitle="Computer Aided Verification (CAV)",
year="2019",
publisher="Springer International Publishing",
pages="180--199",
isbn="978-3-030-25540-4"
}




@InProceedings{junkil_tacas19,
author="Park, Junkil
and Pajic, Miroslav
and Sokolsky, Oleg
and Lee, Insup",
title="LCV: A Verification Tool for Linear Controller Software",
booktitle="Tools and Algorithms for the Construction and Analysis of Systems (TACAS)",
year="2019",
publisher="Springer International Publishing",
pages="213--225",
}


@ARTICLE{jakovljevic_tcst19, 
author={Z. {Jakovljevic} and V. {Lesi} and S. {Mitrovic} and M. {Pajic}}, 
journal={IEEE Transactions on Control Systems Technology}, 
title={Distributing Sequential Control for Manufacturing Automation Systems}, 
year={2019}, 
volume={}, 
number={}, 
pages={1-9}, 
keywords={Petri nets;Decentralized control;Intelligent sensors;Manufacturing;Actuators;Discrete-event systems;distributed control;petri nets;sequential systems.}, 
doi={10.1109/TCST.2019.2912776}, 
ISSN={2374-0159}, 
month={},}



@inproceedings{lesi_sm2n19,
  title={Towards Resilient and Reliable Distributed Automation for Smart Manufacturing Systems},
  author={V. Lesi and Z. Jakovljevic and M. Pajic},
  booktitle={Workshop on Smart Manufacturing Modeling and Analysis (SM²N),  part of CPS-IoT Week},
  location = {Montreal, Quebec, Canada},
  year={2019},
  month = {April}
}

@inproceedings{elfar2017platform,
  title={Platform for security-aware design of human-on-the-loop cyber-physical systems},
  author={Elfar, Mahmoud and Zhu, Haibei and Raghunathan, Adithya and Tay, Yi Y and Wubbenhorst, Jeffrey and Cummings, ML and Pajic, Miroslav},
  booktitle={Proceedings of the 8th International Conference on Cyber-Physical Systems (ICCPS)},
  year={2017},
  organization={ACM}
}

######################################################################################################################################################################################################################################################################################################################################################################################################

@misc{arsc19,
author = {{CPSL@Duke}},
  title = {{ARSC -- Tool for Synthesis of Attack-Resilient Supervisory Controllers}},
  howpublished = {\url{https://github.com/alperkamil/arsc}},
  year = {2019},
}


@misc{csrl19,
author = {{CPSL@Duke}},
  title = {{CSRL -- Control synthesis for LTL objectives via model-free reinforcement learning,}},
  howpublished = {\url{https://gitlab.oit.duke.edu/cpsl/csrl}},
  year = {2019},
  note = {Accessed: 2019-03-08}
}



@misc{hypersmc19,
author = {{CPSL@Duke}},
  title = {{HyperSMC -- Statistical model checker for hyper probabilistic temporal logics}},
  howpublished = {\url{https://gitlab.oit.duke.edu/cpsl/hypersmc}},
  year = {2019},
}



@misc{smclearning19,
author = {{CPSL@Duke}},
  title = {{SMCLearning -- Statistical model checker for deep-neural-network-based cyber-physical systems}},
  howpublished = {\url{https://gitlab.oit.duke.edu/cpsl/smclearning}},
  year = {2019},
}

@misc{mp_hyper19,
author = {{CPSL@Duke}},
  title = {{MPHyper -- Symbolic motion planner for HyperLTL objectives}},
  howpublished = {\url{https://gitlab.oit.duke.edu/cpsl/mp_hyper}},
  year = {2019},
}




######################################################################################################################################################################################################################################################################################################################################################################################################

@inproceedings{wang_tacas20,
  author    = {Y. Wang and S. Nalluri  and B. Bonakdarpour and M. Pajic},
  title     = {Statistical Model Checking for Probabilistic Hyperproperties},
  booktitle ={26th International Conference on Tools and Algorithms for the Construction and Analysis of Systems (TACAS)},
  year      = {2020},
  note       = {under review, available at \url{https://arxiv.org/abs/1902.04111}},
}

@ARTICLE{wang_tac19, 
author={Y. Wang and A. Bozkurt and M. {Pajic}}, 
journal={IEEE Transactions on Automatic Control}, 
title={Attack-Resilient Supervisory Control of Discrete-Event Systems}, 
year={2019}, 
volume={}, 
number={}, 
numpages={14}, 
note ={under review, available at \url{https://arxiv.org/abs/1904.03264}}
}


@ARTICLE{lesi_tcps20, 
author={V. Lesi and I. Jovanov and M. {Pajic}}, 
journal={ACM Trans. on Cyber-Physical Systems}, 
title={Integrating Security in Resource-Constrained Cyber-Physical Systems}, 
year={2020}, 
volume={}, 
number={}, 
numpages={25}, 
note ={\url{https://arxiv.org/abs/1811.03538}}
}


@ARTICLE{luo_automatica19, 
author={X. Luo and M. Pajic and M. Zavlanos}, 
journal={Automatica}, 
title={A Scalable and Optimal Graph-Search Method for Secure State Estimation}, 
year={2019}, 
volume={}, 
number={}, 
note ={under review, available at \url{https://arxiv.org/abs/1903.10620}}
}

@ARTICLE{lesi_tase19, 
author={V. Lesi and S. Nalluri and Z. Jakovljevic and M. Pajic}, 
journal={IEEE Transactions on Automation Science and Engineering}, 
title={Security-Analysis for Distributed IoT-Based Industrial Automation}, 
year={2019}, 
volume={}, 
number={}, 
note ={under review, available at \url{http://people.duke.edu/~mp275/publications.html}}
}


@ARTICLE{lesi_tii19, 
author={V. Lesi and Z. Jakovljevic and M. Pajic}, 
journal={IEEE Transactions on Industrial Informatics}, 
title={Distributing Numerical Control for Reconfigurable Manufacturing Systems}, 
year={2019}, 
volume={}, 
number={}, 
note ={under review, available at \url{http://people.duke.edu/~mp275/publications.html}}
}

@ARTICLE{jakovljevic_tii19, 
author={Z. Jakovljevic and V. Lesi and  M. Pajic}, 
journal={IEEE Transactions on Industrial Informatics}, 
title={Attacks on Distributed Sequential Control in Manufacturing Automation}, 
year={2019}, 
volume={}, 
number={}, 
note ={under review, available at \url{http://people.duke.edu/~mp275/publications.html}}
}



@article{wang_tecs19,
 author = {Wang, Yu and Zarei, Mojtaba and Bonakdarpour, Borzoo and Pajic, Miroslav},
 title = {Statistical Verification of Hyperproperties for Cyber-Physical Systems},
 journal = {ACM Trans. Embed. Comput. Syst.},
 issue_date = {October 2019},
 volume = {18},
 number = {5s},
 month = oct,
 year = {2019},
 issn = {1539-9087},
 pages = {92:1--92:23},
 articleno = {92},
 numpages = {23},
 url = {http://doi.acm.org/10.1145/3358232},
 doi = {10.1145/3358232},
 acmid = {3358232},
 publisher = {ACM},
 address = {New York, NY, USA},
 keywords = {Cyber-physical systems, embedded control software, hyperproperties, statistical model checking},
}


@INPROCEEDINGS{wang_cdc19a, 
author={Y. Wang and M. {Pajic}}, 
booktitle={2019 IEEE Conference on Decision and Control (CDC)}, 
title={Supervisory Control of Discrete Event Systems in the Presence of Sensor and Actuator Attacks}, 
year={2019},
}

@INPROCEEDINGS{wang_cdc19b, 
author={Y. Wang and M. {Pajic}}, 
booktitle={2019 IEEE Conference on Decision and Control (CDC)}, 
title={Attack-Resilient Supervisory Control with Intermittently Secure Communication}, 
year={2019},
}


@INPROCEEDINGS{lesi_etfa19, 
author={V. {Lesi} and Z. {Jakovljevic} and M. {Pajic}}, 
booktitle={2019 24th IEEE International Conference on Emerging Technologies and Factory Automation (ETFA)}, 
title={Synchronization of Distributed Controllers in Cyber-Physical Systems}, 
year={2019}, 
volume={}, 
number={}, 
pages={710-717}, 
keywords={distributed control;manufacturing systems;motion control;synchronisation;distributed controllers;Cyber-Physical Systems;misaligned clock sources;distributed control;synchronous execution;control algorithms;distributed system components;cyber-synchronization;physical-synchronization;general requirements;specific real-world application-distributed motion control;reconfigurable manufacturing systems;synchronization challenges;synchronization errors;system functionality;low-cost synchronization scheme;cyber-physical synchronization techniques;system design;Synchronization;Clocks;Motion control;Hardware;Frequency control;Protocols;Robots;cyber-physical synchronization;distributed position control;reconfigurable manufacturing systems}, 
doi={10.1109/ETFA.2019.8869467}, 
ISSN={1946-0740}, 
month={Sep.},}


@inproceedings{bozkurt_icra20,
  author    = {A. Bozkurt and Y. Wang and M. Zavlanos  and M. Pajic},
  title     = {Control Synthesis from Linear Temporal Logic Specifications using Model-Free Reinforcement Learning},
  booktitle ={2020 International Conference on Robotics and Automation (ICRA)},
  year      = {2020},
  note       = {to appear, available at \url{http://people.duke.edu/~mp275/publications.html}},
}

@inproceedings{wang_icra20,
  author    = {Y. Wang and S. Nalluri  and M. Pajic},
  title     = {Hyperproperties  for  Robotics:  Motion  Planning  via HyperLTL},
  booktitle ={2020 International Conference on Robotics and Automation (ICRA)},
  year      = {2020},
  note       = {to appear, available at \url{http://people.duke.edu/~mp275/publications.html}},
}

@inproceedings{gao_icra20,
  author    = {Y. Gao and M. Pajic and M. Zavlanos},
  title     = {Deep Imitative Reinforcement Learning for Temporal Logic Robot Motion Planning with Noisy Semantic ObservationsL},
  booktitle ={2020 International Conference on Robotics and Automation (ICRA)},
  year      = {2020},
  note       = {to appear, available at \url{http://people.duke.edu/~mp275/publications.html}},
}

@inproceedings{khazraei_acc20,
  author    = {A. Khazraei and M. Pajic},
  title     = {Perfect Attackability of Linear Dynamical Systems with Bounded Noise},
  booktitle ={2020 American Control Conference (ACC)},
  year      = {2020}
}

@inproceedings{zarei_hscc20,
  author    = {M. Zarei, Y. Wang, and M. Pajic},
  title     = {Statistical Verification of Learning-Based Cyber-Physical Systems},
  booktitle ={23rd ACM International Conference on Hybrid Systems: Computation and Control (HSCC)},
  year      = {2020},
  note       = {to appear, available at \url{http://people.duke.edu/~mp275/publications.html}},
}

@inproceedings{gao_iccps20,
  author    = {Q. Gao, M. Naumann, I. Jovanov, V. Lesi, K. Kumaravelu, W. Grill, and M. Pajic},
  title     = {Model-based Design of Closed-Loop Deep Brain Stimulation Controllers using Reinforcement Learning},
  booktitle ={11th ACM/IEEE International Conference on Cyber-Physical Systems (ICCPS)},
  year      = {2020},
  note       = {to appear, available at \url{http://people.duke.edu/~mp275/publications.html}},
}





\begin{thebibliography}{25}
\providecommand{\natexlab}[1]{#1}
\providecommand{\url}[1]{\texttt{#1}}
\expandafter\ifx\csname urlstyle\endcsname\relax
  \providecommand{\doi}[1]{doi: #1}\else
  \providecommand{\doi}{doi: \begingroup \urlstyle{rm}\Url}\fi

\bibitem[Fawzi et~al.(2014)Fawzi, Tabuada, and Diggavi]{fawzi2014secure}
H.~Fawzi, P.~Tabuada, and S.~Diggavi.
\newblock Secure estimation and control for cyber-physical systems under
  adversarial attacks.
\newblock \emph{IEEE Transactions on Automatic control}, 59\penalty0
  (6):\penalty0 1454--1467, 2014.

\bibitem[Golub and Van~Loan(2012)]{golub2012matrix}
G.~H. Golub and C.~F. Van~Loan.
\newblock \emph{Matrix computations}, volume~3.
\newblock JHU press, 2012.

\bibitem[Jovanov and Pajic(2019)]{jovanov_tac19}
I.~Jovanov and M.~Pajic.
\newblock Relaxing integrity requirements for attack-resilient cyber-physical
  systems.
\newblock \emph{IEEE Transactions on Automatic Control}, 64\penalty0
  (12):\penalty0 4843--4858, 2019.
\newblock ISSN 2334-3303.

\bibitem[Kerns et~al.(2014)Kerns, Shepard, Bhatti, and
  Humphreys]{kerns2014unmanned}
A.~J. Kerns, D.~P. Shepard, J.~A. Bhatti, and T.~E. Humphreys.
\newblock Unmanned aircraft capture and control via gps spoofing.
\newblock \emph{Journal of Field Robotics}, 31\penalty0 (4):\penalty0 617--636,
  2014.

\bibitem[Khazraei and Pajic(2020)]{khazraei_acc20}
A.~Khazraei and M.~Pajic.
\newblock Perfect attackability of linear dynamical systems with bounded noise.
\newblock In \emph{2020 American Control Conference (ACC)}, 2020.

\bibitem[Khazraei et~al.(2017)Khazraei, Kebriaei, and Salmasi]{khazraei2017new}
A.~Khazraei, H.~Kebriaei, and F.~R. Salmasi.
\newblock A new watermarking approach for replay attack detection in lqg
  systems.
\newblock In \emph{56th IEEE Annual Conf. on Decision and Control (CDC)}, pages
  5143--5148, 2017.

\bibitem[Kwon et~al.(2014)Kwon, Liu, and Hwang]{kwon2014analysis}
C.~Kwon, W.~Liu, and I.~Hwang.
\newblock Analysis and design of stealthy cyber attacks on unmanned aerial
  systems.
\newblock \emph{Journal of Aerospace Information Systems}, 11\penalty0
  (8):\penalty0 525--539, 2014.

\bibitem[Langner(2011)]{langner2011stuxnet}
R.~Langner.
\newblock Stuxnet: Dissecting a cyberwarfare weapon.
\newblock \emph{IEEE Security \& Privacy}, 9\penalty0 (3):\penalty0 49--51,
  2011.

\bibitem[Lesi et~al.(2017)Lesi, Jovanov, and Pajic]{lesi_tecs17}
V.~Lesi, I.~Jovanov, and M.~Pajic.
\newblock Security-aware scheduling of embedded control tasks.
\newblock \emph{ACM Trans. Embed. Comput. Syst.}, 16\penalty0 (5s):\penalty0
  188:1--188:21, 2017.

\bibitem[Lesi et~al.(2020)Lesi, Jovanov, and {Pajic}]{lesi_tcps20}
V.~Lesi, I.~Jovanov, and M.~{Pajic}.
\newblock Integrating security in resource-constrained cyber-physical systems.
\newblock \emph{ACM Trans. on Cyber-Physical Systems}, 2020.
\newblock \url{https://arxiv.org/abs/1811.03538}.

\bibitem[Luo et~al.(2019)Luo, Pajic, and Zavlanos]{luo2019scalable}
X.~Luo, M.~Pajic, and M.~M. Zavlanos.
\newblock A scalable and optimal graph-search method for secure state
  estimation.
\newblock \emph{arXiv preprint arXiv:1903.10620}, 2019.

\bibitem[Mo and Sinopoli(2009)]{mo2009secure}
Y.~Mo and B.~Sinopoli.
\newblock Secure control against replay attacks.
\newblock In \emph{47th Annual Allerton Conference on Communication, Control,
  and Computing}, pages 911--918. IEEE, 2009.

\bibitem[Pajic et~al.(2014)Pajic, Weimer, Bezzo, Tabuada, Sokolsky, Lee, and
  Pappas]{pajic_iccps14}
M.~Pajic, J.~Weimer, N.~Bezzo, P.~Tabuada, O.~Sokolsky, I.~Lee, and G.~Pappas.
\newblock Robustness of attack-resilient state estimators.
\newblock In \emph{ACM/IEEE International Conference on Cyber-Physical Systems
  (ICCPS)}, pages 163--174, April 2014.

\bibitem[{Pajic} et~al.(2017{\natexlab{a}}){Pajic}, {Lee}, and
  {Pappas}]{pajic_tcns17}
M.~{Pajic}, I.~{Lee}, and G.~J. {Pappas}.
\newblock Attack-resilient state estimation for noisy dynamical systems.
\newblock \emph{IEEE Transactions on Control of Network Systems}, 4\penalty0
  (1):\penalty0 82--92, 2017{\natexlab{a}}.

\bibitem[{Pajic} et~al.(2017{\natexlab{b}}){Pajic}, {Weimer}, {Bezzo},
  {Sokolsky}, {Pappas}, and {Lee}]{pajic_csm17}
M.~{Pajic}, J.~{Weimer}, N.~{Bezzo}, O.~{Sokolsky}, G.~J. {Pappas}, and
  I.~{Lee}.
\newblock Design and implementation of attack-resilient cyberphysical systems:
  With a focus on attack-resilient state estimators.
\newblock \emph{IEEE Control Systems Magazine}, 37\penalty0 (2):\penalty0
  66--81, 2017{\natexlab{b}}.

\bibitem[Pasqualetti et~al.(2013)Pasqualetti, D{\"o}rfler, and
  Bullo]{pasqualetti2013attack}
F.~Pasqualetti, F.~D{\"o}rfler, and F.~Bullo.
\newblock Attack detection and identification in cyber-physical systems.
\newblock \emph{IEEE Transactions on Automatic Control}, 58\penalty0
  (11):\penalty0 2715--2729, 2013.

\bibitem[Shoukry and Tabuada(2016)]{shoukry2016event}
Y.~Shoukry and P.~Tabuada.
\newblock Event-triggered state observers for sparse sensor noise/attacks.
\newblock \emph{IEEE Trans. on Aut. Control}, 61\penalty0 (8):\penalty0
  2079--2091, 2016.

\bibitem[Shoukry et~al.(2017)Shoukry, Nuzzo, Puggelli, Sangiovanni-Vincentelli,
  Seshia, and Tabuada]{shoukry2017secure}
Y.~Shoukry, P.~Nuzzo, A.~Puggelli, A.~L. Sangiovanni-Vincentelli, S.~A. Seshia,
  and P.~Tabuada.
\newblock Secure state estimation for cyber-physical systems under sensor
  attacks: A satisfiability modulo theory approach.
\newblock \emph{IEEE Transactions on Automatic Control}, 62\penalty0
  (10):\penalty0 4917--4932, 2017.

\bibitem[Smith(2015)]{smith2015covert}
R.~S. Smith.
\newblock Covert misappropriation of networked control systems: Presenting a
  feedback structure.
\newblock \emph{IEEE Control Systems Magazine}, 35\penalty0 (1):\penalty0
  82--92, 2015.

\bibitem[Sundaram and Hadjicostis(2011)]{sundaram2011distributed}
S.~Sundaram and C.~N. Hadjicostis.
\newblock Distributed function calculation via linear iterative strategies in
  the presence of malicious agents.
\newblock \emph{IEEE Transactions on Automatic Control}, 56\penalty0
  (7):\penalty0 1495--1508, 2011.

\bibitem[Sundaram et~al.(2010)Sundaram, Pajic, Hadjicostis, Mangharam, and
  Pappas]{sundaram_cdc10}
S.~Sundaram, M.~Pajic, C.~Hadjicostis, R.~Mangharam, and G.~Pappas.
\newblock {The Wireless Control Network: Monitoring for Malicious Behavior}.
\newblock In \emph{49th IEEE Conference on Decision and Control (CDC)}, pages
  5979--5984, Dec 2010.

\bibitem[Teixeira et~al.(2012)Teixeira, P{\'e}rez, Sandberg, and
  Johansson]{ncs_attack_model}
A.~Teixeira, D.~P{\'e}rez, H.~Sandberg, and K.~H. Johansson.
\newblock Attack models and scenarios for networked control systems.
\newblock In \emph{1st ACM Int. Conf. on High Confidence Netw. Systems}, pages
  55--64, 2012.

\bibitem[Teixeira et~al.(2015)Teixeira, Shames, Sandberg, and
  Johansson]{teixeira2015secure}
A.~Teixeira, I.~Shames, H.~Sandberg, and K.~H. Johansson.
\newblock A secure control framework for resource-limited adversaries.
\newblock \emph{Automatica}, 51:\penalty0 135--148, 2015.

\bibitem[{Y. Mo and B. Sinopoli}(2010)]{mo2010false}
{Y. Mo and B. Sinopoli}.
\newblock False data injection attacks in control systems.
\newblock In \emph{First workshop on Secure Control Systems}, pages 1--6, 2010.

\bibitem[Zetter(2016)]{zetter2016inside}
K.~Zetter.
\newblock Inside the cunning, unprecedented hack of ukraine’s power grid.
\newblock \emph{Wired}, 2016.

\end{thebibliography}

\end{document}